\documentclass[lettersize,journal]{IEEEtran}
\usepackage{amsmath,amsfonts}
\usepackage{algorithmic}
\usepackage[caption=false,font=normalsize,labelfont=sf,textfont=sf]{subfig}
\usepackage{stfloats}
\usepackage{url}
\usepackage{verbatim}
\def\BibTeX{{\rm B\kern-.05em{\sc i\kern-.025em b}\kern-.08em
    T\kern-.1667em\lower.7ex\hbox{E}\kern-.125emX}}
\usepackage{balance}
\usepackage[normalem]{ulem}
\usepackage{multirow}
\usepackage{booktabs}
\usepackage{graphicx}
\usepackage{tabularx}    
\usepackage{array}       
\usepackage{caption}     
\usepackage{hyperref}
\usepackage{scalerel}
\usepackage{xspace}
\usepackage{enumitem}
\usepackage{makecell}
\usepackage{float}
\usepackage{longtable}
\usepackage{cuted}
\usepackage[most]{tcolorbox}
\usepackage[table,xcdraw]{xcolor}  
\hypersetup{
  colorlinks=true,
  linkcolor=black,
  urlcolor=black,
  citecolor=black,
  unicode=true,
}

\definecolor{fpblue}{HTML}{2F80ED}
\definecolor{fpyellow}{HTML}{B59A00}
\definecolor{fpgreen}{HTML}{2E9B45}
\definecolor{fpred}{HTML}{D64545}

\newcommand{\fpcat}[2]{\textcolor{#1}{\textit{\textbf{#2}}}}
\newcommand{\fptag}[2]{%
  \tcbox[
    on line,
    colback=#1!12,
    colframe=#1!55,
    boxrule=0.3pt,
    arc=1pt,
    left=1.5pt,
    right=1.5pt,
    top=0.2pt,
    bottom=0.2pt
  ]{\textcolor{#1!85!black}{\textbf{#2}}}%
}
\newcommand{\fpitem}[4]{%
  \item[\fptag{#1}{#2}] \textcolor{#1!85!black}{\textbf{#3:}} #4%
}

\newtcolorbox{findingbox}{
  colback=lightgray!20,
  colframe=gray!50,
  boxrule=0.5pt,
  left=10pt,
  right=10pt,
  top=10pt,
  bottom=10pt,
  arc=0pt,
  boxsep=0pt
}

\ifCLASSOPTIONcompsoc
  \usepackage[nocompress]{cite}
\else
  \usepackage{cite}
\fi

\begin{document}

\title{LLM-based Vulnerability Detection at Project Scale: An Empirical Study}


\author{Fengjie Li, Jiajun Jiang, Dongchi Chen, Yingfei Xiong
\thanks{Fengjie Li, Jiajun Jiang are with School of Computer Software, College of Intelligence and Computing, Tianjin University, Tianjin, China; Dongchi Chen is with International Joint Institute of Tianjin University, Fuzhou, China; Yingfei Xiong is with the Key Laboratory of High Confidence
Software Technologies (Peking University), Ministry of Education,
School of Computer Science, Peking University, Beijing, China. (E-mail: fengjie@tju.edu.cn; jiangjiajun@tju.edu.cn; chendongchi@tju.edu.cn; xiongyf@pku.edu.cn).}}

\markboth{Journal of \LaTeX\ Class Files,~Vol.~18, No.~9, October~2025}%
{How to Use the IEEEtran \LaTeX \ Templates}

\maketitle

\newcommand{\repoaudit}{RepoAudit}
\newcommand{\knighter}{KNighter}
\newcommand{\iris}{IRIS}
\newcommand{\llmdfa}{LLMDFA}
\newcommand{\inferroi}{INFERROI}
\newcommand{\codeql}{CodeQL}
\newcommand{\semgrep}{Semgrep}
\newcommand{\csa}{CSA}
\newcommand{\spotbugs}{SpotBugs}
\newcommand{\codex}{Codex}
\newcommand{\cc}{Claude Code}

\newcommand{\gptfour}{GPT-4}
\newcommand{\gptfive}{GPT-5.4}
\newcommand{\othree}{O3-mini}
\newcommand{\sonnet}{Claude Sonnet 3.5}
\newcommand{\sonnetfour}{Claude Sonnet 4.6}

\newcommand{\sonnett}{Sonnet 3.5}
\newcommand{\sonnetfourr}{Sonnet 4.6}
\newcommand{\dsfourflash}{DeepSeek V4 Flash}

\newcommand{\openai}{\scalerel*{\includegraphics{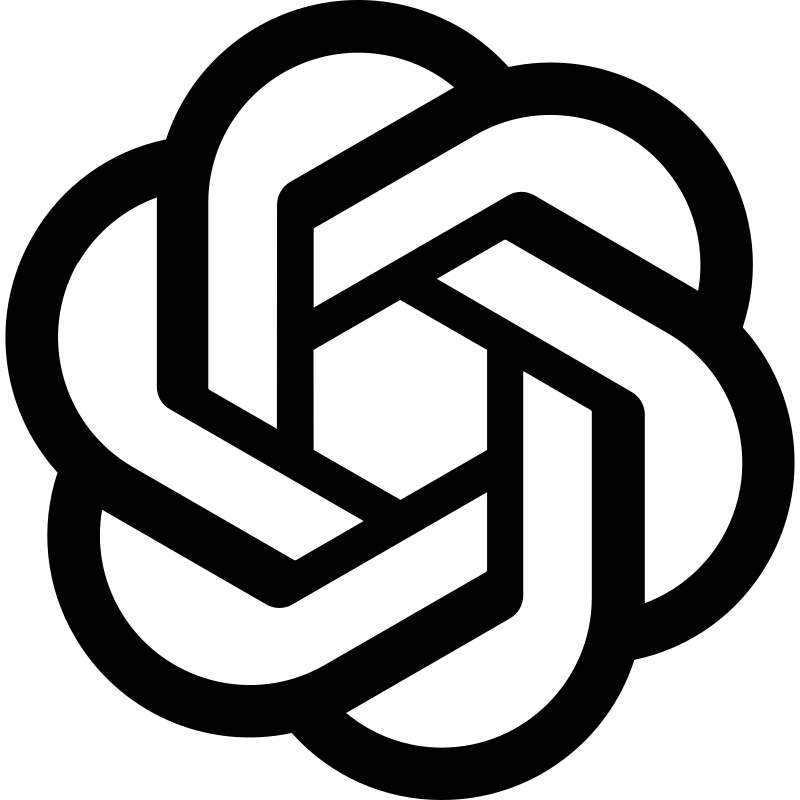}}{\textrm{C}}\xspace}
\newcommand{\claude}{\scalerel*{\includegraphics{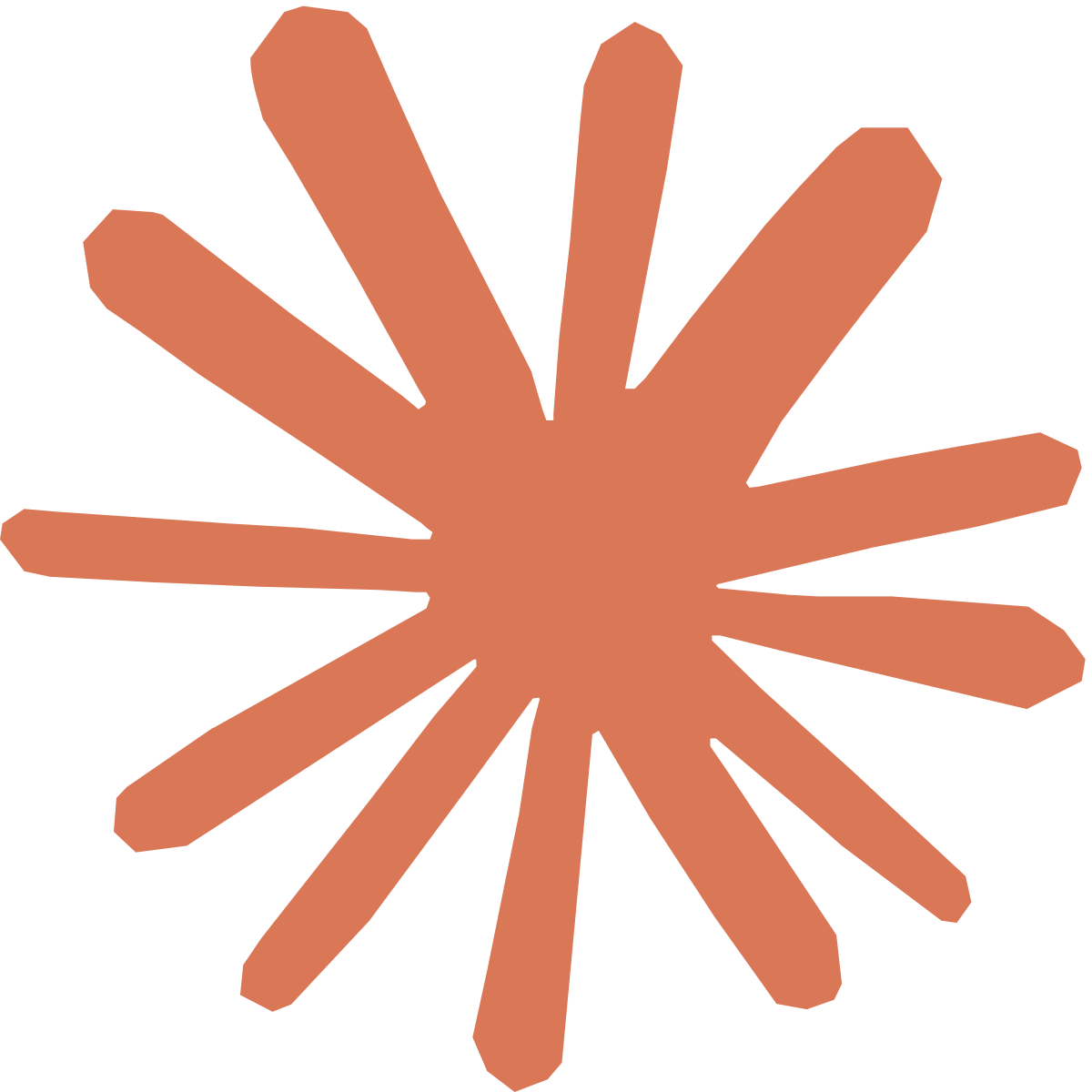}}{\textrm{C}}\xspace}

\renewcommand{\thetable}{\Roman{table}}
\renewcommand{\tablename}{TABLE}

\newcommand{\fj}[1]{\textcolor{orange}{[Fengjie: #1]}}
\newcommand{\jiajun}[1]{\textcolor{cyan}{[Jiajun: #1]}}

\newcommand{\homepage}{\href{https://github.com/Feng-Jay/LLM4Security}{https://github.com/Feng-Jay/LLM4Security}}

\newcommand{\linkcodex}{\href{https://github.com/Feng-Jay/LLM4Security/blob/master/src/core/codex.py}{\add{https://github.com/Feng-Jay/LLM4Security/blob/master/src/core/codex.py}}}

\newcommand{\linkvalidate}{\href{https://github.com/Feng-Jay/reports-filter/blob/master/src/codex.rs}{\add{https://github.com/Feng-Jay/reports-filter/blob/master/src/codex.rs}}}

\newcommand{\todo}[1]{{\color{red}[TODO: #1]}}

\newcommand{\add}[1]{{\color{blue}#1}}
\newcommand{\del}[1]{{\color{red}\sout{#1}}}
\renewcommand{\add}[1]{#1}
\renewcommand{\del}[1]{}

\newcommand{\minoradd}[1]{{\color{blue}#1}}
\newcommand{\minordel}[1]{{\color{red}\sout{#1}}}
\renewcommand{\minoradd}[1]{#1}
\renewcommand{\minordel}[1]{}

\newcommand{\code}[1]{\texttt{#1}}

\begin{abstract}
As software complexity grows, security vulnerabilities become more frequent and harmful, making automated vulnerability detection essential. While recent LLM-based detectors combine semantic reasoning with static analysis for project-scale scanning, their real-world effectiveness remains unclear, and the underlying root causes of their failures are still underexplored. In this paper, we present the first comprehensive empirical study of specialized LLM-based detectors and compare them with traditional static analyzers at the project scale. Specifically, our study evaluates five of the latest and representative \add{specialized} LLM-based methods\add{, two general Agent-based methods} and \del{two}\add{four} traditional tools using: 1) a \minordel{in-house}\minoradd{controlled} benchmark of \del{219}\add{265} known real-world vulnerabilities (C/C++ and Java) to assess detection capability, and 2) 24 active open-source projects\add{/modules}, where we \del{fisrt manually inspected 385 warnings to assess their practical usability and underlying root causes of failures}\add{analyzed 6,442 sampled warnings using \codex{}-assisted labeling with stratified manual validation and classified the root causes of 5,896 false positives using a taxonomy derived from 355 manually inspected reports}.

Our evaluation yields three key findings: First, general agent methods achieve the highest recall, but still struggle with complex code. And both LLM-based and traditional methods suffer from \minordel{source/sink mismatches}\minoradd{incomplete API modeling}. Second, in open-source projects\add{/modules}, both LLM-based and traditional tools generate substantial warnings but suffer from very high false discovery rates, hindering practical use. Our \del{manual }analysis further reveals \del{shallow interprocedural reasoning}\add{truncated interprocedural context} and \minordel{misidentified source/sink pairs}\minoradd{incorrect program-point classification} as primary failure causes, with LLM-based tools exhibiting additional unique failures. Finally, LLM-based methods incur substantial computational costs--hundreds of thousands to hundreds of millions of tokens \add{(cost up to more than \$1,000)} and multi-hour to multi-day runtimes. Overall, our findings underscore critical limitations in the robustness, reliability, and scalability of current LLM-based detectors. We ultimately summarize a set of implications for future research toward more effective and practical project-scale vulnerability detection.
\end{abstract}

\begin{IEEEkeywords}
Vulnerability detection, Large language model, Empirical study.
\end{IEEEkeywords}


\section{Introduction}
\label{sec:intro}
\IEEEPARstart{W}{ith}
the growth of software complexity and deployment scale, the impact of security vulnerabilities becomes more severe. As a result, automated vulnerability detection has become a critical component of modern software development and maintenance. Broadly speaking, existing vulnerability detection approaches can be categorized into static and dynamic techniques~\cite{chess2004static, manes2019art}. Among them, static vulnerability detection has been widely adopted by developers because it can be seamlessly integrated into continuous integration pipelines, operates as long as the code is parsable or compilable, and has good scalability~\cite{li2025automated}.

Traditional static vulnerability detection approaches~\cite{spotbugs, codeql2025, semgrep2025, codeguru, sonarqube, yamaguchi2014modeling, csa} typically encode expert knowledge as manually engineered rules, patterns, and data-flow templates that search for known classes of security problems, such as unsafe data flows from untrusted sources to sensitive sinks~\cite{livshits2005finding} or misuse of security-critical APIs. Although these analyzers are widely used in practice, they require substantial manual effort to design and maintain their rule sets, which is time-consuming and labor-intensive~\cite{smith2015questions, li2025iris, li2025automated}. Moreover, their reliance on predefined patterns inherently limits the generalization and coverage for previously unseen or evolving vulnerability types~\cite{cao2023learning, charoenwet2024empirical}.

Recent advances in Large Language Models (LLMs) have sparked growing interest in LLM-based vulnerability detection. Trained on large corpora of code and natural language, LLMs can interpret code semantics, summarize complex logic, and reason about potential security risks beyond simple syntactic patterns. This has led to an emerging line of LLM-based vulnerability methods~\cite{lu2024grace,sun2024gptscan, zhang2024prompt, jiang2024stagedvulbert, halder2025funcvul}. Unlike traditional tools, these methods feed source code directly to LLMs and rely on the models’ reasoning capabilities instead of manually engineered program analyses.
However, most of these tools focus on function-level or even smaller hunk-level code snippets, where LLMs are given relatively simple prompts to decide whether a vulnerability is present. This limited context and simplistic prompting make the detection particularly sensitive to LLM-induced hallucinations and inconsistent reasoning~\cite{liu2024exploring}.

More recently, a growing line of work has proposed LLM-based vulnerability detectors that combine the semantic reasoning capabilities of LLMs with traditional static analysis workflows to scale vulnerability detection to the project level. These approaches either orchestrate existing static analysis tools with LLMs or use LLMs to emulate static analysis reasoning processes such as dependency and data-flow tracing, including \del{agent-centric frameworks}\add{specially designed agent pipelines} such as \repoaudit{}~\cite{guo2025repoaudit}, \llmdfa{}~\cite{wang2024llmdfa}, LLM-assisted checker generators such as \knighter{}~\cite{yang2025knighter}, \iris{}~\cite{li2025iris}, LLMxCPG~\cite{lekssays2025llmxcpg}, and control-flow guided methods such as \inferroi{}~\cite{wang2025boosting}. Although these tools demonstrate promising results, their evaluations are conducted on disparate datasets using inconsistent and limited settings and metrics, hindering a fair and systematic comparison of their effectiveness in real-world scenarios.

Several studies~\cite{wen2024vuleval, purba2023software, lin2025large, yildiz2025benchmarking, thapa2022transformer, ullah2024llms, gao2023far, steenhoek2024comprehensive, khare2025understanding} have examined the static vulnerability detection capabilities of LLMs, but they either focus primarily on simple prompt-based adaptation of LLMs~\cite{wen2024vuleval, purba2023software, lin2025large} or report detection metrics only on function- or hunk-level datasets~\cite{thapa2022transformer, ullah2024llms, gao2023far, steenhoek2024comprehensive, khare2025understanding}, thereby providing limited evidence of their effectiveness when working at the project scale. The exploration of these more specialized LLM-based detectors across different vulnerability types on real-world projects remains limited, leaving the following several key aspects insufficiently explored: \textbf{1) Effectiveness and complementarity on real-world projects:} These methods were originally evaluated on relatively limited benchmarks and adopt different evaluation settings. As a result, it remains unclear how robust and generalizable these methods perform when applied to large-scale real-world projects that contain diverse vulnerability types and complex project-specific APIs, and how much cross-tool complementarity they can actually provide in practice. \textbf{2) Comprehensive analysis of failure root causes: } Prior studies on LLM-based vulnerability detection primarily report aggregate metrics (e.g., precision, recall, F1) without systematically analyzing the factors that affect their practical usefulness. In particular, the root causes of false positives are rarely categorized or quantified, leaving practitioners with limited guidance on how to improve these tools. \textbf{3) Scalability \& Overhead:} One primary concern for LLM-based methods is their cost and efficiency. However, existing work rarely measures the full computational overhead of LLM-based vulnerability detectors on real-world projects, such as LLM token usage and time overhead across different projects~\cite{wen2024vuleval}. Consequently, there is a pressing need for a systematic study on real-world scenarios that jointly examines the effectiveness, failure root causes, and cost characteristics of these specialized LLM-based vulnerability detectors.

To address these gaps, this paper presents a comprehensive empirical study of specialized project-scale LLM-based vulnerability detectors across multiple CWE types, programming languages, and workflows, aiming to systematically analyze their strengths and weaknesses and to provide insights for more robust and efficient vulnerability detection techniques. Specifically, we evaluate five recent and representative LLM-based vulnerability detectors that are publicly available, span diverse design workflows, together with two widely used \add{general agents, and four }traditional static analyzers. Our evaluation spans two complementary settings: (1) a \minordel{in-house}\minoradd{controlled} benchmark of \textbf{\del{219}\add{265} already-known real-world vulnerabilities \add{curated from existing benchmarks,} covering 8 CWE types in C/C++ and Java}, which we use to assess the vulnerability detection capability of these methods systematically, and (2) \textbf{24} actively maintained open-source projects\add{/modules}, which we use to evaluate their practical usability in real-world scenarios. On the latter, we \del{manually examine \textbf{385} sampled reports with more than \textbf{150} human hours to characterize false positives and their root causes.}\add{first analyzed 6,442 sampled warnings using \codex{}-assisted labeling with stratified manual validation and classified the root causes of 5,896 false positives using a taxonomy derived from 355 manually inspected reports (took more than 150 human hours), and the root-cause labels were further evaluated through stratified manual validation.} For each tool, we systematically study its detection effectiveness, analyze the underlying causes of false positives, and measure the overhead.

Our results reveal several key findings.
\del{\textbf{First, existing LLM-based methods exhibit limited coverage of real-world vulnerabilities.} On the \minordel{in-house}\minoradd{controlled} dataset that contains 219 known vulnerabilities, the evaluated LLM-based methods achieve \textbf{average recalls of 21.09\% and 33.82\%} for C/C++ and Java, respectively, indicating that they miss a large portion of ground-truth vulnerabilities, often due to misaligned source/sink definitions based on manual analysis. Nevertheless, across this \minordel{in-house}\minoradd{controlled} dataset, while different methods present complementary results, the LLM-based methods still uncover more unique vulnerabilities than the traditional methods (including up to 23 unique cases for CWE-722).}\add{\textbf{First, general agent methods achieve the highest recall, but still struggle with complex code.} And both LLM-based and traditional tools suffer from \minordel{source/sink mismatches}\minoradd{incomplete API modeling}, and providing precise \minordel{source/sink specifications}\minoradd{project-specific API specifications} substantially improves detection performance. Nevertheless, across this \minordel{in-house}\minoradd{controlled} dataset, while different methods present complementary results, the LLM-based methods still uncover more unique vulnerabilities than the traditional methods.}
\textbf{Second, both LLM-based and traditional methods exhibit a high false discovery rate.} Both LLM-based and traditional methods produce a large number of false alarms, \del{with even the best-performing tool still reaching an average false discovery rate of \textbf{22.73\%}}\add{with the lowest overall SFDR being 9.09\% for C/C++ and 31.82\% for Java}. \del{Through manual analysis of \textbf{385} sampled reports, we build a taxonomy of false positive reasons and quantify their distribution across tools}\add{We derive the taxonomy from 355 manually inspected reports, use Codex to classify the root causes of 5,896 false positives, and perform stratified manual validation of the generated labels. The results} show that truncated interprocedural context, \minordel{imprecise source/sink identification}\minoradd{incorrect program-point classification}, and overlooked key program points in complex context are the dominant root causes, accounting for \del{136, 69 and 46}\add{861, 3734 and 543} false positives in our study, respectively.
\textbf{Finally, project-scale LLM-based detection remains computationally expensive.} Our cost analysis shows that project-scale LLM-based detection can require between hundreds of thousands and hundreds of millions of tokens per project/module (e.g., up to \textbf{225{,}493.97K} input tokens and \textbf{38{,}078.26K} output tokens \minoradd{with \sonnet{}}), incur substantial monetary costs (up to \$1247.66 per project \minoradd{with the same model}), and take multi-hour to multi-day runtimes (e.g., up to \textbf{4{,}638.00} minutes \minoradd{with \gptfour{}}), making scalability a critical bottleneck. 

In summary, this work makes the following contributions:

\begin{itemize}
\item We conduct, to our knowledge, the first empirical study that compares \textbf{5} specialized LLM-based vulnerability detection tools with \textbf{2} \add{general agents and \textbf{4}} traditional static analyzers at the project scale.

\item We systematically evaluate the detection effectiveness and cross-tool complementarity of these tools on \textbf{\del{219}\add{265}} known vulnerabilities from \textbf{8} CWE types and \textbf{24} open-source projects\add{/modules}.

\item We construct a taxonomy of false positive root causes for both LLM-based and traditional methods\del{, manually label \textbf{\del{385}} sampled reports, and quantify how different design choices (e.g., dataflow construction, source/sink inference, and prompt following) contribute to these failures.}\add{. We first manually label \textbf{\del{385}\add{355}} sampled reports, then use \del{LLMs}\add{\codex{}} to label the remaining reports, and finally perform sampling-based human validation of the \del{LLM-}generated labels. By this process, we quantify how different design choices (e.g., data-flow construction, source/sink inference, and prompt following) contribute to these failures.}

\item We measure the computational overhead of LLM-based methods in terms of token usage\add{, financial cost} and \del{time cost}\add{runtime}, and discuss the implications for practical deployment and future research on more effective and scalable analysis workflows.

\item To foster reproducibility and future research, we release all experimental artifacts, including evaluation scripts, prompts, taxonomy labels, and detailed statistics, on our project homepage: \homepage{}.
\end{itemize}




\section{Background and Related Works}
\label{sec:related}

\subsection{Large Language Models}

Recent advances in Large Language Models (LLMs) have significantly expanded their capability in Software Engineering~(SE) fields. Modern foundation models (e.g., ChatGPT~\cite{openai2023gpt4}, Claude~\cite{anthropic2024claude}, and Gemini~\cite{rohan2023gemini}) are trained on large corpora of source code and natural language, enabling them to generalize across diverse downstream tasks such as code summarization~\cite{ahmed2024automatic}, program synthesis~\cite{dong2025survey}, and vulnerability repair~\cite{fu2022vulrepair}. With in-context learning, LLMs can be adapted to downstream tasks through few-shot examples~\cite{brown2020language} or chain-of-thought (COT) prompting~\cite{wei2022chain}, without requiring fine-tuning. In addition, emerging multi-agent frameworks further enhance LLM utility by decomposing complex tasks into coordinated agents, each responsible for localized subtasks, providing a scalable mechanism to integrate external tools and structured workflows~\cite{wu2024autogen}. These advances have recently motivated a growing number of LLM-based vulnerability detectors that aim to leverage the semantic reasoning capabilities of LLMs for static vulnerability detection.

\subsection{Static Vulnerability Detection}

Static vulnerability detection is a long-standing research topic in software security, which aims to detect vulnerabilities by analyzing source code or intermediate representations without executing the program~\cite{chess2004static}.
Modern static analyzers detect vulnerabilities through pattern matching, dataflow analysis, and \minordel{taint analysis, which traces untrusted inputs (sources) along program paths until they reach security-sensitive operations (sinks).}\minoradd{symbolic execution. Guided by predefined rules/checkers/patterns, they analyze vulnerability-relevant program information, such as data flows from untrusted inputs (sources) to security-sensitive operations (sinks), pointer states, resource lifecycles, and shared-memory accesses and synchronization.}
Representative tools include \codeql{}~\cite{codeql2025}, which treats code as a relational database and detects vulnerabilities using a declarative domain-specific language (DSL). \semgrep{}~\cite{semgrep2025} performs fast pattern-based detection through syntactic and structural matching. SpotBugs~\cite{spotbugs}, which performs bytecode-level pattern and dataflow checks for Java. Other widely used tools include joern~\cite{yamaguchi2014modeling}, SonarQube~\cite{sonarqube}, CodeSonar~\cite{codesonar}, CodeGuru~\cite{codeguru}, Infer~\cite{infer2025}. Despite their good scalability, these tools rely heavily on manually designed rules, \minordel{predefined source and sink specifications}\minoradd{predefined API models}, and language-specific vulnerability templates, which limit their ability to detect previously unseen or evolving vulnerability patterns.

Recently, numerous learning-based and LLM-based vulnerability detectors~\cite{zhou2019devign, li2021vulnerability, fu2022linevul,li2022sysevr,cheng2021deepwukong, lu2024grace, du2024vul, sun2024gptscan, zhang2024prompt, jiang2024stagedvulbert, halder2025funcvul, yang2025dlap} have been proposed to overcome these limitations. \add{These methods include context-enhanced prompting, in-context learning, and RAG-based techniques that retrieve relevant code or vulnerability knowledge to improve LLM-based reasoning. Representative examples include GRACE~\cite{lu2024grace}, prompt-enhanced vulnerability detection~\cite{zhang2024prompt}, and RAG-based methods such as Vul-RAG and DLAP~\cite{du2024vul, yang2025dlap}.}
Although these approaches demonstrate promising performance on benchmark datasets,  they are often evaluated at function- or hunk-level granularity\del{,}\add{. In particular, these methods perform detection on individual functions rather than scanning complete repositories and producing project-level warning reports. Therefore, their results may not directly reflect tool effectiveness, false-positive behavior, or scalability in real-world project-scale settings}\del{and may not reflect their effectiveness in real-world scenarios}~\cite{croft2023data,risse2025top}.
More recently, LLM-based vulnerability detection frameworks have been integrated into static analysis workflows to scale vulnerability detection to the project level. Examples include \repoaudit{}~\cite{guo2025repoaudit} and \llmdfa{}~\cite{wang2024llmdfa}, which use multi-agent workflows or iterative prompting to analyze entire repositories. \knighter{}~\cite{yang2025knighter}, \iris{}~\cite{li2025iris} and LLMxCPG~\cite{lekssays2025llmxcpg}, which utilize LLMs to automatically synthesize static checkers. And \inferroi{}~\cite{wang2025boosting}, which integrates control-flow slicing with LLM-guided root cause inference. These \add{specialized LLM-based} tools indicate a transition toward project-scale vulnerability analysis. However, systematic evaluation of their effectiveness, failure modes, and scalability remains limited.

\subsection{Existing Studies on LLM-based Vulnerability Detection}

Several studies have already explored the use of LLMs for vulnerability detection. Wen et al.~\cite{wen2024vuleval}, Purba et al\minoradd{.}~\cite{purba2023software}, and Lin et al.~\cite{lin2025large} evaluate various LLMs at both function- and project-level using real-world CVEs. However, rather than analyzing full software systems, their evaluation is restricted to dependency-related functions extracted around the specific modification locations of each CVE. The resulting contexts are often incomplete: they cover only a subset of files deemed related to the vulnerability and focus narrowly on code fragments surrounding the known issue. Consequently, such settings cannot faithfully reflect the false-positive behavior and practical effectiveness of LLM-based methods when applied to real-world projects. 
Similarly, Yildiz et al.~\cite {yildiz2025benchmarking} benchmarked multiple LLM-based workflows for vulnerability detection at the commit level. Besides, although some studies~\cite{thapa2022transformer, ullah2024llms, gao2023far, steenhoek2024comprehensive, khare2025understanding} also explored LLM-based vulnerability detection from various perspectives, they primarily operate at function- or hunk-level and lack systematic evaluation in real-world project settings, leaving it unclear how well such methods generalize to large software systems with complex call graphs and project-specific APIs.

\section{Study Design}
\label{sec:study_design}

\subsection{Research Questions}
\label{sec:approach_rqs}

To comprehensively evaluate the effectiveness of existing LLM-based vulnerability detectors and to further understand the key factors that affect their effectiveness, we primarily design and investigate the following four research questions in this study:

\textbf{RQ1: How effective are LLM-based methods in detecting known vulnerabilities in the \minordel{in-house}\minoradd{controlled} benchmark?} In this RQ, we curate a \minordel{in-house}\minoradd{controlled} vulnerability dataset by collecting and filtering existing peer-reviewed benchmarks~\cite{wang2024reposvul, li2025iris, liu2024jleaks} and then evaluate the selected methods on it. Our analysis aims to provide a clear and direct assessment of the detection capabilities of these methods, identifying the key factors that influence their effectiveness.

\textbf{RQ2: How reliable are LLM-based vulnerability detectors when applied to real-world projects\add{/modules} in terms of false discovery rates?} To further investigate the practical effectiveness of LLM-based vulnerability detectors, we apply them to a set of recent real-world projects. In this RQ, rather than comparing the vulnerability detection capability (recall), we focus on the number of warnings these methods report and the number of false positives. This provides an empirical understanding of how these tools perform in realistic development environments.

\textbf{RQ3: What are the root causes of false positives?} In this RQ, we \minordel{manually analyze}\minoradd{combine manual taxonomy construction, LLM-assisted classification, and manual validation to analyze }the causes behind incorrect vulnerability reports produced by these tools in real-world projects\add{/modules}. This analysis aims to better understand the limitations of current LLM-based detectors in real-world scenarios and to offer implications for future research and practical tool development.

\textbf{RQ4: What is the overhead of applying LLM-based vulnerability detectors at the project scale?} Compared with traditional tools, a key concern for LLM-based methods lies in their cost and efficiency. In this RQ, we investigate the runtime and token usage of LLM-based vulnerability detectors to evaluate their scalability and practicality for real-world project-level analysis.

\subsection{Evaluated Methods}
\label{sec:approach_methods}

\begin{table*}[htbp]
  \centering
  \caption{Studied vulnerability detection methods in this paper}
  \resizebox{\textwidth}{!}{
    \begin{tabular}{ccccccc}
    \toprule
          Category & Method & LLM Backbone  & Languages & Evaluated CWE Type & Workflow & \minordel{Source\&Sink Source}\minoradd{API Model Source}\\
    \midrule
    \multirow{5}[0]{*}{\del{LLM-based}\makecell{\add{Specialized}\\\add{LLM-based}}} 
    & RepoAudit~\cite{guo2025repoaudit}& \makecell{\claude{} \sonnet{}\\/ \add{\openai{} \gptfive{}}} & C/C++ & CWE-401, CWE-416, CWE-476 & Multi-Agent & Manual\\
    & Knighter~\cite{yang2025knighter}& \openai{} \othree{} / \add{\openai{} \gptfive{}}  & C/C++ & CWE-401, CWE-416, CWE-476 &  LLM+CSA~\cite{csa} & LLM-inferred\\
    & IRIS~\cite{li2025iris}     & \openai{} \gptfour{} / \add{\openai{} \gptfive{}} & Java  & CWE-022, CWE-078, CWE-079, CWE-094 & LLM+\codeql{} & LLM-inferred\\
    & LLMDFA~\cite{wang2024llmdfa}   & \openai{} \gptfour{} / \add{\openai{} \gptfive{}} & Java  & CWE-078, CWE-079 & Multi-Agent & LLM-inferred\\
    & INFERROI~\cite{wang2025boosting} & \openai{} \gptfour{} / \add{\openai{} \gptfive{}} & Java  & CWE-772 & LLM+CFG & LLM-inferred\\
    \midrule
    \multirow{2}[0]{*}{\add{General Agents}}
    & \add{\codex{}~\cite{codex}} & \add{\openai{} \gptfive{}} & \add{C/C++, Java} & \add{All above} & \add{Multi-Agent} & \add{LLM-inferred}\\
    & \add{\cc{}~\cite{cc}} & \add{\claude{} \sonnetfour{}} & \add{C/C++, Java} & \add{All above} & \add{Multi-Agent} & \add{LLM-inferred}\\
    \midrule
    \multirow{4}[0]{*}{Traditional} 
    & \add{\spotbugs{}~\cite{spotbugs}} &   \add{-}   & \add{Java}  & \add{CWE-022, CWE-078, CWE-079, CWE-094, CWE-772} & \minordel{Checker-driven}\minoradd{Bytecode analysis} & \add{Built-in}\\
    & \add{\csa{}~\cite{csa}} &  \add{-}   & \add{C/C++} & \add{CWE-401, CWE-416, CWE-476} & \minordel{Checker-driven}\minoradd{Symbolic execution} & \add{Built-in}\\
    & CodeQL~\cite{codeql2025}  &   -   & C/C++, Java & All above & \minordel{Checker-driven}\minoradd{Predicate-based querying} & Built-in\\
    & \multirow{2}{*}{Semgrep~\cite{semgrep2025}} & \multirow{2}{*}{-} & \multirow{2}{*}{C/C++, Java} & CWE-401, CWE-416, CWE-476 & \multirow{2}{*}{\minordel{Pattern-driven}\minoradd{Pattern matching}} & \multirow{2}{*}{Built-in} \\ 
    & & & & CWE-022, CWE-078, CWE-079, CWE-094 & & \\
    \bottomrule
    \end{tabular}%
    }
  \label{tab:methods}%
\end{table*}

As shown in Table~\ref{tab:methods}, we selected five of the latest and most representative \add{specialized} LLM-based methods\add{, two general agents} and \del{two}\add{four} traditional static analysis methods that (i) are publicly available, (ii) cover a diverse range of workflows, \add{(iii) support project-scale analysis and can produce concrete vulnerability warnings} and (\del{iii}\add{iv}) can be readily applied across different projects and programming languages. This selection enables us to assess the effectiveness of LLM-based approaches as well as their advantages, limitations, and potential complementarity with existing traditional tools in real-world scenarios. \add{Based on the above criteria, we focus on methods that can directly scan complete repositories and produce project-level warning reports. Therefore, we do not include prompt-only, in-context learning, or RAG-based methods such as GRACE, Vul-RAG, and DLAP as direct baselines, because they are mainly evaluated at the function level and do not directly provide comparable repository-level warning reports in our setting. We also do not include LLMxCPG as a direct baseline because its workflow requires local model finetuning, inference, and non-trivial adaptation to our CWE-specific evaluation setting. Moreover, its CPG-based checker-generation design is close to the checker-synthesis family already represented by \knighter{}.} \add{For specialized LLM-based methods, considering that different backbone LLMs may affect detection capability, we evaluate each method using both the model adopted in its original paper and the strongest recent model available at the time of our study (\gptfive{}) to control for backbone-model effects.} These methods are selected to represent a diverse range of detection workflows, LLM backbones, supported programming languages, and targeted CWE types\footnote{CWE categories and definitions referenced in this paper are based on the official MITRE classification at \url{https://cwe.mitre.org/}.}. In the following, we briefly introduce each of them.

\subsubsection{\add{Specialized} LLM-based Methods}
\begin{itemize}[leftmargin=*]
    \item \textbf{\repoaudit{}~\cite{guo2025repoaudit}} A multi-agent framework initiates scanning from predefined source points and systematically explores program paths by identifying critical callee or caller functions with the assistance of LLMs. When a path reaches a predefined sink point, it reports a potential vulnerability. The reported paths are then subsequently validated by LLMs to reduce false positives and improve the overall performance. 
    \item \textbf{\knighter{}~\cite{yang2025knighter}} An LLM-based checker generation framework that automatically generates Clang Static Analyzer~(CSA)~\cite{csa} checkers from a vulnerability fixing commit. The generation process includes few-shot prompting, iterative regeneration with syntax validation and repair, and refinement steps to ensure high detection precision. Since \knighter{} relies on the CSA, it can detect multiple types of vulnerabilities in C/C++. 
    \item \textbf{\iris{}~\cite{li2025iris}} \iris{} utilizes LLMs to automatically identify and label all methods and their corresponding parameters within a project as potential sources or sinks. This information is then embedded into predefined \codeql{} query templates to detect possible vulnerabilities. The paths flagged as potentially vulnerable by these queries are then re-evaluated by LLMs to reduce false positives. 
    \item \textbf{\llmdfa{}~\cite{wang2024llmdfa}} An agent-centric method initiates scanning from source program points \minordel{defined}\minoradd{identified} by the LLM. Unlike \repoaudit{}, which will explore program paths on demand, \llmdfa{} will explore all possible paths. When a path reaches a sink point \minordel{defined}\minoradd{identified} by the LLM, it reports a potential vulnerability path. These paths are further validated for reachability by using LLMs or the Z3 solver~\cite{de2008z3}. 
    \item \textbf{\inferroi{}~\cite{wang2025boosting}} This method is specifically designed to detect resource leak vulnerability~(CWE-772). It first employs LLMs to infer the intent of program statements~(resource acquisition, release, or check). Then it explores program paths based on the control flow graph~(CFG). If a path exists where a resource is acquired but never released, it is reported as a potential vulnerability. 
\end{itemize}

\subsubsection{\add{General Agents}}
\begin{itemize}[leftmargin=*]
\item \add{\textbf{\codex{}}~\cite{codex} and \textbf{\cc{}}~\cite{cc} are general-purpose coding agents capable of repository inspection, command execution, file editing, and multi-step reasoning from natural-language instructions. We use them as general-agent baselines to evaluate whether broadly capable coding agents can perform vulnerability detection without a specialized security-analysis pipeline.}
\end{itemize}

\subsubsection{Traditional Static Methods}
\begin{itemize}[leftmargin=*]
    \item \textbf{\codeql{}~\cite{codeql2025}} A semantic, \minordel{query-based}\minoradd{predicate-based querying} static analysis framework widely used in industry for vulnerability detection. It models source code as a relational dataset and uses logic-style queries to identify relevant patterns across control-flow and data-flow paths. It supports multiple programming languages, including C/C++ and Java, and provides a rich collection of built-in security queries covering a broad range of CWEs. In this study, we apply \codeql{} to C/C++ and Java vulnerability types supported by the evaluated LLM-based vulnerability detectors, enabling a consistent and comparable assessment across all methods.
    
    \item \textbf{\semgrep{}~\cite{semgrep2025}} A lightweight, pattern-based static analysis tool that identifies vulnerabilities using rule-based matching over Abstract Syntax Trees~(ASTs). It also supports many programming languages, including C/C++ and Java, and offers an extensive rule ecosystem maintained by the community and security researchers. Similarly, we apply \semgrep{} to C/C++ and Java vulnerabilities that are supported by both the LLM-based tools and the available \semgrep{} rule set.
    
    \item \add{\textbf{\spotbugs{}~\cite{spotbugs}} A bytecode-level static analysis tool for Java that detects bug patterns and security-relevant defects through a collection of predefined detectors. \spotbugs{} analyzes compiled Java bytecode and reports potential issues such as null dereferences, resource leaks, and unsafe API usages. In this study, we apply \spotbugs{} together with Find Security Bugs~\cite{findsecuritybugs}, its recommended security plugin, to Java vulnerability types covered by their detector sets, providing a traditional Java-specific baseline.}

    \item \add{\textbf{Clang Static Analyzer~(CSA)~\cite{csa}} is a path-sensitive symbolic-execution-based static analyzer for C/C++ that tracks program states along feasible execution paths and reports defects such as null dereferences, use-after-free errors, and memory leaks. We include CSA as a C/C++ static-analysis baseline and apply it to vulnerability types supported by its analysis rules.}
\end{itemize}

\subsection{Datasets}
\label{sec:approach_datasets}



\begin{table*}[t]
\centering
\caption{Dataset overview and illustrative CWE examples}
\label{tab:datasets_in}
\small
\setlength{\tabcolsep}{3pt}
\renewcommand{\arraystretch}{1.15}
\begin{tabular}{@{}ccclc@{}}
\toprule
Language & CWE & Cases
& \minoradd{Illustrative Example}
& Benchmark \\
\midrule

\multirow{3}{*}{C/C++}
& CWE-401 & \del{40}\add{50}
& \minoradd{Missing \code{free} after \code{malloc}}
& \multirow{3}{*}{ReposVul~\cite{wang2024reposvul}} \\

& CWE-416 & \del{10}\add{25}
& \minoradd{Reading a freed buffer via \code{memcpy}}
& \\

& CWE-476 & \del{14}\add{35}
& \minoradd{Dereferencing \code{p} when \code{p == NULL}}
& \\

\midrule

\multirow{5}{*}{Java}
& CWE-022 & 49
& \minoradd{Path traversal through \code{FileInputStream}}
& \multirow{4}{*}{CWE-Bench-Java~\cite{li2025iris}} \\

& CWE-078 & 11
& \minoradd{Injected commands executed by \code{Runtime.exec}}
& \\

& CWE-079 & 28
& \minoradd{Unescaped HTML output via \code{JspWriter.print}}
& \\

& CWE-094 & 17
& \minoradd{Injected code executed by \code{ScriptEngine.eval}}
& \\

& CWE-772 & 50
& \minoradd{Missing \code{close} after \code{FileReader} creation}
& JLeaks~\cite{liu2024jleaks} \\

\bottomrule
\end{tabular}%
\end{table*}

\begin{table}[tbp]
  \centering
    \caption{Real-world projects\add{/modules} used in this study}
    \resizebox{\columnwidth}{!}{
    \begin{tabular}{ccccc}
    
    \toprule
    CWE   & Repository Name & Commit & Size (LOC) & Stars \\
    \midrule
          & linux/sound & 6093a68 & 1,253,234 & 207K \\
    CWE-401 & linux/mm & 6093a68 & 133,952 & 207K \\
          & ImageMagick & 3bf1076 & 680,878 & 14.9K \\
    \midrule
          & linux/net & 6093a68 & 962,143 & 207K \\
    CWE-416 & linux/drivers/net & 6093a68 & 3,924,621 & 207K \\
          & vim   & 8feaa94 & 1,192,925 & 39.3K \\
    \midrule
          & linux/drivers/peci & 6093a68 & 1,715 & 207K \\
    CWE-476 & gpac  & c6a72c3 & 859,528 & 3.1K \\
          & bitlbee & 8af06ca & 33,650 & 632 \\
    \midrule
          & OpenOLAT & df53b85 & 1,907,083 & 394 \\
    CWE-022 & spark & 1973e40 & 11,941 & 9.7K \\
          & Dspace & 3b24801 & 448,030 & 1K \\
    \midrule
          & xstream & a22d3af & 76,339 & 752 \\
    CWE-078 & workflow-cps-plugin & e4b9517 & 238,588 & 179 \\
          & tika  & 2b38ed1 & 228,309 & 3.4K \\
    \midrule
          & xwiki-platform & fc73498 & 5,553,098 & 1.2K \\
    CWE-079 & jenkins & 69d5a54 & 300,491 & 24.7K \\
          & keycloak & 4f55b9b6 & 6,033,855 & 31K \\
    \midrule
          & onedev & b8a4d7cd & 552,486 & 14.5K \\
    CWE-094 & activemq & 56ea235e & 611,964 & 2.4K \\
          & cron-utils & bac6e866 & 16,244 & 1.2K \\
    \midrule
          & sql2o & 744b85b3 & 8,027 & 1.2K \\
    CWE-\del{722}\add{772} & RxJava & f07765ed & 325,394 & 48.4K \\
          & jsoup & acafbcf3 & 38,128 & 11.3K \\
    \bottomrule
    \end{tabular}%
    }
  \label{tab:datasets_real}%
\end{table}%

As illustrated in our research questions, we evaluate the selected methods in two scenarios: a \minordel{in-house}\minoradd{controlled} dataset containing projects that are verified to contain known vulnerabilities, and a real-world dataset composed of the latest open-source projects\add{/modules}. 
\minordel{Since different methods support diverse CWE types, we choose a subset of CWE-categories~(as shown in Table~\ref{tab:methods}) covered by most of them, to enable a fair comparison and meaningful analysis across methods. This choice is also practical in real-world settings, where extending these LLM-based tools (many of which target only a limited set of CWEs) to new CWE types often requires substantial engineering effort (e.g., redesigning rules, prompts, or analysis pipelines), rather than being a simple configuration change. The details of the studied datasets are presented in Table~\ref{tab:datasets_in} and Table~\ref{tab:datasets_real}.}\minoradd{We select eight CWE categories based on implementation support (Table~\ref{tab:methods}) and benchmark availability. They cover path traversal and injection, memory safety, and resource management. We evaluate methods within their supported categories: for example, \inferroi{} targets resource leaks, while \iris{} uses CWE-specific query templates. Extending these tools may require substantial effort and new analysis logic, which we leave to future work. Dataset details appear in Table~\ref{tab:datasets_in} and Table~\ref{tab:datasets_real}.}

\begin{itemize}[leftmargin=*]
    \item \textbf{\minordel{In-house}\minoradd{Controlled} dataset:} We construct a \minordel{in-house}\minoradd{controlled} dataset by consolidating vulnerabilities from three peer-reviewed benchmarks, providing a controlled setting to assess the detection effectiveness of the evaluated methods: ReposVul~\cite{wang2024reposvul}, CWE-Bench-Java~\cite{li2025iris}, and JLeaks~\cite{liu2024jleaks}.\minordel{ReposVul is an automatically curated dataset built using LLMs and static analysis tools, containing 6{,}134 CVE entries covering 236 CWE types across 1{,}491 open-source projects.}
    
    For C/C++, \minordel{portion of ReposVul, we select, for each CWE, vulnerabilities from the three repositories containing the most cases. This strategy preserves broad vulnerability coverage while limiting compilation overhead, as different versions of a repository often require separate build and configuration effort.}\minoradd{we use ReposVul, curated using LLMs and static analysis and containing 6{,}134 CVEs across 236 CWE types and 1{,}491 projects. We select vulnerabilities from the three repositories with the most cases for each evaluated CWE, limiting build and configuration effort across repository versions.}
    
    For Java, CWE-Bench-Java provides manually curated Java vulnerabilities from real-world projects. From this benchmark, we select vulnerable cases from Maven projects to guarantee successful building and analysis. JLeaks is a high-quality benchmark focusing on Java resource leak vulnerabilities and has been manually validated. From JLeaks, we randomly sample 50 representative cases. All selected cases from these three sources are further manually checked to ensure correctness, validity, and successful compilation.
    
    \item \textbf{Real-world projects\add{/modules}:} To complement the \minordel{in-house}\minoradd{controlled} dataset\minordel{ and avoid data leakage}, we construct a real-world dataset by selecting actively maintained open-source projects\add{/modules} that have historically exhibited vulnerabilities of the studied CWE types. As summarized in Table~\ref{tab:datasets_real}, these projects\add{/modules} span both C/C++ and Java and cover diverse domains such as operating systems, multimedia processing, networking, data platforms, and web frameworks. \minoradd{We use the latest versions of actively maintained software with historical vulnerabilities in the studied CWE types, including Linux, ImageMagick, and Vim for C/C++~\cite{wang2024reposvul}, and XStream, Jenkins, and Keycloak for Java~\cite{li2025iris}.}\minordel{For C/C++ projects\add{/modules}, we select the latest versions of widely used software (e.g., the Linux kernel, ImageMagick, Vim) that historically contain vulnerabilities of the corresponding CWE types, ensuring that the chosen commits are recent and the projects\add{/modules} remain actively maintained. Similarly, for Java projects, we select the latest versions of well-maintained projects (e.g., XStream, Jenkins, Keycloak) that historically contain vulnerabilities of the studied CWE types, and ensure that each project can be successfully built and analyzed.} \minoradd{Of the 24 project/module entries, 21 use commits postdating the reported August 2025 knowledge cutoffs of \gptfive{}~\cite{gpt-5.4} and \sonnetfour{}~\cite{claude46}.} This dataset provides a realistic and diverse foundation for evaluating the scalability and practical effectiveness of the evaluated methods.
\end{itemize}

\subsection{Evaluation Metrics}
\label{sec:approach_metrics}

To evaluate the effectiveness of the studied methods, we adopt several standard detection metrics by following existing studies~\cite{purba2023software, wen2024vuleval, lin2025large, yildiz2025benchmarking, thapa2022transformer, ullah2024llms, gao2023far, steenhoek2024comprehensive, khare2025understanding}, and additionally consider tool costs in terms of token usage and time overhead.

\noindent\textbf{\minordel{In-house}\minoradd{Controlled} dataset.} For the curated \minordel{in-house}\minoradd{controlled} dataset, each vulnerable project is labeled with the actual vulnerable program point(s). Consequently, we consider a vulnerability successfully detected by a tool as long as it reports the associated vulnerable point(s) (\textit{a.k.a}, true positive: TP), otherwise, the vulnerability is missed by the tool (\textit{a.k.a}, false negative: FN). To measure precision, we classify reported warnings as false positives (FPs) if their associated data flows involve known vulnerable files but do not match the data-flow patterns of known vulnerabilities, as defined by the IRIS setup~\cite{li2025iris}. \minoradd{We made this decision for two reasons. First, as our experiments will show, the studied methods have relatively high false positive rates. Consequently, we are confident that many of these FPs are actually not vulnerabilities while in theory they have the potential to be other unknown genuine vulnerabilities. Second, in a later experiment, we will analyze a large set of warnings reported by the same tools on other real-world projects or modules. Therefore, to keep the experiments manageable and avoid repetitive work, we focus on the known vulnerabilities in this controlled dataset.}
\del{In particular, since the non-vulnerable code space in the projects is huge and not fully labeled, we focus on recall rather than full precision on this dataset:}\add{We then compute recall and precision as follows:} \[
\mathrm{Recall} = \frac{\mathrm{TP}}{\mathrm{TP} + \mathrm{FN}} \times 100\%.
\quad
\mathrm{\add{Precision}} = \frac{\mathrm{\add{TP}}}{\mathrm{\add{TP}} + \mathrm{\add{FP}}} \times \add{100\%}.
\]

\noindent\textbf{Real-world projects\add{/modules}.} For these real-world open-source projects\add{/modules}, it is unknown whether there are vulnerabilities in them. For each tool and project\add{/module}, we therefore report:  
\textbf{1) \#Reports}, the total number of warnings produced, \textbf{2) \#Files}, the number of distinct source files that contain at least one warning, and \textbf{3) \#Sampled FPs}, the number of FPs (benign instances incorrectly reported as vulnerabilities by the tools) among the sampled warnings. Since manually validating a warning is time-consuming and labor-intensive, it often requires inspecting project-specific APIs, transitive dependencies, and relevant control and data flows across the entire project\add{/module}. Therefore, \add{we utilize \codex{} to automatically label whether each reported warning is a true vulnerability or not.}\del{two authors first sampled up to 10 warnings per tool per project and independently labeled each as a true positive or false positive, resolving disagreements through discussion. In total, we manually examined \textbf{{385}} sampled warnings across all tools and projects, which required more than \textbf{150} human hours.} Based on these labels, we estimate the Sampled False Discovery Rate (SFDR) as
\[
\mathrm{SFDR} = \frac{\mathrm{\#Sampled\;FPs}}{\mathrm{\#Sampled\;Reports}}
\]

\add{To facilitate a comprehensive root cause analysis of false positives reported by existing approaches, up to 10 warnings per tool per project were initially sampled by two authors. The identified causes were then summarized to formulate an initial taxonomy and corresponding classification codebook. This manual effort involved the examination of \textbf{\add{355}} sampled warnings, consuming more than \textbf{150} person-hours. Based on this taxonomy, \codex{} was subsequently used to label the remaining warnings. To verify the reliability of these automated results, \codex{}'s outputs, for both the labeling and the root-cause classification phases, were manually checked against a sampled subset of instances. The details will be introduced in Section~\ref{sec:results_rq2} and \ref{sec:results_rq3}.}

\noindent\textbf{Token usage and time overhead.}
For RQ4, we quantify the computational overhead of LLM-based methods by measuring the input and output token consumption of each tool, \add{financial cost,} as well as the end-to-end detection time per project\add{/module} (Section~\ref{sec:results_rq4}).

\section{Experimental Setup}
\label{sec:setup}

For specialized LLM-based vulnerability detection methods, we directly use their released replication packages and follow hyperparameters (e.g., function call-chain exploration depth, temperature, and top-p) recommended in the original papers\minoradd{, imposing no additional timeouts, token or monetary budgets}. \add{To assess the impact of model choice, we evaluate each method with both its original backbone model and the latest model, \gptfive{}.} All experiments are conducted via the official OpenAI and Anthropic Claude APIs~\cite{o3mini,gpt-4,claude35, gpt-5.4, claude46}. \minordel{For tools that provide default source--sink configurations (e.g., \repoaudit{}), we consistently use the tool's default source--sink configuration for both the in-house dataset and the real-world projects/modules. For tools whose sources and sinks are inferred by LLMs, we directly adopt the definitions generated by the corresponding models.}\minoradd{For \repoaudit{} that predefined vulnerability-related program points, we use their default configurations on both datasets. For \iris{}, \llmdfa{}, and \inferroi{}, we use the API lists generated by their respective LLMs. For \knighter{}, we generate CSA checkers for each CWE using the corresponding bug-fixing patches from its original study (18 patches in total).}

\add{For the general-agent baselines, \codex{} and \cc{}, we use their official CLI interfaces with built-in tools and skills enabled, and disable all Internet search capabilities to ensure a controlled evaluation setting. All experiments are conducted through the official OpenAI and Anthropic APIs. The evaluation prompt is available at our project\footnote{\linkcodex{}}.}

For traditional static vulnerability detection methods, we manually curate the predefined rule set by selecting only those rules that are relevant to the target CWE types evaluated in this study. Specifically, for \codeql{}, we finally selected 48 predefined rules that are relevant to our target CWE types from the official \codeql{} rule set, such as \path{CWE-401/MemoryLeakOnFailedCallToRealloc.ql} and \path{Critical/MemoryMayNotBeFreed.ql} for CWE-401. For \semgrep{}, we finally selected 59 predefined rules that are relevant to our target CWE types from the official rule sets (e.g., the official \texttt{XSS} rules for CWE-079). \add{For \spotbugs{}, together with the Find Security Bugs~\cite{findsecuritybugs} plugin, we iterated over the detectors and bug patterns from its official list and the Find Security Bugs plugin, mapped them to our target Java CWE types, and finally obtained \textbf{21} relevant predefined patterns. For CSA, we selected \textbf{12} built-in C/C++ checkers relevant to our target CWE types, including memory allocation/deallocation-related checkers, nullness checkers, and \texttt{StdCLibraryFunctions} checkers.} More details about the selected rules\add{/patterns/checkers} are available in our
Appendix~B
and on our homepage~\cite{homepage}. Although different rules\add{/patterns/checkers} may produce reports with varying severity levels (e.g., some are marked as errors and others as recommendations), we treat all reports equally in this study.

All experiments are conducted on a local server equipped with dual Intel Xeon 6388 CPUs, 512 GB of RAM, running Ubuntu 20.04.6 LTS. All LLM API requests, as well as \codeql{}\del{and }\add{,} \semgrep{}\add{, \spotbugs{} and \csa{}} analyses, are executed on this server in our experiments.
\section{Result Analysis}
\label{sec:results}

\begin{table}[tbp]
  \centering
  \caption{Effectiveness of evaluated tools on the C/C++ \minordel{in-house}\minoradd{controlled} dataset.}
  \small
  \setlength{\tabcolsep}{4pt}
  \renewcommand{\arraystretch}{1}
    \resizebox{\columnwidth}{!}{
    \begin{tabular}{@{}llrrrr@{}}
    \toprule
    \add{CWE}   & \add{Method} & \add{\#Warn.} & \add{\#Det.} & \add{Prec.} & \add{Rec.} \\
    \midrule
    \multicolumn{1}{c}{\multirow{9}[6]{*}{\makecell{\add{CWE-401}\\\add{Total: 50}}}}
          & \add{\repoaudit{}$_{\claude{}\sonnett{}}$} & \add{0}     & \add{0}     & \add{--}     & \add{0.00} \\
          & \add{\repoaudit{}$_{\openai{}\gptfive{}}$} & \add{0}     & \add{0}     & \add{--}     & \add{0.00} \\
          & \add{\knighter{}$_{\openai{}\othree{}}$} & \add{0}     & \add{0}     & \add{--}     & \add{0.00} \\
          & \add{\knighter{}$_{\openai{}\gptfive{}}$} & \add{0}     & \add{0}     & \add{--}     & \add{0.00} \\
\cmidrule{2-6}          
          & \add{\codex{}$_{\openai{}\gptfive{}}$} & \add{81}    & \add{34}    & \add{41.98}    & \add{68.00} \\
          & \add{\cc{}$_{\claude{}\sonnetfourr{}}$} & \add{100}    & \add{40}    & \add{40.00}    & \add{80.00} \\
\cmidrule{2-6}          
          & \add{\codeql{}} & \add{21}     & \add{3}     & \add{14.29}     & \add{6.00} \\
          & \add{\semgrep{}}& \add{0}     & \add{0}     & \add{--}     & \add{0.00} \\
          & \add{\csa{}} & \add{1}     & \add{0}     & \add{0.00}     & \add{0.00} \\
\midrule
    \multicolumn{1}{c}{\multirow{9}[6]{*}{\makecell{\add{CWE-416}\\\add{Total: 25}}}} 
          & \add{\repoaudit{}$_{\claude{}\sonnett{}}$} & \add{33}     & \add{0}     & \add{0.00}     & \add{0.00} \\
          & \add{\repoaudit{}$_{\openai{}\gptfive{}}$} & \add{23}     & \add{0}     & \add{0.00}     & \add{0.00} \\
          & \add{\knighter{}$_{\openai{}\othree{}}$} & \add{0}     & \add{0}     & \add{--}     & \add{0.00} \\
          & \add{\knighter{}$_{\openai{}\gptfive{}}$} & \add{0}     & \add{0}     & \add{--}     & \add{0.00} \\
\cmidrule{2-6}          
          & \add{\codex{}$_{\openai{}\gptfive{}}$} & \add{22} & \add{10} &\add{45.45}& \add{40.00} \\
          & \add{\cc{}$_{\claude{}\sonnetfourr{}}$} & \add{51} & \add{9} & \add{17.65}& \add{36.00} \\
\cmidrule{2-6}
          & \add{\codeql{}}  & \add{0}     & \add{0}     & \add{--}     & \add{0.00} \\
          & \add{\semgrep{}}  & \add{2}     & \add{0}     & \add{0.00}     & \add{0.00} \\
          & \add{\csa{}}  & \add{11}     & \add{0}     & \add{0.00}     & \add{0.00} \\
\midrule
    \multicolumn{1}{c}{\multirow{9}[6]{*}{\makecell{\add{CWE-476}\\\add{Total: 35}}}}
          & \add{\repoaudit{}$_{\claude{}\sonnett{}}$} & \add{7} & \add{0} & \add{0.00} & \add{0.00} \\
          & \add{\repoaudit{}$_{\openai{}\gptfive{}}$} & \add{55} & \add{2} & \add{3.64} & \add{5.71} \\
          & \add{\knighter{}$_{\openai{}\othree{}}$} & \add{0}     & \add{0}     & \add{--}     & \add{0.00} \\
          & \add{\knighter{}$_{\openai{}\gptfive{}}$} & \add{0}     & \add{0}     & \add{--}     & \add{0.00} \\
\cmidrule{2-6}          
          & \add{\codex{}$_{\openai{}\gptfive{}}$} & \add{60} & \add{22} & \add{36.67} & \add{62.86} \\
          & \add{\cc{}$_{\claude{}\sonnetfourr{}}$} & \add{93} & \add{16} & \add{17.20} & \add{45.71} \\
\cmidrule{2-6}          
          & \add{\codeql{}} & \add{1110}     & \add{12}     & \add{1.08}     & \add{34.29} \\
          & \add{\semgrep{}} & \add{0}     & \add{0}     & \add{--}     & \add{0.00} \\
          & \add{\csa{}} & \add{11}    & \add{0}     & \add{0.00}    & \add{0.00} \\
    \bottomrule
    \end{tabular}%
    }
  \label{tab:inhouse_results_c}%
\end{table}%

\subsection{RQ1: \minordel{Effectiveness in Detecting In-house Vulnerabilities}\minoradd{Detection Effectiveness on the Controlled Dataset}}
\label{sec:results_rq1}

\noindent\textbf{\textit{Results on C/C++ vulnerabilities.}} Table~\ref{tab:inhouse_results_c} demonstrates the detection results of evaluated methods on different CWE types. \del{Overall, LLM-based methods achieve an average recall of 21.09\% (27/128). In particular, \repoaudit{} consistently achieves the highest recall across all C/C++ vulnerabilities, detecting 55.00\% of CWE-401 cases, 10.00\% of CWE-416 cases, and 28.57\% of CWE-476 cases. In contrast, \knighter{}, \codeql{}, and \semgrep{} fail to detect any instances in these three categories.}\add{Overall, the two general agent methods achieve an average recall of 59.55\% (131/220) and an average precision of 32.19\% (131/407). In particular, \codex{} detects the largest number of vulnerabilities\minordel{, achieving recalls of 68.00\% for CWE-401, 40.00\% for CWE-416, and 62.86\% for CWE-476}. 
In contrast, among the two specialized LLM-based methods, \knighter{} fails to detect any vulnerabilities in these three CWE categories, while \repoaudit{}$_{\openai{}\gptfive{}}$ detects only two. Their performance is worse than that of some traditional tools, such as \codeql{}, which detects 3 CWE-401 and 12 CWE-476 vulnerabilities.} This consistently low recall across \del{tools}\add{specialized LLM-based methods} suggests limited generalizability beyond their original evaluation settings. In particular, for linux kernel–related vulnerabilities, \repoaudit{} was evaluated on only 4 known intra-function cases in its original study, while \knighter{} did not report recall against known ground truth\del{, }\add{. These limitations}\del{which} motivate\del{s} us further to investigate the underlying causes of their missed detections \add{and the substantial performance gap between specialized LLM-based methods and general-purpose agents} on this dataset.

\begin{figure}[tbp] 
\centering 
\includegraphics[width=\columnwidth]{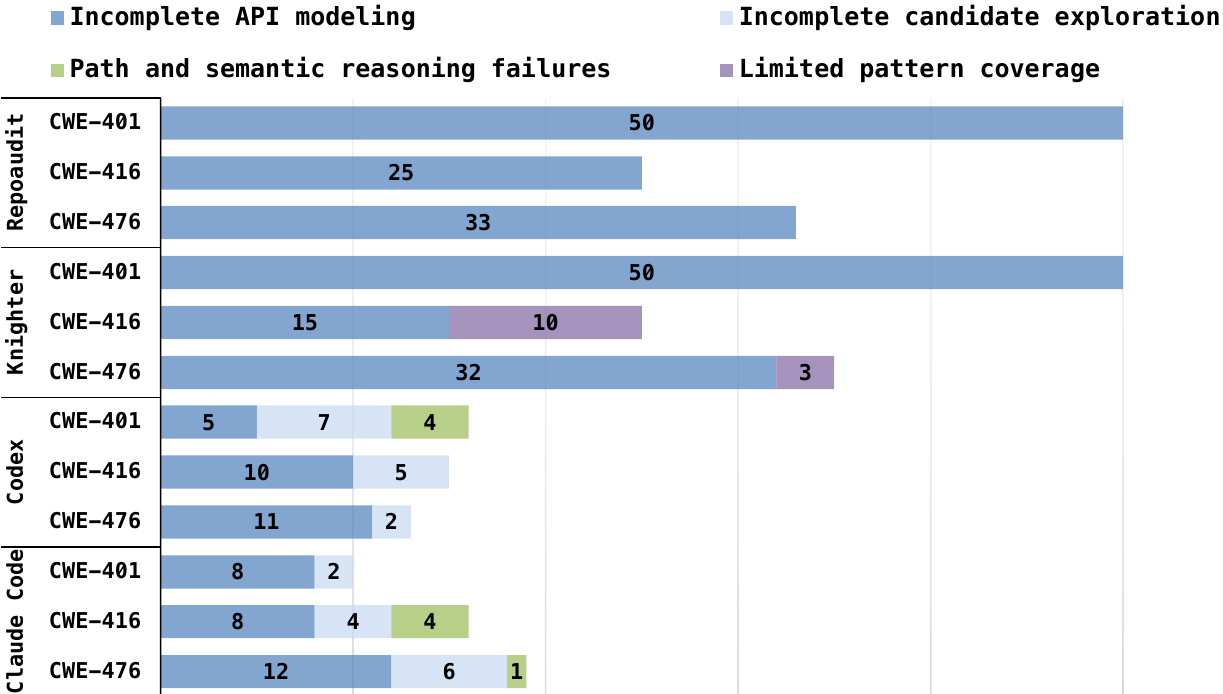} 
\caption{Reasons for undetected C/C++ vulnerabilities.} 
\label{fig:rq1_reasons} 
\end{figure}

Manual analysis (Fig.~\ref{fig:rq1_reasons}) \minordel{shows that \textbf{most missed detections stem from incorrect source/sink API specifications or incomplete source/sink discovery.} Specifically, the source/sink specifications used during detection often fail to match the APIs used in the target projects. Even when the LLM correctly identifies these APIs, it may fail to examine all candidate sites, thereby overlooking the vulnerability location.}\minoradd{identifies \textbf{incomplete API modeling and incomplete candidate exploration as major causes of missed detections}. Incomplete API modeling occurs when rules, generated checkers, or LLM reasoning omit relevant APIs or their roles, such as resource allocation and release. For source/sink-based analyses, this includes omitting a source, a sink, or both. Incomplete candidate exploration occurs when relevant APIs are recognized, but vulnerability-relevant call sites are not examined.} Our analysis attributes 108/108, 97/110, 40/44, and 40/45 missed vulnerabilities for \repoaudit{}$_{\openai{}\gptfive{}}$, \knighter{}$_{\openai{}\gptfive{}}$, \codex{}, and \cc{}, respectively, to these two issues. Moreover, since \knighter{} generates CSA checkers from existing patches, its checkers encode not only data-flow information but also patch-specific patterns. \minordel{Despite \knighter{}'s generalization process, 13 vulnerabilities are missed because the relevant source APIs are matched, but the encoded patterns are not.}\minoradd{\knighter{} additionally misses 13 vulnerabilities due to limited pattern coverage. Its generated checkers include the relevant APIs, but their encoded pattern conditions do not cover the vulnerable usage in these cases.} For \cc{} and \codex{}, several additional misses arise from incorrectly handling complex control-flow graphs.

\begin{figure*}[htbp] 
\centering 
\includegraphics[width=0.8\textwidth]{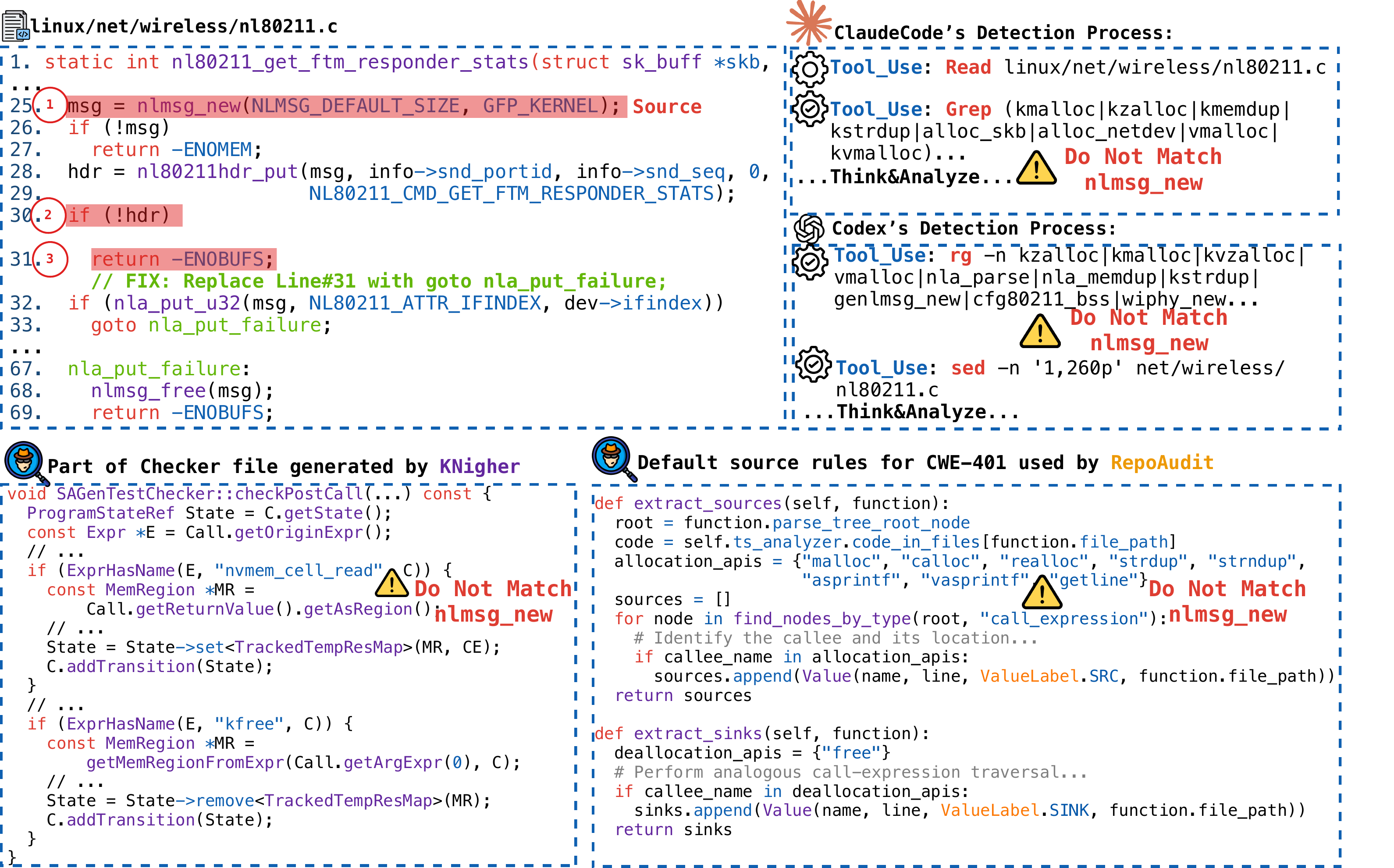} 
\caption{\add{Illustrating example of why these tools fail to detect the vulnerability.}}
\label{fig:rq1_example} 
\end{figure*}

\begin{figure}[tbp] 
\centering 
\includegraphics[width=0.8\columnwidth]{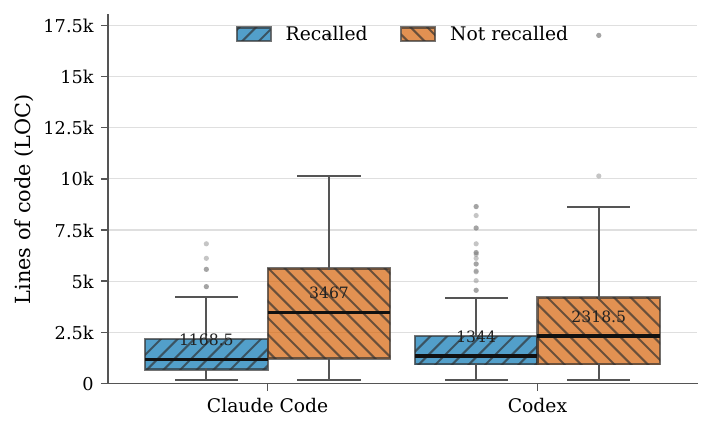} 
\caption{\add{LOC distribution for recalled vs. not-recalled C/C++ vulnerabilities}}
\label{fig:rq1_box_plot_c} 
\end{figure}

Here is an example shown in Fig.~\ref{fig:rq1_example}, the target CWE-401 vulnerability is triggered because the function directly returns in the \texttt{if-true} branch (line \del{17}\add{31}) without releasing the \texttt{\del{skb}\add{msg}} object allocated by the project-specific API \texttt{\del{dev\_alloc\_skb}\add{nlmsg\_new}} (line \del{7}\add{25}). However, all \del{four selected tools}\add{four methods shown in the figure} cannot detect it. 
For example, \cc{} and \codex{} first inspect the source file and execute commands such as \texttt{grep} and \texttt{rg} to search for potential APIs. However, our analysis of their sessions shows that neither identifies the project-specific allocation API \texttt{nlmsg\_new}, likely because of the file's complexity and size (more than 17,000 lines of code). Consequently, their subsequent reasoning fails to detect the vulnerability. More generally, we collect the files containing the source and sink APIs of both recalled and not recalled vulnerabilities and measure their LOC. As shown in Fig.~\ref{fig:rq1_box_plot_c}, missed vulnerabilities tend to involve larger files, suggesting that file size and complexity hinder general-purpose agents’ ability to detect vulnerabilities through simple file inspection and API searches.

\begin{figure}[tbp] 
\centering 
\includegraphics[width=0.8\columnwidth]{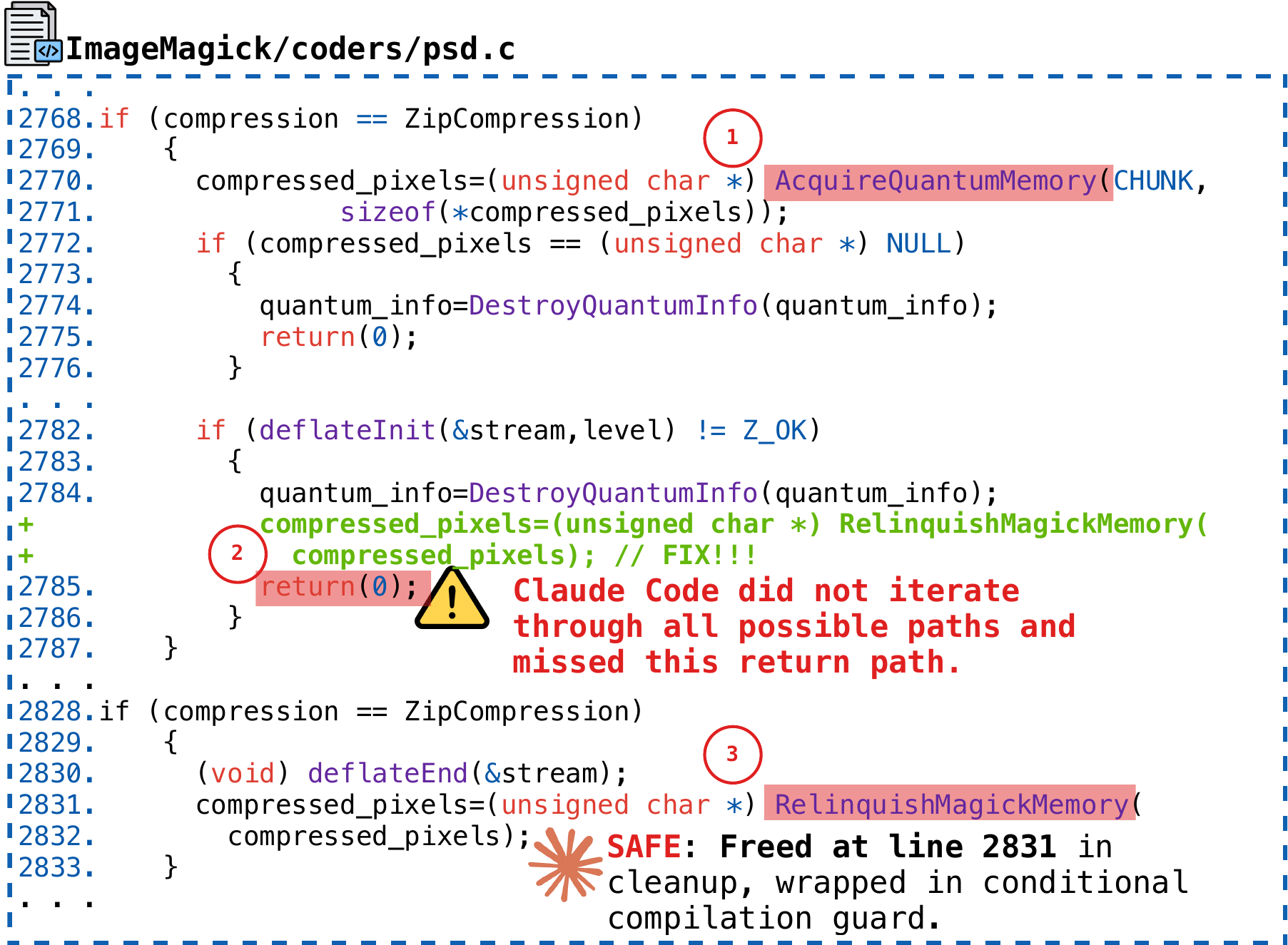} 
\caption{\add{\cc{} missed vulnerability due to incomplete path exploration.}} 
\label{fig:rq1_cc_not_recalled} 
\end{figure}

Similarly, CSA checkers generated by \knighter{} focus exclusively on matching limited deallocation functions (e.g., \texttt{kfree}), while this design enables customization, extending these checkers in practice requires familiarity with CSA libraries, macros, and the tool’s analysis workflow, and typically involves writing and debugging substantial C code (often more than 100 lines). By default, \repoaudit{} models only a limited set of common allocation and deallocation APIs. Supporting project-specific APIs requires developers to understand the relevant AST structures and libraries and to configure and refine the corresponding modeling rules, thereby increasing the barrier to practical adoption.

\add{Fig.~\ref{fig:rq1_cc_not_recalled} shows another example of path and semantic reasoning failures. The vulnerability is triggered because \texttt{compressed\_pixels}, allocated at Line~2770, is not cleaned up before the early \texttt{return (0)} at Line~2785. However, \cc{} incorrectly treats the path as safe by relying on a later release at Line~2831, which is unreachable once the function returns early.}

\begin{table}[t]
  \centering
  \caption{\add{CWE-401 detection performance with and without
  perfect \minordel{source/sink}\minoradd{allocation/deallocation} API specifications.}}
  \label{tab:api_guided}
  \resizebox{\columnwidth}{!}{
  \begin{tabular}{lrrr}
    \toprule
    \textbf{\add{Method}} &
    \textbf{\add{Default Det./Rec.}} &
    \textbf{\add{Perfect Det./Rec.}} &
    \textbf{\add{$\Delta$ Rec.}} \\
    \midrule
    \add{\repoaudit{}$_{\openai{}\gptfive{}}$} & \add{0 / 0.00\%}  & \add{24 / 48.00\%} & \add{+48.00} \\
    \add{\codex{}$_{\openai{}\gptfive{}}$}     & \add{34 / 68.00\%} & \add{45 / 90.00\%} & \add{+22.00} \\
    \add{\cc{}$_{\claude{}\sonnetfourr{}}$}    & \add{40 / 80.00\%} & \add{48 / 96.00\%} & \add{+16.00} \\
    \add{\codeql{}}                            & \add{3 / 6.00\%}  & \add{28 / 56.00\%} & \add{+50.00} \\
    \bottomrule
  \end{tabular}
  }
\end{table}

\noindent\textbf{\textit{\minoradd{Results with API guidance.}}} \add{Given the substantial impact of \minordel{source/sink API identification}\minoradd{API modeling}, we conduct an additional API-guided experiment on CWE-401 vulnerabilities. Specifically, we provide \repoaudit{}$_{\openai{}\gptfive{}}$, \codex{}, \cc{}, and \codeql{} with project-specific allocation and deallocation APIs while keeping all other settings unchanged. We exclude \knighter{}, \csa{}, and \semgrep{} because their checkers encode hard-coded, pattern-specific information and cannot be readily adapted to new API specifications. As shown in Table~\ref{tab:api_guided}, perfect API specifications substantially improve all four methods. \codeql{} exhibits the largest absolute recall improvement, increasing recall by 50.00 percentage points, followed by \repoaudit{}$_{\openai{}\gptfive{}}$ with 48.00 percentage points. The recall of \codex{} and \cc{} increases to 90.00\% and 96.00\%, respectively. These results confirm that \minordel{inaccurate or incomplete source/sink API identification}\minoradd{incomplete API modeling} is a major cause of missed detections. 

Nevertheless, \repoaudit{}$_{\openai{}\gptfive{}}$ and \codeql{} still miss 26 and 22 vulnerabilities, respectively, indicating that accurate API specifications alone cannot address limitations in subsequent data-flow and control-flow analysis. For \repoaudit{}$_{\openai{}\gptfive{}}$, the remaining misses mainly arise because it formulates vulnerability detection as a path-feasibility problem and uses a \textbf{fixed few-shot prompt} to ask the LLM to generate formatted path descriptions. Our manual inspection shows that these extracted paths are often incomplete or fail to be parsed correctly by the downstream analysis logic, causing true vulnerabilities to be missed. For \codeql{}, although the supplied APIs improve \minordel{source/sink coverage}\minoradd{the completeness of API modeling}, the built-in queries are primarily centered on \textbf{stack-local} ownership and treat field or global assignments as ownership escapes. As a result, it often fails to report cases where a field-owned object is not properly cleaned up along a specific error path. Thus, these results show that accurate \minordel{source/sink API specifications are}\minoradd{API modeling is} critical, but \minordel{they are}\minoradd{it is} not sufficient on \minordel{their}\minoradd{its} own. Existing methods with relatively fixed detection logic can still constrain the capability of the underlying model. For example, \repoaudit{}$_{\openai{}\gptfive{}}$ and \codex{} use the same model, yet \codex{} achieves higher recall in our setting. This suggests that workflow design, such as flexible agentic exploration versus fixed prompt-and-parsing pipelines, can substantially affect how effectively the model's reasoning capability is used.}

\begin{table*}[t]
  \centering
  \caption{\add{Effectiveness of evaluated tools on the Java \minordel{in-house}\minoradd{controlled} dataset.}}
  \label{tab:inhouse_results_java}
  \small
  \setlength{\tabcolsep}{4pt}
  \renewcommand{\arraystretch}{1.05}
  \resizebox{\textwidth}{!}{
  \begin{tabular}{clrrrr|clrrrr}
    \toprule
    \add{CWE} & \add{Method} & \add{Warn.} & \add{Det.} & \add{Prec.} & \add{Rec.}
      & \add{CWE} & \add{Method} & \add{Warn.} & \add{Det.} & \add{Prec.} & \add{Rec.} \\
    \midrule
    \multirow{11}{*}{\makecell{\add{CWE-078}\\\add{Total: 11}}}
     & \multirow{2}{*}{\add{\iris{}$_{\openai{}\gptfour{}}$}} & \multirow{2}{*}{\add{33}} & \multirow{2}{*}{\add{2}} & \multirow{2}{*}{\add{6.06}} & \multirow{2}{*}{\add{18.18}} & \multirow{7}{*}{\makecell{\add{CWE-022}\\\add{Total: 49}}}
      & \add{\iris{}$_{\openai{}\gptfour{}}$} & \add{381} & \add{24} & \add{6.30} & \add{48.98} \\
      
      &                                &   &   &   & 
      & & \add{\iris{}$_{\openai{}\gptfive{}}$} & \add{305} & \add{23} & \add{7.54} & \add{46.94} \\
      
      & \add{\iris{}$_{\openai{}\gptfive{}}$} & \add{5} & \add{1} & \add{20.00} & \add{9.09}
      & & \add{\codex{}$_{\openai{}\gptfive{}}$} & \add{63} & \add{37} & \add{58.73} & \add{75.51} \\
      
      & \add{\llmdfa{}$_{\openai{}\gptfour{}}$} & \add{21} & \add{0} & \add{0.00} & \add{0.00}
      & & \add{\cc{}$_{\claude{}\sonnetfourr{}}$} & \add{118} & \add{44} & \add{37.29} & \add{89.80} \\

      & \add{\llmdfa{}$_{\openai{}\gptfive{}}$} & \add{0} & \add{0} & \add{--} & \add{0.00}
      & & \add{\codeql{}} & \add{199} & \add{16} & \add{8.04} & \add{32.65} \\
      
      & \add{\codex{}$_{\openai{}\gptfive{}}$} & \add{5} & \add{2} & \add{40.00} & \add{18.18}
      & & \add{\semgrep{}} & \add{2075} & \add{6} & \add{0.29} & \add{12.24} \\
      
      & \add{\cc{}$_{\claude{}\sonnetfourr{}}$} & \add{12} & \add{5} & \add{41.67} & \add{45.45}
      & & \add{\spotbugs{}} & \add{910} & \add{16} & \add{1.76} & \add{32.65} \\
      \noalign{\vskip 0.3ex}
     \cline{7-12}
     \noalign{\vskip 0.3ex}

      & \add{\codeql{}} & \add{1} & \add{0} & \add{0.00} & \add{0.00} & \multirow{7}{*}{\makecell{\add{CWE-094}\\\add{Total: 17}}}
      & \add{\iris{}$_{\openai{}\gptfour{}}$} & \add{238} & \add{6} & \add{2.52} & \add{35.29} \\
      
      & \add{\semgrep{}} & \add{5} & \add{1} & \add{20.00} & \add{9.09}
      & & \add{\iris{}$_{\openai{}\gptfive{}}$} & \add{51} & \add{4} & \add{7.84} & \add{23.53} \\
      
      & \multirow{2}{*}{\add{\spotbugs{}}} & \multirow{2}{*}{\add{5}} & \multirow{2}{*}{\add{2}} & \multirow{2}{*}{\add{40.00}} & \multirow{2}{*}{\add{18.18}}
      & & \add{\codex{}$_{\openai{}\gptfive{}}$} & \add{11} & \add{7} & \add{63.64} & \add{41.18} \\
      
      &                                &   &   &   & 
      & & \add{\cc{}$_{\claude{}\sonnetfourr{}}$} & \add{38} & \add{15} & \add{39.47} & \add{88.24} \\
      
    \cline{1-6}
    \noalign{\vskip 0.3ex}\multirow{9}{*}{\makecell{\add{CWE-079}\\\add{Total: 28}}}
      & \add{\iris{}$_{\openai{}\gptfour{}}$}& \add{149} & \add{11} & \add{7.38} & \add{39.29}
      & & \add{\codeql{}} & \add{0} & \add{0} & \add{--} & \add{0.00} \\
      
      & \add{\iris{}$_{\openai{}\gptfive{}}$} & \add{82} & \add{5} & \add{6.10} & \add{17.86}
      & & \add{\semgrep{}} & \add{50} & \add{0} & \add{0.00} & \add{0.00} \\
      
      & \add{\llmdfa{}$_{\openai{}\gptfour{}}$} & \add{9} & \add{0} & \add{0.00} & \add{0.00}
      & & \add{\spotbugs{}} & \add{15} & \add{2} & \add{13.33} & \add{11.76} \\
      \noalign{\vskip 0.3ex}
      \cline{7-12}
      \noalign{\vskip 0.3ex}
       & \add{\llmdfa{}$_{\openai{}\gptfive{}}$} & \add{5} & \add{1} & \add{20.00} & \add{3.57} 
       & \multirow{6}{*}{\makecell{\add{CWE-772}\\\add{Total: 50}}} 
       & \add{\inferroi{}$_{\openai{}\gptfour{}}$} & \add{197} & \add{31} & \add{15.74} & \add{62.00} \\
      
      & \add{\codex{}$_{\openai{}\gptfive{}}$} & \add{38} & \add{14} & \add{36.84} & \add{50.00}
      & & \add{\inferroi{}$_{\openai{}\gptfive{}}$} & \add{161} & \add{33} & \add{20.50} & \add{66.00} \\
      
      & \add{\cc{}$_{\claude{}\sonnetfourr{}}$} & \add{107} & \add{18} & \add{16.82} & \add{64.29}
      & & \add{\codex{}$_{\openai{}\gptfive{}}$} & \add{75} & \add{39} & \add{52.00} & \add{78.00} \\
    
      & \add{\codeql{}} & \add{11} & \add{1} & \add{9.09} & \add{3.57}
      & & \add{\cc{}$_{\claude{}\sonnetfourr{}}$} & \add{107} & \add{45} & \add{42.06} & \add{90.00} \\
      
      & \add{\semgrep{}} & \add{42} & \add{0} & \add{0.00} & \add{0.00}
      & & \add{\codeql{}} & \add{10} & \add{5} & \add{50.00} & \add{10.00} \\

      & \add{\spotbugs{}} & \add{16} & \add{0} & \add{0.00} & \add{0.00}
      & & \add{\spotbugs{}} & \add{4} & \add{3} & \add{75.00} & \add{6.00} \\
    \bottomrule
  \end{tabular}
  }
\end{table*}

\noindent\textbf{\textit{Results on Java vulnerabilities.}} For Java vulnerabilities, effectiveness varies across tools and CWE types. \add{Overall, the general-agent methods, \codex{} and \cc{}, detect the most vulnerabilities, achieving average recalls of 63.87\% (99/155) and 81.94\% (127/155), respectively. Traditional analyzers (\codeql{}, \semgrep{}, and \spotbugs{}) perform worse, achieving average recalls of 14.19\%, 6.67\%, and 14.84\%, respectively. For specialized LLM-based methods, \iris{}$_{\openai{}\gptfive{}}$ achieves strong performance\minordel{, consistent with the results reported in its original paper, achieving recalls of 46.94\% on CWE-022, 9.09\% on CWE-078, 17.86\% on CWE-079, and 23.53\% on CWE-094}. In contrast, \llmdfa{}$_{\openai{}\gptfive{}}$ yields very few detections, identifying only 1 vulnerability. Manual inspection shows that, similar to the C/C++ results, most of \llmdfa{}'s missed cases (38/39) arise because the source and sink APIs inferred by the LLMs do not match the actual project-specific APIs. The fundamental reason is that \textbf{LLMs struggle to generate accurate project-specific \minordel{source-sink definitions}\minoradd{API models} from limited prompt context.} In \llmdfa{}, few-shot prompting is used to guide the LLM to predict potential source-sink APIs, so \minordel{the model tends to produce definitions that are similar to the examples}\minoradd{the inferred API models tend to resemble the examples} rather than fully tailored to the project.} In contrast, \iris{} systematically enumerates and analyzes all imported and locally defined APIs within the project, leading to more accurate and \minordel{comprehensive source-sink coverage}\minoradd{complete API models}. Besides, for resource-leak vulnerabilities~(CWE-772), \inferroi{}$_{\openai{}\gptfive{}}$ achieves a recall of 66.00\%, which is in line with its original paper and outperforms the traditional analyzers \codeql{} and \spotbugs{}, which achieve recalls of 10.00\% and 6.00\%, respectively. It also performs comparably to \codex{} in recall.

\add{Moreover, we observe that replacing the models originally used by these specialized LLM-based methods with a stronger recent model (\gptfive{}) does not substantially improve recall, similar to the results on C/C++ vulnerabilities. This suggests that their recall is largely constrained by the detection workflows designed by the methods themselves, rather than by model capability alone. However, stronger models do lead to notable precision improvements. For example, with \gptfive{}, \iris{} improves its precision on CWE-078 from 6.06\% to 20.00\%, while \inferroi{} improves from 15.74\% to 20.50\%.}

\begin{figure}[t] 
\centering 
\includegraphics[width=0.8\columnwidth]{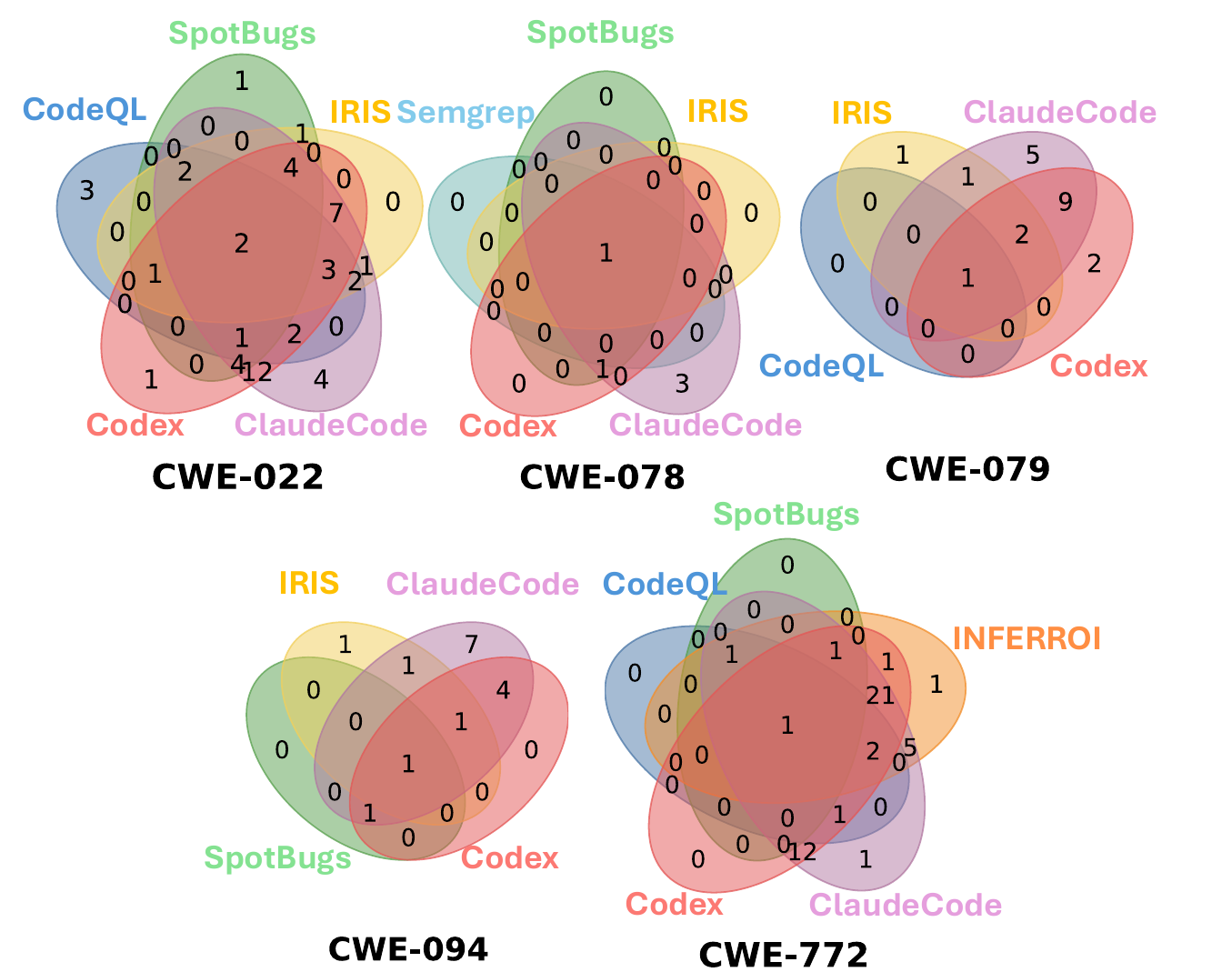} 
\caption{\add{Unique vulnerabilities detected by each tool}}
\label{fig:rq1_venn} 
\end{figure}

\begin{figure}[t]
\centering 
\includegraphics[width=0.9\columnwidth]{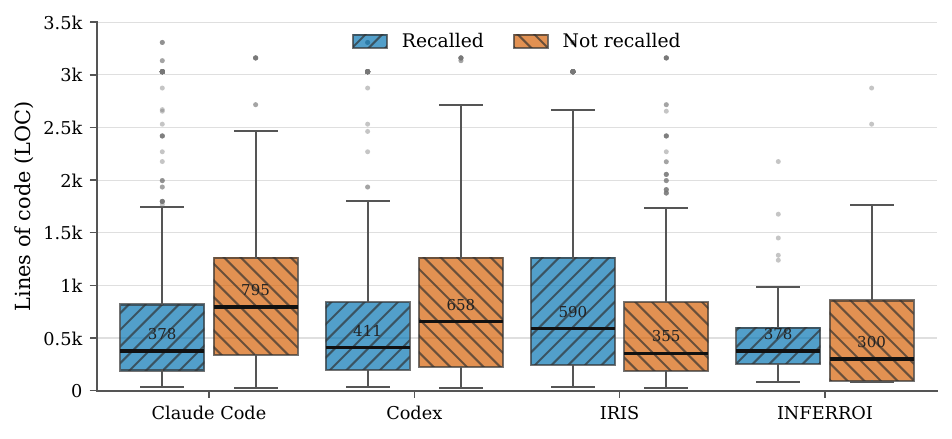}
\caption{\add{LOC distribution for recalled vs. not-recalled Java vulnerabilities}}
\label{fig:rq1_box_plot_java} 
\end{figure}

\noindent\textbf{\textit{Complementary Analysis.}} To further investigate how different tools complement or overlap with each other, we analyze the overlap of \del{detected vulnerabilities across tools}\add{vulnerabilities detected by traditional tools and LLM-based methods equipped with \gptfive{} and \sonnetfour{}}.\del{For CWE types that are detected by only one tool, this complementary analysis is omitted.} As shown in Fig.~\ref{fig:rq1_venn}, LLM-based methods contribute a substantially larger portion of unique detections, highlighting their ability to surface project-specific or unconventional patterns that traditional tools do not cover. \add{In particular, the two general-agent methods, \cc{} and \codex{}, account for most detections and exhibit substantial overlap, indicating that they often identify similar vulnerable patterns. Meanwhile, specialized LLM-based methods such as \iris{} and \inferroi{} provide complementary coverage for specific CWE categories; for example, \inferroi{} detects many CWE-772 cases and overlaps substantially with both general agents.} In contrast, traditional tools \del{(\codeql{} and \semgrep{})}\add{such as \codeql{}, \semgrep{}, and \spotbugs{}} yield very limited unique detections\del{(at most 3 unique detections on CWE-022), and their findings largely fall within the detection sets of LLM-based methods.}\add{. For example, \codeql{} contributes at most three unique detections on CWE-022, while most findings from traditional tools fall within the detection sets of LLM-based methods.} These results suggest that, in our evaluated setting, these LLM-based methods \del{often cover a substantial portion of the vulnerabilities reported by traditional static tools and can additionally uncover vulnerabilities beyond their coverage}\add{show partial overlap with traditional static tools and may identify some vulnerabilities not detected by the traditional baselines}.

\add{Although \codex{} and \cc{} achieve strong recall, they still exhibit limitations when analyzing complex files. As shown in Fig.~\ref{fig:rq1_box_plot_java}, we collect the LOC of files containing the source and sink APIs for both detected and missed vulnerabilities across \codex{}, \cc{}, \iris{}, and \inferroi{}. For general-agent methods, files associated with detected vulnerabilities tend to have lower LOC than those associated with missed vulnerabilities, suggesting that file size and complexity remain key obstacles for agents that rely on direct source-code inspection and search. In contrast, specialized LLM-based methods are largely unaffected by this issue. This is because \iris{} and \inferroi{} leverage traditional analysis tools or algorithms to reason over more compact representations, such as CFGs, rather than directly analyzing full source files, thereby mitigating the impact of large file size.}

\begin{findingbox}
\textbf{Finding 1:} General-agent methods achieve the highest recall and precision, but still struggle with complex code. Both LLM-based and traditional tools suffer from \minordel{source/sink mismatches}\minoradd{incomplete API modeling}, and providing \minordel{precise source/sink specifications}\minoradd{project-specific API specifications} substantially improves detection performance.
\end{findingbox}

\begin{findingbox}
\textbf{Finding 2:} \add{Specialized LLM-based methods rely on rigid analysis workflows, which can limit stronger models' capabilities while making their performance less sensitive to weaker backbones.}
\end{findingbox}

\newsavebox{\warnsamplebox}
\newlength{\warnsamplewidth}
\newcommand{\warnsample}[2]{%
  \begingroup
  \settowidth{\warnsamplewidth}{\add{\#Warn.}}%
  \sbox{\warnsamplebox}{\add{#1}\minoradd{\textsuperscript{(#2)}}}%
  \ifdim\wd\warnsamplebox>\warnsamplewidth
    \resizebox{\warnsamplewidth}{!}{\usebox{\warnsamplebox}}%
  \else
    \usebox{\warnsamplebox}%
  \fi
  \endgroup
}

\begin{table*}[htbp]
  \centering
  \caption{\add{Detection results of the evaluated tools on real-world C/C++ projects. In this table, \#Warn. denotes the number of warnings}\minoradd{, superscripts indicate SFDR sample sizes.} \add{-- indicates that the SFDR cannot be calculated.}}
  \resizebox{\textwidth}{!}{
  
    \begin{tabular}{cccccccccccccccc}
    \toprule
    \multirow{2}[4]{*}{\add{\textbf{CWE}}} & \multirow{2}[4]{*}{\add{\textbf{Project}}} & \multicolumn{2}{c}{\add{\textbf{RepoAudit$_{\claude{}\sonnett{}}$}}} & \multicolumn{2}{c}{\add{\textbf{Knighter$_{\openai{}\othree{}}$}}} 
    & \multicolumn{2}{c}{\add{\textbf{Codex$_{\openai{}\gptfive{}}$}}} &
    \multicolumn{2}{c}{\add{\textbf{Claude Code$_{\claude{}\sonnetfourr{}}$}}}
    & \multicolumn{2}{c}{\add{\textbf{CodeQL}}} & \multicolumn{2}{c}{\add{\textbf{Semgrep}}} 
    & \multicolumn{2}{c}{\add{\textbf{CSA}}}  \\
    
    \cmidrule{3-16} & & \add{\textbf{\#Warn.}} & \add{\textbf{SFDR}}
                      & \add{\textbf{\#Warn.}} & \add{\textbf{SFDR}}
                      & \add{\textbf{\#Warn.}} & \add{\textbf{SFDR}}
                      & \add{\textbf{\#Warn.}} & \add{\textbf{SFDR}}
                      & \add{\textbf{\#Warn.}} & \add{\textbf{SFDR}}
                      & \add{\textbf{\#Warn.}} & \add{\textbf{SFDR}}
                      & \add{\textbf{\#Warn.}} & \add{\textbf{SFDR}}\\
    \midrule
            & \add{linux/sound} & \warnsample{0}{0} & \add{--}
                          & \warnsample{0}{0}     & \add{--} 
                          & \warnsample{2}{2}     & \add{0.00\%} 
                          & \warnsample{2}{2}     & \add{100.00\%}
                          & \warnsample{0}{0}     & \add{--}
                          & \warnsample{0}{0}     & \add{--}
                          & \warnsample{0}{0}     & \add{--}\\
    \add{CWE-401} & \add{linux/mm} & \warnsample{0}{0} & \add{--}  
                       & \warnsample{0}{0}     & \add{--}
                       & \warnsample{6}{6}     & \add{33.33\%}
                       & \warnsample{2}{2}     & \add{100.00\%}
                       & \warnsample{0}{0}     & \add{--}
                       & \warnsample{0}{0}     & \add{--}
                       & \warnsample{0}{0}     & \add{--}\\
            & \add{ImageMagick} & \warnsample{0}{0}  & \add{--} 
                         & \warnsample{0}{0}        & \add{--}    
                         & \warnsample{3}{3} & \add{0.00\%}
                         & \warnsample{3}{3}   & \add{66.67\%}
                         & \warnsample{141}{141}     & \add{89.36\%}
                         & \warnsample{37}{37}     & \add{100.00\%}
                         & \warnsample{0}{0}     & \add{--}\\
    \midrule
    \multicolumn{2}{c}{\add{\textbf{Average of CWE-401}}} & \add{0} & \minordel{0}\minoradd{--}
                                                    & \add{0} & \add{--} 
                                                    & \add{3.67} & \add{18.18\%} 
                                                    & \add{2.33} & \add{85.71\%}
                                                    & \add{47}     & \add{89.36\%}
                                                    & \add{12.33}     & \add{100.00\%}
                                                    & \add{0}     & \add{--}\\ 
    \midrule[1.01pt]
            & \add{linux/net} & \warnsample{3}{3}      & \add{66.67\%}
                         & \warnsample{0}{0}    & \add{--}
                         & \warnsample{1}{1}        & \add{0.00\%}
                         & \warnsample{5}{5}        & \add{80.00\%}
                         & \warnsample{0}{0}        & \add{--}
                         & \warnsample{0}{0}        & \add{--}
                         & \warnsample{72}{72}        & \add{97.22\%}\\
    \add{CWE-416} & \add{linux/drivers/net} & \warnsample{1}{1}   & \add{100.00\%} 
                                & \warnsample{48}{48} & \add{81.25\%} 
                                & \warnsample{3}{3}   & \add{0.00\%}
                                & \warnsample{9}{9}       & \add{100.00\%}
                                & \warnsample{1}{1}        & \add{100.00\%}
                                & \warnsample{0}{0}        & \add{--}
                                & \warnsample{74}{74}        & \add{100.00\%}\\
            & \add{vim}   & \warnsample{31}{31} & \add{100.00\%}
                    & \warnsample{0}{0}      & \add{--}
                    & \warnsample{0}{0}      & \add{--}
                    & \warnsample{6}{6}  & \add{100.00\%}
                    & \warnsample{0}{0}        & \add{--}
                    & \warnsample{2}{2}        & \add{100.00\%}
                    & \warnsample{0}{0}        & \add{--}\\
    \midrule
    \multicolumn{2}{c}{\add{\textbf{Average of CWE-416}}} & \add{11.67} & 97.14\%
                                           & \add{16.00} & \add{81.25\%} 
                                           & \add{1.33}  & \add{0.00\%}
                                           & \add{6.67}  & \add{95.00\%} 
                                           & \add{0.33}        & \add{100.00\%}
                                           & \add{0.67}        & \add{100.00\%}
                                           & \add{48.67}        & \add{98.63\%}\\ 
    \midrule[1.01pt]
            & \add{linux/drivers/peci} & \warnsample{3}{3} & \add{100.00\%}
                                 & \warnsample{0}{0}     & \add{--}
                                 & \warnsample{0}{0} & \add{--}
                                 & \warnsample{4}{4}     & \add{100.00\%}
                                 & \warnsample{3}{3}        & \add{100.00\%}
                                 & \warnsample{0}{0}        & \add{--}
                                 & \warnsample{0}{0}        & \add{--}\\
    \add{CWE-476} & \add{gpac}  & \warnsample{77}{77}    & \add{97.40\%}
                    & \warnsample{0}{0}          & \add{--}
                    & \warnsample{4}{4} & \add{0.00\%}
                    & \warnsample{15}{15}      & \add{26.67\%}
                    & \warnsample{6345}{300}        & \add{76.00\%}
                    & \warnsample{1}{1}        & \add{100.00\%}
                    & \warnsample{194}{194}        & \add{75.77\%}\\
            & \add{bitlbee} & \warnsample{374}{300} & \add{91.00\%}
                      & \warnsample{0}{0}        & \add{--}
                      & \warnsample{3}{3} & \add{0.00\%}
                      & \warnsample{8}{8}        & \add{87.50\%}
                      & \warnsample{284}{284}        & \add{94.72\%}
                      & \warnsample{0}{0}        & \add{--}
                      & \warnsample{5}{5}        & \add{80.00\%}\\
    \midrule
    \multicolumn{2}{c}{\add{\textbf{Average of CWE-476}}} & \add{151.33} & \add{92.37\%}
                                                    & \add{0} & \add{--}
                                                    & \add{2.33} & \add{0.00\%}
                                                    & \add{9.00} & \add{55.56\%}
                                                    & \add{2210.67}    & \add{85.18\%}
                                                    & \add{0.33}       & \add{100.00\%}
                                                    & \add{66.33}    & \add{75.88\%}\\ 
    \midrule[1.01pt]
    \multicolumn{2}{c}{\add{\textbf{Overall Average}}} & \add{54.33} & 92.77\%
                                         & \add{5.33}    & \add{81.25\%} 
                                         & \add{2.44} & \add{9.09\%}
                                         & \add{6.00}    & \add{74.07\%} 
                                         & \add{752.67}        & \add{86.01\%}
                                         & \add{4.44}        & \add{100.00\%}
                                         & \add{38.33}        & \add{85.51\%}\\
                                         
    \bottomrule
    \end{tabular}%
  }
  \label{tab:results_real_world_c}%
\end{table*}%

\begin{table}[t]
\centering
\small
\setlength{\tabcolsep}{2pt}
\caption{\add{Detection results of the evaluated tools using \gptfive{} on real-world C/C++ projects. In this table, \#Warn. denotes the number of warnings}\minoradd{, superscripts indicate SFDR sample sizes.} \add{-- indicates that the SFDR cannot be calculated.}}
\resizebox{\columnwidth}{!}{
\begin{tabular}{lccccc}
\toprule
\multirow{2}{*}{\add{CWE}} & \multirow{2}{*}{\add{Project}}
& \multicolumn{2}{c}{\add{\repoaudit{}$_{\openai{}\gptfive{}}$}}
& \multicolumn{2}{c}{\add{\knighter{}$_{\openai{}\gptfive{}}$}}\\
\cmidrule(lr){3-4}\cmidrule(lr){5-6}
& & \add{\#Warn.} & \add{SFDR} & \add{\#Warn.} & \add{SFDR} \\
\midrule
\multirow{3}{*}{\add{CWE-401}}
& \add{linux/sound} & \warnsample{0}{0} & \add{--} & \warnsample{0}{0} & \add{--} \\
& \add{linux/mm} & \warnsample{0}{0} & \add{--} & \warnsample{0}{0} & \add{--} \\
& \add{ImageMagick} & \warnsample{0}{0} & \add{--} & \warnsample{0}{0} & \add{--} \\
\midrule
\multicolumn{2}{c}{\add{Average of CWE-401}} & \add{0} & \add{--} & \add{0} & \add{--} \\
\midrule
\multirow{3}{*}{\add{CWE-416}}
& \add{linux/net} & \warnsample{2}{2} & \add{0.00\%} & \warnsample{5}{5} & \add{100.00\%} \\
& \add{linux/drivers/net} & \warnsample{2}{2} & \add{100.00\%} & \warnsample{265}{265} & \add{95.47\%} \\
& \add{vim} & \warnsample{14}{14} & \add{100.00\%} & \warnsample{0}{0} & \add{--} \\
\midrule
\multicolumn{2}{c}{\add{Average of CWE-416}} & \add{6} & \add{88.89\%} & \add{90.00} & \add{95.56\%} \\
\midrule
\multirow{3}{*}{\add{CWE-476}}
& \add{linux/drivers/peci} & \warnsample{1}{1} & \add{100.00\%} & \warnsample{0}{0} & \add{--} \\
& \add{gpac} & \warnsample{95}{95} & \add{95.79\%} & \warnsample{0}{0} & \add{--} \\
& \add{bitlbee} & \warnsample{339}{300} & \add{95.67\%} & \warnsample{0}{0} & \add{--} \\
\midrule
\multicolumn{2}{c}{\add{Average of CWE-476}} & \add{145} & \add{95.71\%} & \add{0} & \add{--} \\
\midrule
\multicolumn{2}{c}{\add{Overall Average}} & 50.33 & \add{95.41\%} & \add{30.00} & \add{95.56\%} \\
\bottomrule
\end{tabular}
}
\label{tab:results_real_world_c_new}
\end{table}

\subsection{RQ2: Effectiveness in Real-world Deployment Scenarios}
\label{sec:results_rq2}

Table~\ref{tab:results_real_world_c}\add{,~\ref{tab:results_real_world_c_new}} and Table~\ref{tab:results_real_world_java}\add{,~\ref{tab:results_real_world_java_new}} summarize the effectiveness of the evaluated methods on large and latest real-world projects\add{/modules}. \add{Since most tool-project/module detection results contain no more than 300 warnings (185 out of 195 experimental setups), for each tool's detection results on each project or module, we sampled up to \textbf{300} reports for each tool on each project or module (resulting in \textbf{6,442} reports in total) and provided them to \codex{} with the prompt\footnote{\linkvalidate{}}. Specifically, for each warning, we provided both the warning report and the corresponding repository, enabling \codex{} to inspect the relevant files and determine whether the warning was a TP or FP. For traditional tools, the report includes checker information, involved code locations, and data-flow information when available. For LLM-based tools, it includes the dataflow and code locations related to the reported vulnerability, as well as the method's original vulnerability analysis. Finally, to validate the reliability of this process, we performed stratified sampling on the labeled results by sampling up to five TPs and five FPs for each tool on each repository. In total, we manually validated 197 TPs and 424 FPs, and the \codex{} labels achieved an accuracy of 94.36\%.} \minoradd{We also estimated the population-weighted TP/FP labeling accuracy as 98.16\%. We evaluated its precision at the 95\% confidence level using a normal approximation based on stratified sampling variance with finite-population correction~\cite{cochran1977sampling}.} \textbf{Overall, the results reveal that LLM-based detectors still struggle when applied in real-world scenarios, with high false discovery rates that limit their practical effectiveness.} 

For C/C++ projects\add{/modules}, specialized LLM-based methods show unsatisfactory performance. \repoaudit{}$_{\claude{}\sonnett{}}$ produces a number of warnings for CWE-476\minordel{(e.g., 374 warnings on \texttt{bitlbee})}, while the vast majority are FPs, resulting in an average SFDR of 92.37\% for CWE-476. \knighter{}$_{\openai{}\othree{}}$ produces 48 warnings in \texttt{linux/drivers/net}, achieving a high SFDR (81.25\%). Traditional tools often generate substantially more warnings, but most of them are false positives. For example, \codeql{} reports 6,345 warnings on \texttt{gpac} for CWE-476 with an SFDR of 76.00\%. Most of these reports (4{,}827 of them) stem from the coding manners of this project, in which the relevant APIs guarantee non-\texttt{NULL} return values under their usage conventions, so the callers omit explicit null checks. Nevertheless, these safe call sites match the \texttt{``Returned pointer not checked’’} pattern, and are flagged as warnings. \csa{} reports 72 and 74 warnings on the two Linux CWE-416 modules, both with SFDRs above 97\%. Similarly, \semgrep{} emits fewer warnings overall, but its warnings are frequently unreliable, reaching 100.00\% SFDR on several projects. Moreover, most methods still suffer from the \minordel{source/sink mismatch}\minoradd{incomplete API modeling} issue. \minordel{For example, \repoaudit{}$_{\claude{}\sonnett{}}$ reports no warnings for CWE-401, \knighter{}$_{\openai{}\othree{}}$'s generated checker is only effective on \texttt{linux/drivers/net}, and traditional tools such as \codeql{}, \semgrep{}, and \csa{} produce no warnings on many repositories or modules.}\minoradd{\repoaudit{}$_{\claude{}\sonnett{}}$ reports no CWE-401 warnings, while \knighter{}$_{\openai{}\othree{}}$ reports warnings only in \texttt{linux/drivers/net}. \codeql{}, \semgrep{}, and \csa{} also produce no warnings on many projects or modules.}

In contrast, the general-agent methods produce far fewer warnings. \codex{} is the most conservative, with only 2.44 warnings on average and the lowest overall SFDR of 9.09\%, suggesting that it avoids large-scale warning floods. \cc{} reports more warnings than \codex{} and detects more candidate issues, but its overall SFDR is much higher at 74.07\%. Our analysis shows that this is because \cc{} tends to report suspicious code locations more aggressively, often before fully validating the related data structure or feasible data-flow path. Which is further explained in Section~\ref{sec:results_rq3}.

\begin{table*}[htbp]
  \centering
  \caption{\add{Detection results of evaluated tools on real-world Java projects. In this table, \#Warn. denotes the number of warnings\minoradd{, superscripts indicate SFDR sample sizes.} -- indicates that the tool does not support the corresponding CWE type or that the SFDR cannot be calculated.}}
  \resizebox{\textwidth}{!}{
    \begin{tabular}{cccccccccccccccccc}
    \toprule
    \multirow{2}[4]{*}{\add{\textbf{CWE}}} 
    & \multirow{2}[4]{*}{\add{\textbf{Project}}}
    & \multicolumn{2}{c}{\add{\textbf{\iris{}$_{\openai{}\gptfour{}}$}}} 
    & \multicolumn{2}{c}{\add{\textbf{\llmdfa{}$_{\openai{}\gptfour{}}$}}}
    & \multicolumn{2}{c}{\add{\textbf{\codex{}$_{\openai{}\gptfive{}}$}}} 
    & \multicolumn{2}{c}{\add{\textbf{\cc{}$_{\claude{}\sonnetfourr{}}$}}}
    & \multicolumn{2}{c}{\add{\textbf{\codeql{}}}}
    & \multicolumn{2}{c}{\add{\textbf{\semgrep{}}}} 
    & \multicolumn{2}{c}{\add{\textbf{\spotbugs{}}}}
    & \multicolumn{2}{c}{\add{\textbf{\inferroi{}$_{\openai{}\gptfour{}}$}}} \\
    
    \cmidrule{3-18} & & \add{\textbf{\#Warn.}} & \add{\textbf{SFDR}} &
    \add{\textbf{\#Warn.}} & \add{\textbf{SFDR}} &
    \add{\textbf{\#Warn.}} & \add{\textbf{SFDR}} &
    \add{\textbf{\#Warn.}} & \add{\textbf{SFDR}} &
    \add{\textbf{\#Warn.}} & \add{\textbf{SFDR}} &
    \add{\textbf{\#Warn.}} & \add{\textbf{SFDR}} &
    \add{\textbf{\#Warn.}} & \add{\textbf{SFDR}} &
    \add{\textbf{\#Warn.}} & \add{\textbf{SFDR}} \\
    \midrule
        & \add{OpenOLAT} & \warnsample{\textbf{709}}{300} & \add{89.67\%}
                   & \add{--}     & \add{--}
                   & \warnsample{1}{1} & \add{0.00\%}
                   & \warnsample{2}{2} & \add{50.00\%}
                   & \warnsample{63}{63}     & \add{71.43\%}
                   & \warnsample{183}{183}     & \add{90.71\%}
                   & \warnsample{677}{300}     & \add{77.33\%} 
                   & \add{--}      &\add{--}\\
    \add{CWE-022} & \add{spark} & \warnsample{7}{7} & \add{42.86\%}
                    & \add{--}     & \add{--}
                    & \warnsample{3}{3} & \add{0.00\%}
                    & \warnsample{5}{5} & \add{60.00\%}
                    & \warnsample{5}{5}     & \add{40.00\%}
                    & \warnsample{5}{5}     & \add{80.00\%}
                    & \warnsample{8}{8}     & \add{62.50\%} 
                    & \add{--}      &\add{--}\\
            & \add{Dspace} & \warnsample{552}{300} & \add{98.00\%}
                     & \add{--}         & \add{--}
                     & \warnsample{3}{3}         & \add{66.67\%}
                     & \warnsample{1}{1}         & \add{0.00\%}
                     & \warnsample{2}{2}         & \add{100.00\%}
                     & \warnsample{107}{107}       & \add{96.26\%}
                     & \warnsample{290}{290}       & \add{94.83\%} 
                     & \add{--}      &\add{--}\\
    \midrule
    \multicolumn{2}{c}{\add{\textbf{Average of CWE-022}}} &\add{422.67} & \add{93.25\%}
                                                    & \add{--} & \add{--}
                                                    & \add{2.33} & \add{28.57\%}
                                                    & \add{2.67} & \add{50.00\%}
                                                    & \add{23.33}     & \add{70.00\%}
                                                    & \add{98.33}     & \add{92.54\%}
                                                    & \add{325}     & \add{85.62\%} 
                                                    & \add{--}      &\add{--}\\
    \midrule[1.01pt]
            & \add{xstream} & \warnsample{17}{17} & \add{100.00\%}
            & \warnsample{0}{0}     & \add{--}
            & \warnsample{2}{2}     & \add{100.00\%}
            & \warnsample{4}{4}     & \add{100.00\%}
            & \warnsample{0}{0}     & \add{--}
            & \warnsample{0}{0}     & \add{--}
            & \warnsample{0}{0}     & \add{--} 
            & \add{--}      &\add{--}\\
    \add{CWE-078} & \add{workflow-cps-plugin} & \warnsample{5}{5} & \add{100.00\%}
                                  & \warnsample{0}{0}     & \add{--}
                                  & \warnsample{0}{0}     & \add{--}
                                  & \warnsample{4}{4}     & \add{100.00\%}
                                  & \warnsample{0}{0}     & \add{--}
                                  & \warnsample{0}{0}     & \add{--}
                                  & \warnsample{0}{0}     & \add{--} 
                                  & \add{--}      &\add{--}\\
            & \add{tika}  & \warnsample{71}{71} & \add{100.00\%}
                    & \warnsample{4}{4}   & \add{100.00\%}
                    & \warnsample{0}{0}  & \add{--}
                    & \warnsample{5}{5} & \add{100.00\%}
                    & \warnsample{13}{13}       & \add{100.00\%}
                    & \warnsample{32}{32}       & \add{96.88\%}
                    & \warnsample{31}{31}       & \add{96.77\%} 
                    & \add{--}      &\add{--}\\
    \midrule
    \multicolumn{2}{c}{\add{\textbf{Average of CWE-078}}} & \add{31} & \add{100.00\%}
                                                    & \add{1.33}   & \add{100.00\%}
                                                    & \add{0.67}   & \add{100.00\%}
                                                    & \add{4.33}  & \add{100.00\%}
                                                    & \add{4.33}     & \add{100.00\%}
                                                    & \add{10.67}     & \add{96.88\%}
                                                    & \add{10.33}     & \add{96.77\%} 
                                                    & \add{--}      &\add{--}\\
    \midrule[1.01pt]
            & \add{xwiki-platform} & \warnsample{\textbf{458}}{300} & \add{99.67\%}
                             & \warnsample{5}{5} & \add{100.00\%}
                             & \warnsample{2}{2} & \add{0.00\%}
                             & \warnsample{7}{7} & \add{85.71\%}
                             & \warnsample{1}{1}     & \add{100.00\%}
                             & \warnsample{1}{1}     & \add{100.00\%}
                             & \warnsample{2}{2}     & \add{100.00\%} 
                             & \add{--}      &\add{--}\\
    \add{CWE-079} & \add{jenkins} & \warnsample{208}{208}& \add{100.00\%}
                      & \warnsample{4}{4}     & \add{100.00\%}
                      & \warnsample{2}{2}         & \add{50.00\%}
                      & \warnsample{11}{11}     & \add{100.00\%}
                      & \warnsample{0}{0}     & \add{--}
                      & \warnsample{1}{1}     & \add{0.00\%}
                      & \warnsample{0}{0}     & \add{--} 
                      & \add{--}      &\add{--}\\
            & \add{keycloak} & \warnsample{358}{300} & \add{100.00\%}
                       & \warnsample{2}{2}    & \add{100.00\%}
                       & \warnsample{2}{2}    & \add{0.00\%}
                       & \warnsample{5}{5}     & \add{80.00\%}
                       & \warnsample{6}{6}     & \add{100.00\%}
                       & \warnsample{0}{0}     & \add{--}
                       & \warnsample{0}{0}     & \add{--} 
                       & \add{--}      &\add{--}\\
    \midrule
    \multicolumn{2}{c}{\add{\textbf{Average of CWE-079}}} &\add{341.33} & \add{99.88\%}
                                                    & \add{3.67} & \add{100.00\%}
                                                    & \add{2.00} & \add{16.67\%}
                                                    & \add{7.67} & \add{91.30\%}
                                                    & \add{2.33}     & \add{100.00\%}
                                                    & \add{0.67}     & \add{50.00\%}
                                                    & \add{0.67}     & \add{100.00\%} 
                                                    & \add{--}      &\add{--}\\
    \midrule[1.01pt]
            & \add{onedev} & \warnsample{128}{128}   & \add{100.00\%}
                     & \add{--}        & \add{--}
                     & \warnsample{0}{0}        & \add{--}
                     & \warnsample{5}{5}        & \add{100.00\%}
                     & \warnsample{0}{0}     & \add{--}
                     & \warnsample{0}{0}     & \add{--}
                     & \warnsample{0}{0}     & \add{--} 
                     & \add{--}      &\add{--}\\
    \add{CWE-094} & \add{activemq} & \warnsample{271}{271} & \add{100.00\%}
                       & \add{--}        & \add{--}
                       & \warnsample{0}{0}    & \add{--}
                       & \warnsample{6}{6}        & \add{100.00\%}
                       & \warnsample{1}{1}     & \add{100.00\%}
                       & \warnsample{0}{0}     & \add{--}
                       & \warnsample{0}{0}     & \add{--} 
                       & \add{--}      &\add{--}\\
            & \add{cron-utils} & \warnsample{5}{5} & \add{100.00\%}
                         & \add{--}     & \add{--}
                         & \warnsample{1}{1}     & \add{100.00\%}
                         & \warnsample{1}{1}     & \add{100.00\%}
                         & \warnsample{0}{0}     & \add{--}
                         & \warnsample{0}{0}     & \add{--}
                         & \warnsample{0}{0}     & \add{--} 
                         & \add{--}      &\add{--}\\
    \midrule
    \multicolumn{2}{c}{\add{\textbf{Average of CWE-094}}} &\add{134.67} &\add{100.00\%}
                                           & \add{--} & \add{--}
                                           & \add{0.33} & \add{100.00\%}
                                           & \add{4.00} & \add{100.00\%}
                                           & \add{0.33}     & \add{100.00\%}
                                           & \add{0}     & \add{--}
                                           & \add{0}     & \add{--} 
                                           & \add{--}      &\add{--}\\
    \midrule[1.01pt]
            & \add{sql2o}  & \add{--}      &\add{--}
                     & \add{--}      &\add{--}
                     & \warnsample{3}{3}    & \add{0.00\%}
                     & \warnsample{5}{5}    & \add{80.00\%}
                     & \warnsample{0}{0}     & \add{--}
                     & \add{--}     & \add{--}
                     & \warnsample{0}{0}     & \add{--} 
                     & \warnsample{23}{23}     & \add{95.65\%}\\
    \add{CWE-772} & \add{RxJava} & \add{--}    & \add{--}
                     & \add{--}    & \add{--}
                     & \warnsample{2}{2}    & \add{50.00\%}
                     & \warnsample{10}{10}    & \add{60.00\%}
                     & \warnsample{0}{0}     & \add{--}
                     & \add{--}     & \add{--}
                     & \warnsample{0}{0}     & \add{--} 
                     & \warnsample{890}{300}     & \add{98.00\%}\\
            & \add{jsoup}  & \add{--}    & \add{--}
                     & \add{--}    & \add{--}
                     & \warnsample{1}{1} & \add{0.00\%}
                     & \warnsample{6}{6}     & \add{50.00\%}
                     & \warnsample{9}{9}     & \add{77.78\%}
                     & \add{--}     & \add{--}
                     & \warnsample{0}{0}     & \add{--} 
                     & \warnsample{66}{66}     & \add{89.39\%}\\
    \midrule
    \multicolumn{2}{c}{\add{\textbf{Average of CWE-772}}} & \add{--} & \add{--}
                                                    & \add{--} & \add{--}
                                                    & \add{2.00} & \add{16.67\%}
                                                    & \add{7.00} & \add{61.90\%}
                                                    & \add{3} & \add{77.78\%}
                                                    & \add{--}     & \add{--}
                                                    & \add{0.00}     & \add{--}
                                                & \add{326.33}     & \add{96.40\%} \\
    \midrule[1.01pt]
    \multicolumn{2}{c}{\add{\textbf{Average}}} & \add{232.42} & \add{97.80\%}
                                         & \add{2.50} & \add{100.00\%}
                                         & \add{1.47} & \add{31.82\%}
                                         & \add{5.13} & \add{81.82\%}
                                         & \add{6.67}     & \add{77.00\%}
                                         & \add{27.42}     & \add{92.71\%}
                                         & \add{67.20}     & \add{86.21\%}
                                         & \add{326.33}     & \add{96.40\%}\\
    \bottomrule
    \end{tabular}%
  }
  \label{tab:results_real_world_java}%
\end{table*}%

\begin{table}[htbp]
\centering
\small
\setlength{\tabcolsep}{2pt}
\caption{\add{Detection results of the evaluated tools using \gptfive{} on real-world Java projects. In this table, \#Warn. denotes the number of warnings\minoradd{, superscripts indicate SFDR sample sizes.} -- indicates that the SFDR cannot be calculated.}}
\resizebox{\columnwidth}{!}{
\begin{tabular}{lccccccc}
\toprule
\multirow{2}{*}{\add{CWE}} & \multirow{2}{*}{\add{Project}}
& \multicolumn{2}{c}{\add{\iris{}$_{\openai{}\gptfive{}}$}}
& \multicolumn{2}{c}{\add{\llmdfa{}$_{\openai{}\gptfive{}}$}}
& \multicolumn{2}{c}{\add{\inferroi{}$_{\openai{}\gptfive{}}$}} \\
\cmidrule(lr){3-4}\cmidrule(lr){5-6}\cmidrule(lr){7-8}
& & \add{\#Warn.} & \add{SFDR} & \add{\#Warn.} & \add{SFDR} & \add{\#Warn.} & \add{SFDR}\\
\midrule
\multirow{3}{*}{\add{CWE-022}}
& \add{OpenOLAT} & \warnsample{131}{131} & \add{68.70\%} & \add{--} & \add{--} & \add{--} & \add{--} \\
& \add{spark} & \warnsample{11}{11} & \add{0.00\%} & \add{--} & \add{--} & \add{--} & \add{--} \\
& \add{Dspace} & \warnsample{19}{19} & \add{73.68\%} & \add{--} & \add{--} & \add{--} & \add{--} \\
\midrule
\multicolumn{2}{c}{\add{Average of CWE-022}} & \add{53.67} & \add{64.60\%} & \add{--} & \add{--} & \add{--} & \add{--} \\
\midrule
\multirow{3}{*}{\add{CWE-078}}
& \add{xstream} & \warnsample{0}{0} & \add{--} & \warnsample{0}{0} & \add{--} & \add{--} & \add{--} \\
& \add{workflow-cps-plugin} & \warnsample{0}{0} & \add{--} & \warnsample{0}{0} & \add{--} & \add{--} & \add{--} \\
& \add{tika} & \warnsample{12}{12} & \add{100.00\%} & \warnsample{1}{1} & \add{100.00\%} & \add{--} & \add{--} \\
\midrule
\multicolumn{2}{c}{\add{Average of CWE-078}}
& \add{4} & \add{100.00\%} & \add{0.33} & \add{100.00\%} & \add{--} & \add{--} \\
\midrule
\multirow{3}{*}{\add{CWE-079}}
& \add{xwiki-platform} & \warnsample{26}{26} & \add{100.00\%} & \warnsample{3}{3} & \add{100.00\%} & \add{--} & \add{--} \\
& \add{jenkins} & \warnsample{14}{14} & \add{100.00\%} & \warnsample{4}{4} & \add{100.00\%} & \add{--} & \add{--} \\
& \add{keycloak} & \warnsample{7}{7} & \add{100.00\%} & \warnsample{2}{2} & \add{100.00\%} & \add{--} & \add{--} \\
\midrule
\multicolumn{2}{c}{\add{Average of CWE-079}} & \add{15.67} & \add{100.00\%} & \add{3.00} & \add{100.00\%} & \add{--} & \add{--} \\
\midrule
\multirow{3}{*}{\add{CWE-094}}
& \add{onedev} & \warnsample{1}{1} & \add{100.00\%} & \add{--} & \add{--} & \add{--} & \add{--} \\
& \add{activemq} & \warnsample{47}{47} & \add{100.00\%} & \add{--} & \add{--} & \add{--} & \add{--} \\
& \add{cron-utils} & \warnsample{0}{0} & \add{--} & \add{--} & \add{--} & \add{--} & \add{--} \\
\midrule
\multicolumn{2}{c}{\add{Average of CWE-094}} & \add{16.00} & \add{100.00\%} & \add{--} & \add{--} & \add{--} & \add{--} \\
\midrule
\multirow{3}{*}{\add{CWE-772}}
& \add{sql2o} & \add{--} & \add{--} & \add{--} & \add{--} & \warnsample{17}{17} & \add{94.12\%} \\
& \add{RxJava} & \add{--} & \add{--} & \add{--} & \add{--} & \warnsample{1016}{300} & \add{98.00\%} \\
& \add{jsoup} & \add{--} & \add{--} & \add{--} & \add{--} & \warnsample{35}{35} & \add{80.00\%} \\
\midrule
\multicolumn{2}{c}{\add{Average of CWE-772}} & \add{--} & \add{--} & \add{--} & \add{--} & \add{356.00} & \add{96.02\%} \\
\midrule
\multicolumn{2}{c}{\add{Overall Average}} & \add{22.33} & \add{78.73\%} & \add{1.67} & \add{100.00\%} & \add{356.00} & \add{96.02\%} \\
\bottomrule
\end{tabular}
}
\label{tab:results_real_world_java_new}
\end{table}

For Java projects, a similar pattern emerges. \iris{}$_{\openai{}\gptfour{}}$ consistently produces the largest number of warnings among all evaluated tools, yet \del{manual}\add{our} inspection again reveals almost no true positives, with an average SFDR of \textbf{\del{94.4\%}\add{97.80\%}}. In contrast, \llmdfa{}$_{\openai{}\gptfour{}}$ generates only a small number of reports, and \del{among them, only a single case is}\add{all of them are} identified as \del{a true positive}\add{FPs}, yielding an SFDR of 100.00\%. \del{\inferroi{} also detects only a limited number of true positives (e.g., one in \texttt{sql2o} and one in \texttt{RxJava}), although it can produce up to 360 reports across the three evaluated projects.}\add{Similarly, \inferroi{}$_{\openai{}\gptfour{}}$ reports many warnings for CWE-772, with an average SFDR of 96.40\%.} \add{Traditional tools also suffer from high SFDRs. \semgrep{} and \spotbugs{} reach 92.71\% and 86.21\%, respectively, while \codeql{} has a lower overall SFDR of 77.00\%. And due to the predefined rule sets of these tools, they can generate warnings in only about half of the selected projects\minordel{ (8 out of 15 projects for \codeql{}, 6 out of 12 projects for \semgrep{}, 5 out of 15 projects for \spotbugs{}}. In contrast, general-agent methods produce substantially fewer warnings. \codex{} is the most conservative, reporting only 1.47 warnings on average and achieving the lowest overall SFDR of 31.82\%, whereas \cc{} reports more candidate issues, with 5.13 warnings on average, but has a higher SFDR of 81.82\%.} 

Collectively, these results indicate that both current LLM-based and traditional static vulnerability detectors struggle to be directly used in real-world scenarios, typically failing either due to near-zero recall, as explained in Section~\ref{sec:results_rq1}, or overwhelming numbers of false positives in our setting. These observations motivate our deeper investigation in RQ3, where we examine the underlying causes of these false positives and analyze why existing techniques break down.

\begin{findingbox}
\textbf{Finding \del{2}\add{3}:} On real-world projects, \del{both LLM-based and traditional tools generate many warnings yet still exhibit very high SFDR (even the best-performing tool averages 85.3\%), severely limiting their practical usefulness.}\add{specialized LLM-based and traditional tools suffer from high SFDRs and limited coverage, while general agents reduce warning volume but still face false-positive risks.}
\end{findingbox}

\subsection{RQ3: Causes of False Positives}
\label{sec:results_rq3}


\begin{figure*}[htbp]
\centering
\begin{tcolorbox}[
  enhanced,
  colback=gray!2,
  colframe=gray!35,
  boxrule=0.4pt,
  arc=1pt,
  left=5pt,
  right=5pt,
  top=5pt,
  bottom=5pt
]
\footnotesize
\setlength{\parindent}{0pt}

\begin{minipage}[t]{0.49\textwidth}
\fpcat{fpblue}{A. Limited Context \& Dataflow Construction}
\begin{itemize}[leftmargin=2.5em, itemsep=0.35ex, topsep=0.4ex]
  \fpitem{fpblue}{A1}{Truncated Interprocedural Context}
  fails to analyze project-level dataflow across functions, files, or modules.
  \fpitem{fpblue}{A2}{Incorrect control flow}
  constructs or follows \minordel{infeasible}\minoradd{structurally unreachable control-flow} paths, e.g., by combining mutually exclusive branches.
\end{itemize}

\vspace{0.5ex}

\fpcat{fpyellow}{B. Misidentification of \minordel{Sources or Sinks}\minoradd{vulnerability-relevant program points.}}
\begin{itemize}[leftmargin=2.5em, itemsep=0.35ex, topsep=0.4ex]
   \fpitem{fpyellow}{B1}{Incorrect \minordel{source/sink selection}\minoradd{program-point classification}}
  \minordel{assigns}\minoradd{classifies} benign program points as vulnerability\minordel{sources or sinks}\minoradd{-relevant}.
\end{itemize}

\vspace{0.5ex}

\fpcat{fpgreen}{C. Prompt Misalignment}
\begin{itemize}[leftmargin=2.5em, itemsep=0.35ex, topsep=0.4ex]
   \fpitem{fpgreen}{C1}{Prompt deviations}
  fails to follow the specified prompt constraints or the target CWE definition.
\end{itemize}
\end{minipage}
\hfill
\begin{minipage}[t]{0.49\textwidth}
\fpcat{fpred}{D. Code Semantics Misinterpretation}
\begin{itemize}[leftmargin=2.5em, itemsep=0.35ex, topsep=0.4ex]
   \fpitem{fpred}{D1}{Missed key program points}
  overlooks or fails to model guards, sanitizers, initialization, or resource-release operations.
   \fpitem{fpred}{D2}{Poor language-feature support}
  misunderstands language-specific constructs such as constructors, destructors, exceptions, or try-with-resources.
  \fpitem{fpred}{D3}{Path-insensitive analysis}
  reports flows that are infeasible \minordel{or unreachable at runtime}\minoradd{under branch or value constraints}.
   \fpitem{fpred}{D4}{Type/structure blindness}
  ignores type, class, field, or data-structure constraints needed to judge feasibility.
\end{itemize}
\end{minipage}

\end{tcolorbox}
\caption{Taxonomy of FP reasons that are related to the cases detected by the selected methods.}
\label{fig:rq3_taxonomy}
\end{figure*}


As described in Section~\ref{sec:approach_metrics}, \add{we first derived a taxonomy of false-positive causes from manual inspection.} \del{two}\add{Two} authors with a background in security and software engineering manually analyzed the sampled false positives. We applied an inductive open-coding procedure with double coding~\cite{denzin2011sage}. Each author first reviewed all examples to derive a coherent taxonomy of FP causes and iteratively refined the codebook. After consolidating and resolving disagreements on category definitions, the experts independently assigned the primary FP reason for each case. For each warning, our goal is to identify the most direct cause of the false positive. The annotation follows a hierarchical decision procedure\minoradd{, assigning each case to the first applicable category in the following order:}

We first examine whether the \minordel{reported source and sink are security-relevant and correctly selected}\minoradd{selected program points are correctly classified as
vulnerability-relevant} (Category B, shown in Fig.~\ref{fig:rq3_taxonomy}). If they are correct, we then inspect the reported code region or path to determine whether the analysis relies on incomplete or \minordel{infeasible}\minoradd{theoretically invalid} data/control-flow reasoning (Category A). If the analyzed path or context is reasonable, we further check whether the tool misinterprets program semantics when drawing the vulnerability conclusion (Category D). Finally, for LLM-based tools, we examine whether the final judgment deviates from the prompt or target CWE definition (Category C). \minordel{When multiple factors are involved, we assign the label according to the dominant cause that most directly leads to the false positive}\minoradd{This ordering reflects dependencies in the report's justification: flow analysis is assessed relative to the selected program points, and subsequent semantic judgments depend on adequate context and a valid control-flow structure. An earlier failure must therefore be corrected before later reasoning can validate the report. For example, semantic reasoning cannot justify a vulnerability claim based on a structurally invalid path; we therefore assign A as the primary cause, even when semantic errors (D) also occur.} Finally, they reconciled disagreements and consolidated the labels into a unified annotation set. \add{We then used \codex{} with the prompt\footnote{\linkvalidate{}} to label the \textbf{5,896} false positives generated from Section~\ref{sec:results_rq2} according to this taxonomy, while allowing it to propose a new reason if none of the existing categories fit. In the prompt, for each warning from each tool, we provided the agent with the tool's workflow description, the taxonomy and category definitions derived from the previous manual analysis, the corresponding repository code, and the warning information produced by the tool. The warning information typically includes the vulnerability type, reported location, and data-flow/path information when available. And for LLM-based tools, it also includes the LLM-generated explanation or analysis. After manually reviewing and consolidating the labels produced by \codex{}, we found that no additional FP category was needed, and the final annotations followed the established taxonomy. To assess labeling reliability, we further performed stratified sampling on the \codex{}-generated labels for manual validation. Specifically, for each tool and each FP reason, we sampled up to five cases separately for C/C++ and Java, resulting in 425 validation cases in total. The \codex{}-generated labels achieved an overall accuracy of 87.76\%, and 77 out of 91 tool-reason groups achieved an accuracy of at least 80\%.} \minoradd{Most misclassifications involved explanations with multiple contributing factors, where \codex{} did not consistently apply the criteria for selecting the primary cause. We also estimated the population weighted FP-reason classification accuracy as 92.52\%. We evaluated its precision at the 95\% confidence level using a normal approximation based on stratified sampling variance with finite-population correction~\cite{cochran1977sampling}.}

\begin{table*}[t]
  \centering
  \caption{Distribution of false-positive reasons by tool. Each cell reports the reason count over the total number of false positives for that tool, followed by the corresponding percentage. Color intensity is rank-scaled independently within each tool row.}
  \label{tab:rq3_heat}
  \begingroup
  \footnotesize
  \setlength{\tabcolsep}{0pt}
  \renewcommand{\arraystretch}{1.0}
  \newlength{\fptoolw}
  \newlength{\fpcellw}
  \setlength{\fptoolw}{3.3cm}
  \setlength{\fpcellw}{2cm}
  \newcolumntype{T}{>{\centering\arraybackslash}m{\fptoolw}}
  \newcolumntype{H}{>{\centering\arraybackslash}m{\fpcellw}}
  \newcommand{\fpheat}[3]{\cellcolor{#1}\makebox[\fpcellw][c]{\makecell[c]{#2\\#3}}}
  \newcommand{\fptool}[1]{\bfseries\makebox[\fptoolw][c]{\small\makecell[c]{#1}}}
  \resizebox{\textwidth}{!}{%
  \begin{tabular}{@{}T@{}*{8}{H@{}}}
    \textbf{Tool} & \cellcolor{fpblue!10}\textbf{A1} & \cellcolor{fpblue!10}\textbf{A2} & \cellcolor{fpyellow!10}\textbf{B1} & \cellcolor{fpgreen!10}\textbf{C1} & \cellcolor{fpred!10}\textbf{D1} & \cellcolor{fpred!10}\textbf{D2} & \cellcolor{fpred!10}\textbf{D3} & \cellcolor{fpred!10}\textbf{D4} \\
    \midrule
\fptool{\repoaudit{}$_{\claude{}\sonnett{}}$} & \fpheat{fpblue!40}{34 / 385}{(8.83\%)} & \fpheat{fpblue!48}{58 / 385}{(15.06\%)} & \fpheat{fpyellow!61}{226 / 385}{(58.70\%)} & \fpheat{white}{0 / 385}{(0.00\%)} & \fpheat{fpred!47}{53 / 385}{(13.77\%)} & \fpheat{white}{0 / 385}{(0.00\%)} & \fpheat{fpred!30}{12 / 385}{(3.12\%)} & \fpheat{fpred!20}{2 / 385}{(0.52\%)} \\
    \fptool{\repoaudit{}$_{\openai{}\gptfive{}}$} & \fpheat{fpblue!38}{36 / 395}{(9.11\%)} & \fpheat{fpblue!36}{31 / 395}{(7.85\%)} & \fpheat{fpyellow!76}{296 / 395}{(74.94\%)} & \fpheat{white}{0 / 395}{(0.00\%)} & \fpheat{fpred!29}{14 / 395}{(3.54\%)} & \fpheat{white}{0 / 395}{(0.00\%)} & \fpheat{fpred!28}{13 / 395}{(3.29\%)} & \fpheat{fpred!23}{5 / 395}{(1.27\%)} \\
    \fptool{\knighter{}$_{\openai{}\othree{}}$} & \fpheat{white}{0 / 39}{(0.00\%)} & \fpheat{fpblue!38}{4 / 39}{(10.26\%)} & \fpheat{fpyellow!80}{31 / 39}{(79.49\%)} & \fpheat{white}{0 / 39}{(0.00\%)} & \fpheat{white}{0 / 39}{(0.00\%)} & \fpheat{white}{0 / 39}{(0.00\%)} & \fpheat{fpred!31}{2 / 39}{(5.13\%)} & \fpheat{fpred!31}{2 / 39}{(5.13\%)} \\
    \fptool{\knighter{}$_{\openai{}\gptfive{}}$} & \fpheat{fpblue!21}{2 / 258}{(0.78\%)} & \fpheat{fpblue!36}{22 / 258}{(8.53\%)} & \fpheat{fpyellow!80}{204 / 258}{(79.07\%)} & \fpheat{white}{0 / 258}{(0.00\%)} & \fpheat{fpred!21}{2 / 258}{(0.78\%)} & \fpheat{white}{0 / 258}{(0.00\%)} & \fpheat{fpred!31}{13 / 258}{(5.04\%)} & \fpheat{fpred!32}{15 / 258}{(5.81\%)} \\
    \fptool{\iris{}$_{\openai{}\gptfour{}}$} & \fpheat{fpblue!30}{96 / 1870}{(5.13\%)} & \fpheat{fpblue!24}{36 / 1870}{(1.93\%)} & \fpheat{fpyellow!88}{1640 / 1870}{(87.70\%)} & \fpheat{fpgreen!17}{3 / 1870}{(0.16\%)} & \fpheat{fpred!29}{80 / 1870}{(4.28\%)} & \fpheat{fpred!16}{1 / 1870}{(0.05\%)} & \fpheat{fpred!17}{4 / 1870}{(0.21\%)} & \fpheat{fpred!19}{10 / 1870}{(0.53\%)} \\
    \fptool{\iris{}$_{\openai{}\gptfive{}}$} & \fpheat{fpblue!35}{11 / 211}{(5.21\%)} & \fpheat{fpblue!40}{17 / 211}{(8.06\%)} & \fpheat{fpyellow!58}{116 / 211}{(54.98\%)} & \fpheat{white}{0 / 211}{(0.00\%)} & \fpheat{fpred!56}{45 / 211}{(21.33\%)} & \fpheat{fpred!27}{4 / 211}{(1.90\%)} & \fpheat{fpred!23}{2 / 211}{(0.95\%)} & \fpheat{fpred!39}{16 / 211}{(7.58\%)} \\
    \fptool{\inferroi{}$_{\openai{}\gptfour{}}$} & \fpheat{fpblue!47}{57 / 375}{(15.20\%)} & \fpheat{fpblue!18}{1 / 375}{(0.27\%)} & \fpheat{fpyellow!68}{248 / 375}{(66.13\%)} & \fpheat{white}{0 / 375}{(0.00\%)} & \fpheat{fpred!49}{65 / 375}{(17.33\%)} & \fpheat{fpred!21}{3 / 375}{(0.80\%)} & \fpheat{white}{0 / 375}{(0.00\%)} & \fpheat{fpred!18}{1 / 375}{(0.27\%)} \\
    \fptool{\inferroi{}$_{\openai{}\gptfive{}}$} & \fpheat{fpblue!57}{74 / 338}{(21.89\%)} & \fpheat{fpblue!21}{2 / 338}{(0.59\%)} & \fpheat{fpyellow!56}{181 / 338}{(53.55\%)} & \fpheat{white}{0 / 338}{(0.00\%)} & \fpheat{fpred!58}{76 / 338}{(22.49\%)} & \fpheat{fpred!21}{2 / 338}{(0.59\%)} & \fpheat{fpred!19}{1 / 338}{(0.30\%)} & \fpheat{fpred!21}{2 / 338}{(0.59\%)} \\
    \fptool{\llmdfa{}$_{\openai{}\gptfour{}}$} & \fpheat{white}{0 / 15}{(0.00\%)} & \fpheat{white}{0 / 15}{(0.00\%)} & \fpheat{fpyellow!81}{12 / 15}{(80.00\%)} & \fpheat{white}{0 / 15}{(0.00\%)} & \fpheat{fpred!48}{3 / 15}{(20.00\%)} & \fpheat{white}{0 / 15}{(0.00\%)} & \fpheat{white}{0 / 15}{(0.00\%)} & \fpheat{white}{0 / 15}{(0.00\%)} \\
    \fptool{\llmdfa{}$_{\openai{}\gptfive{}}$} & \fpheat{white}{0 / 10}{(0.00\%)} & \fpheat{white}{0 / 10}{(0.00\%)} & \fpheat{fpyellow!81}{8 / 10}{(80.00\%)} & \fpheat{white}{0 / 10}{(0.00\%)} & \fpheat{fpred!48}{2 / 10}{(20.00\%)} & \fpheat{white}{0 / 10}{(0.00\%)} & \fpheat{white}{0 / 10}{(0.00\%)} & \fpheat{white}{0 / 10}{(0.00\%)} \\
    \fptool{\codex{}$_{\openai{}\gptfive{}}$} & \fpheat{white}{0 / 9}{(0.00\%)} & \fpheat{fpblue!70}{2 / 9}{(22.22\%)} & \fpheat{fpyellow!38}{3 / 9}{(33.33\%)} & \fpheat{fpgreen!53}{1 / 9}{(11.11\%)} & \fpheat{fpred!82}{3 / 9}{(33.33\%)} & \fpheat{white}{0 / 9}{(0.00\%)} & \fpheat{white}{0 / 9}{(0.00\%)} & \fpheat{white}{0 / 9}{(0.00\%)} \\
    \fptool{\cc{}$_{\claude{}\sonnetfourr{}}$} & \fpheat{fpblue!60}{20 / 103}{(19.42\%)} & \fpheat{fpblue!54}{15 / 103}{(14.56\%)} & \fpheat{fpyellow!46}{43 / 103}{(41.75\%)} & \fpheat{white}{0 / 103}{(0.00\%)} & \fpheat{fpred!59}{19 / 103}{(18.45\%)} & \fpheat{fpred!24}{1 / 103}{(0.97\%)} & \fpheat{fpred!32}{3 / 103}{(2.91\%)} & \fpheat{fpred!29}{2 / 103}{(1.94\%)} \\
    \fptool{\csa{}} & \fpheat{fpblue!82}{87 / 295}{(29.49\%)} & \fpheat{fpblue!68}{54 / 295}{(18.31\%)} & \fpheat{fpyellow!34}{86 / 295}{(29.15\%)} & \fpheat{white}{0 / 295}{(0.00\%)} & \fpheat{fpred!38}{11 / 295}{(3.73\%)} & \fpheat{white}{0 / 295}{(0.00\%)} & \fpheat{fpred!64}{47 / 295}{(15.93\%)} & \fpheat{fpred!37}{10 / 295}{(3.39\%)} \\
    \fptool{\codeql{}} & \fpheat{fpblue!82}{252 / 704}{(35.80\%)} & \fpheat{fpblue!59}{110 / 704}{(15.62\%)} & \fpheat{fpyellow!22}{114 / 704}{(16.19\%)} & \fpheat{white}{0 / 704}{(0.00\%)} & \fpheat{fpred!49}{66 / 704}{(9.38\%)} & \fpheat{fpred!31}{15 / 704}{(2.13\%)} & \fpheat{fpred!65}{142 / 704}{(20.17\%)} & \fpheat{fpred!24}{5 / 704}{(0.71\%)} \\
    \fptool{\semgrep{}} & \fpheat{fpblue!64}{93 / 345}{(26.96\%)} & \fpheat{fpblue!36}{17 / 345}{(4.93\%)} & \fpheat{fpyellow!53}{170 / 345}{(49.28\%)} & \fpheat{white}{0 / 345}{(0.00\%)} & \fpheat{fpred!53}{57 / 345}{(16.52\%)} & \fpheat{fpred!23}{3 / 345}{(0.87\%)} & \fpheat{white}{0 / 345}{(0.00\%)} & \fpheat{fpred!26}{5 / 345}{(1.45\%)} \\
    \fptool{\spotbugs{}} & \fpheat{fpblue!50}{99 / 544}{(18.20\%)} & \fpheat{fpblue!34}{32 / 544}{(5.88\%)} & \fpheat{fpyellow!67}{356 / 544}{(65.44\%)} & \fpheat{white}{0 / 544}{(0.00\%)} & \fpheat{fpred!39}{47 / 544}{(8.64\%)} & \fpheat{fpred!18}{1 / 544}{(0.18\%)} & \fpheat{fpred!24}{8 / 544}{(1.47\%)} & \fpheat{fpred!18}{1 / 544}{(0.18\%)} \\
  \end{tabular}%
  }
  \endgroup
\end{table*}

\begin{figure}[t]
    \centering
    \includegraphics[width=0.9\columnwidth]{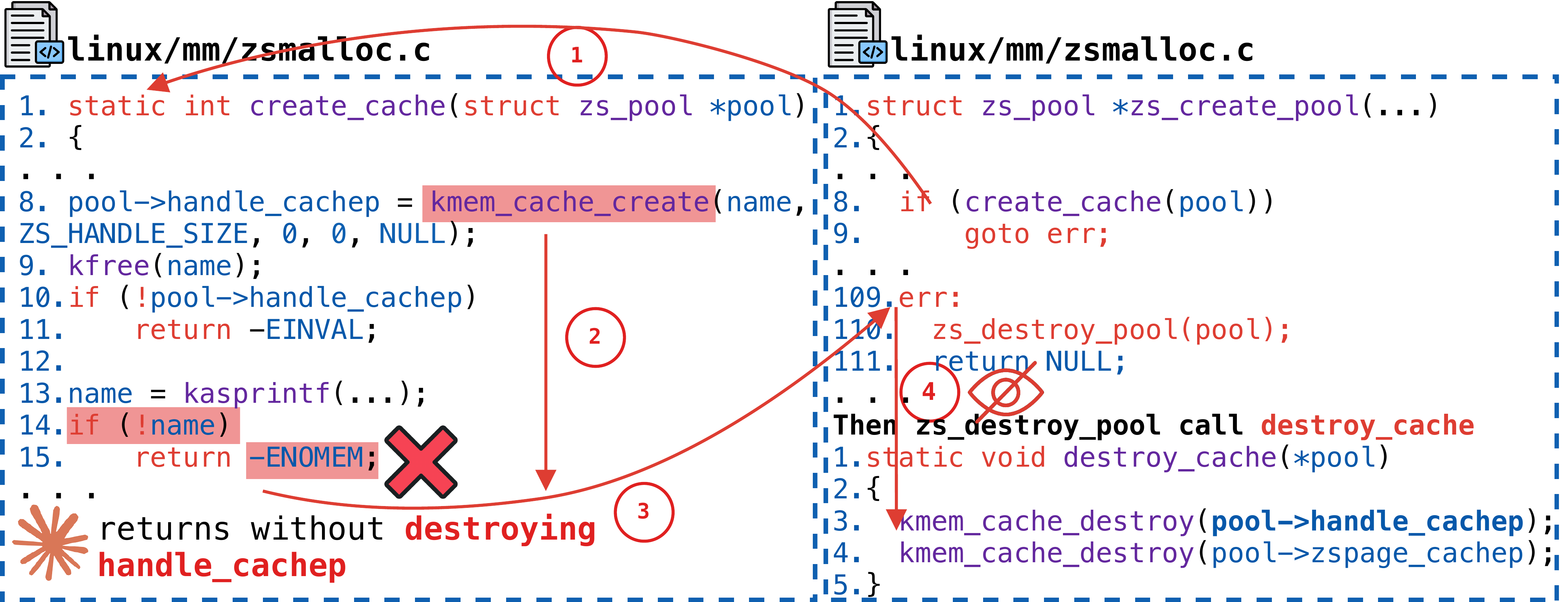}
    \caption{\add{An illustrating FP example of reason A1}}
    \label{fig:case_0}
\end{figure}

\begin{figure}[t]
    \centering
    \includegraphics[width=0.9\columnwidth]{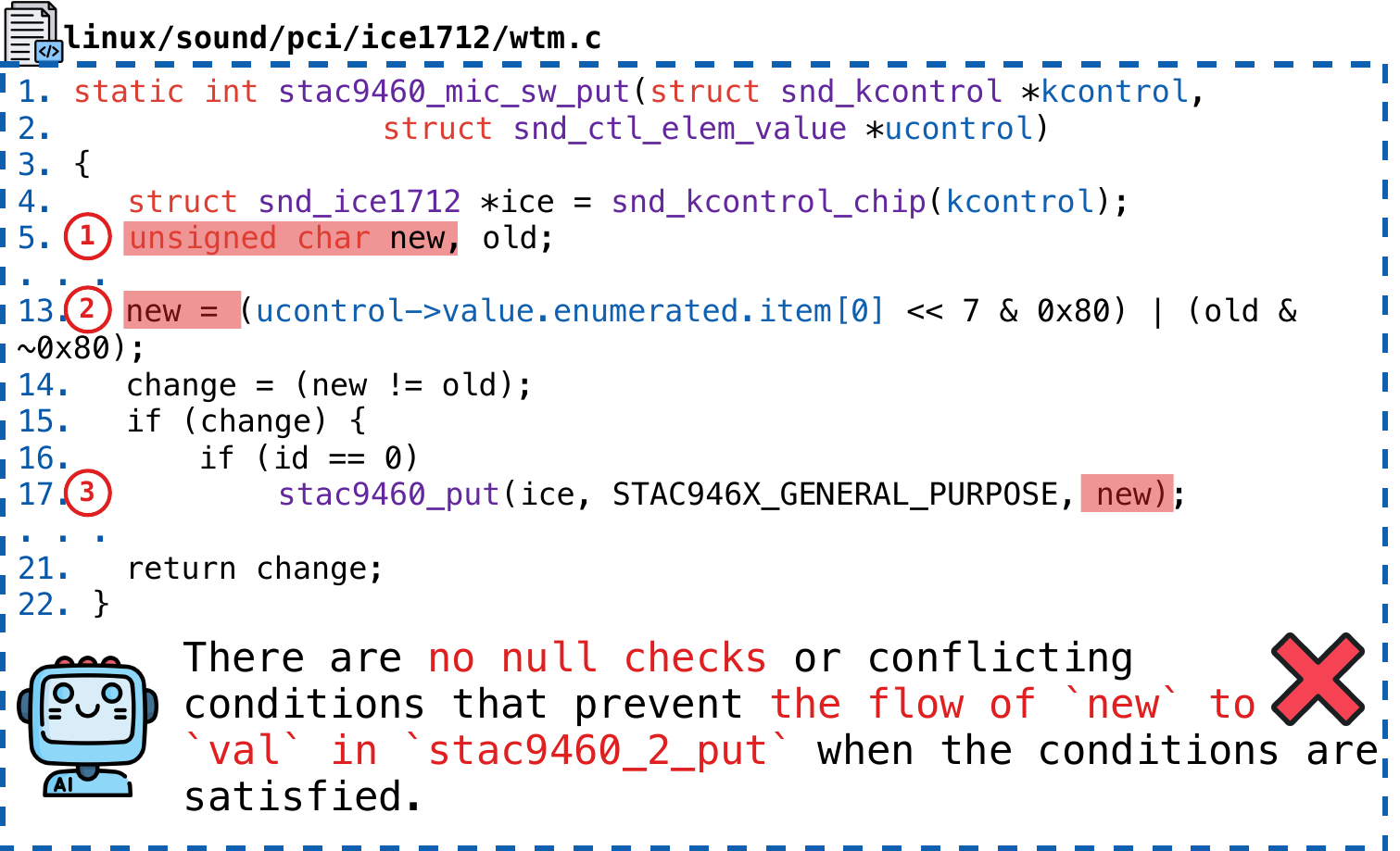}
    \caption{An illustrating FP example of reason B1}
    \label{fig:case_1}
\end{figure}

In the following, we analyze several representative reasons from this taxonomy. The taxonomy and the definition of each reason are summarized in Fig.~\ref{fig:rq3_taxonomy} and \del{Figure}\add{Table}~\ref{tab:rq3_heat}. Due to page limits, we present \del{two}\add{three} representative illustrative examples here, and include more detailed examples for the remaining FP reasons in 
Appendix~A.
As shown in the figures, LLM-based and traditional methods share two major sources of false positives: 
\textbf{(1) Truncated Interprocedural Context (\add{Reason} A1).} To limit analysis complexity, most tools rely either on source–sink style reasoning or on fixed-depth call-graph exploration. \del{For example, \repoaudit{} explores three layers of function calls from each specified source by default,}\add{As shown in Fig.~\ref{fig:case_0}, \cc{} reports a memory-leak warning because it observes that \texttt{create\_cache} may return at Line~15 without directly releasing the allocated \texttt{pool->handle\_cachep} from Line~8. However, this is a false positive. \cc{} fails to recognize that \texttt{pool} is passed by pointer and that ownership is handled by the caller. When \texttt{zs\_create\_pool} calls \texttt{create\_cache} and the call fails, it jumps to the error-handling path, invokes \texttt{zs\_destroy\_pool}, and eventually releases \texttt{pool->handle\_cachep} through \texttt{destroy\_cache} via \texttt{kmem\_cache\_destroy}. Thus, the warning is caused by incomplete interprocedural reasoning over caller-side cleanup logic.} Similarly, \minordel{traditional static analyzers such as \codeql{} are also configured with explicit source and sink definitions and track data flows only along the extracted source–sink paths.}\minoradd{data-flow analyses in traditional tools such as \codeql{} track dataflows between configured source and sink points.} However, in real-world projects, where data may originate from diverse entry points and propagate through rich, multi-hop control-data paths, such restricted flow modeling often fails to capture the full chain from input to output, leading to extensive false positives.
\textbf{(2) \minordel{Inaccurate identification of sources and sinks}\minoradd{Incorrect program-point classification} (\add{Reason} B1)}, where tools either misclassify benign program points as security-relevant or fail to correctly interpret the semantics of API usages. As shown in Fig.~\ref{fig:case_1}, the LLM misidentifies the variable \texttt{new} as a pointer to newly allocated memory in C++ and assumes it may be \texttt{NULL}. Even if it were uninitialized, using it would not constitute an NPD. At most, it would be an uninitialized-value issue. Therefore, the warning \minordel{is caused by an incorrect source/sink match}\minoradd{results from incorrect program-point classification (B1)}. However, throughout the entire reasoning process on this path, the LLM fails to recognize this and continues to treat the use of \texttt{new} as a potentially NPD vulnerability at line 17.

\begin{figure}[t]
    \centering
    \includegraphics[width=0.9\columnwidth]{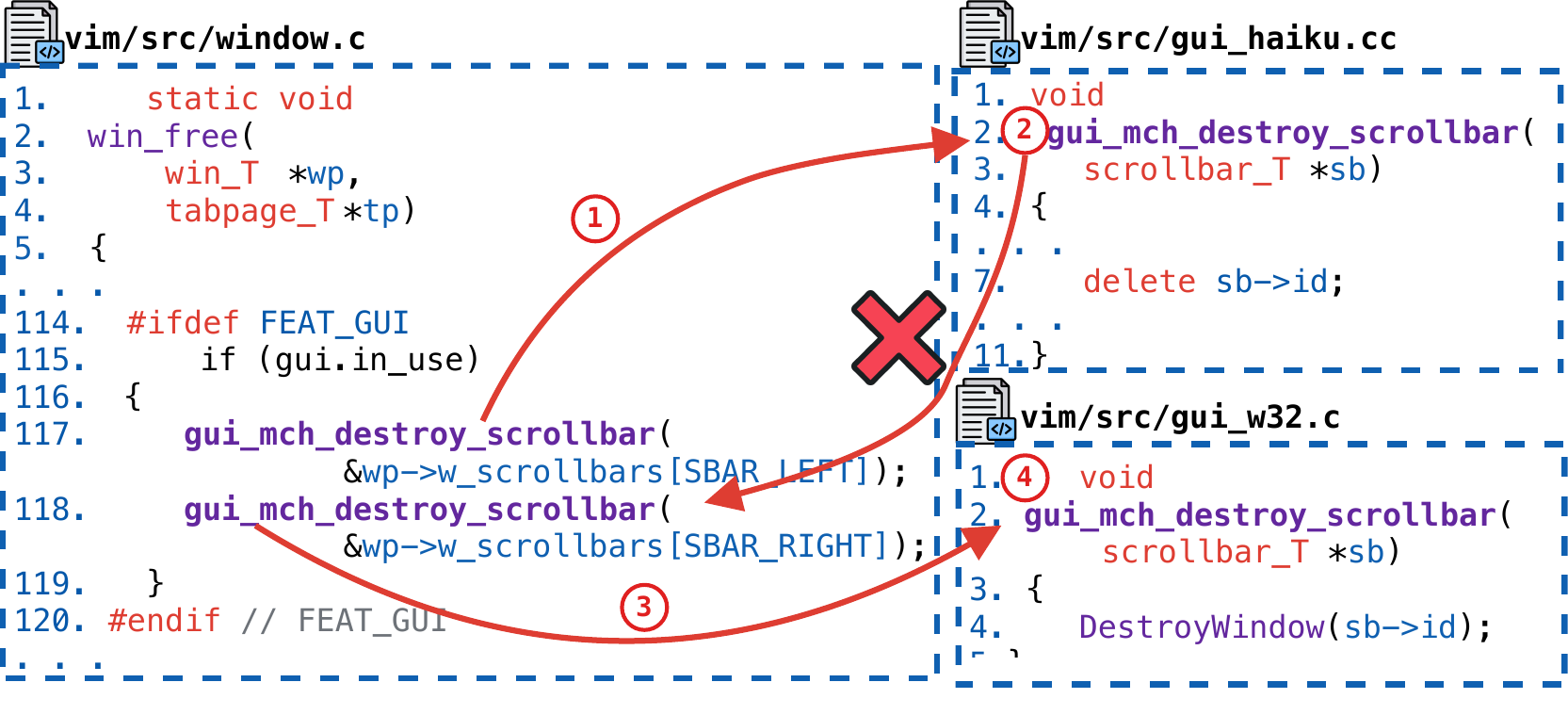}
    \caption{An illustrating FP example of reason A2}
    \label{fig:case_2}
\end{figure}

Beyond the shared sources of false positives, we further analyze the reasons \del{unique to}\add{that are particularly prevalent in} LLM-based vulnerability detection methods. Although LLMs exhibit strong code analysis abilities, their performance on vulnerability detection still faces several inherent limitations: \textbf{First, LLMs often fail to recover the correct execution flow in the presence of complex control structures (Reason A2).} For example, \del{24.6}\add{15.06}\% of \repoaudit{}$_{\claude{}\sonnett{}}$’s false positives arise because it overlooks key control-flow branches or constructs an incorrect data-flow graph, causing the detector to analyze paths that do not actually exist in the program. There is another interesting example in Fig.~\ref{fig:case_2}, in the \texttt{vim} project, there are two \texttt{gui\_mch\_destroy\_scrollbar} functions with the same signature, defined in \texttt{gui\_haiku.cc} and \texttt{gui\_w32.c} to support different operating systems. In practice, these two functions cannot be invoked together in a single build. However, when the LLM analyzes the two consecutive invocations of \texttt{gui\_mch\_destroy\_scrollbar} in \texttt{window.c}, it incorrectly concludes that the first call resolves to the implementation in \texttt{gui\_haiku.cc} and the second to the implementation in \texttt{gui\_w32.c}, merging mutually exclusive execution paths and leading to an incorrect understanding of the program behavior. \textbf{Second, LLMs frequently overlook implicit sanitization logic within a dataflow (Reason D1).} \del{Four out of five}\add{Most} selected LLM-based methods suffer noticeably from this issue\add{, with 7 out of 12 settings exceeding 15.00\%}, leading them to flag dataflows as vulnerable even when proper sanitizers exist along the path. \textbf{Finally, LLMs sometimes do not follow the prompt or CWE definition well (Reason C1).} Both \codex{} and \iris{} flag benign code as vulnerable or report issues that do not conform to the CWE definitions specified in the prompt. The corresponding examples are shown in our 
Appendix~A.

\add{Furthermore, for specialized LLM-based methods, the impact of stronger models differs depending on where the LLM is used in the detection workflow. For methods where the LLM directly participates in vulnerability judgment, such as \repoaudit{} and \knighter{}, where vulnerable paths or patterns are analyzed by LLMs, the proportions of Category A and D errors decrease, while \minordel{source/sink-related errors}\minoradd{program-point classification errors} increase substantially. In contrast, for methods where LLMs are mainly used for \minordel{source/sink identification}\minoradd{API-role inference} and the final detection is performed by a more constrained analysis procedure, such as \iris{} with \codeql{} and \inferroi{} over CFGs, \minordel{source/sink-related errors}\minoradd{program-point classification errors} decrease more clearly. This suggests that stronger models can reduce errors in the parts of the workflow they directly improve, but the overall effectiveness remains constrained by each method's workflow design and validation mechanism.}

\add{For \repoaudit{}, \knighter{}, and \csa{}, which rely on either LLMs or static path-enumeration mechanisms to extract path information before applying fixed vulnerability-checking logic, the proportion of A2 errors is much higher. This suggests current path extraction results remain unreliable under complex control flow, and errors in the extracted path can directly propagate to the final vulnerability judgment. For general-agent methods, our session analysis of \codex{} and \cc{} shows that, across 24 real-world projects/modules, \cc{} invoked tools 2,702 times and reported 131 warnings, while \codex{} invoked tools 1,415 times and reported 44 warnings. On average, this corresponds to 20.63 tool invocations per \minordel{detected vulnerability}\minoradd{reported warning} for \cc{} and 32.16 for \codex{}. This suggests that \codex{} spends more analysis effort on each reported \minordel{vulnerability}\minoradd{warning}, which helps explain its much lower SFDR and the lower frequency of analysis-related FP reasons compared with \cc{}.}

These findings suggest that although LLMs are capable of reasoning about code semantics, reliably capturing precise and complete program behavior in complex real-world projects, as well as strictly adhering to prompt specifications, remain major challenges.

\begin{findingbox}
\textbf{Finding \del{3}\add{4}:} False positives are dominated by truncated interprocedural context (A1), \minordel{imprecise source–sink identification}\minoradd{incorrect program-point classification} (B1), which account for \del{37.47\% (136/363)}\add{14.60\% (861 / 5,896)} and \del{19.00\% (69/363)}\add{63.33\% (3,734 / 5,896)} of the total FPs, respectively. LLM-based tools further incur errors due to hallucinated code understanding/reasoning, and imperfect adherence to prompts (Reason A2, C1, D1).
\end{findingbox}

\subsection{RQ4: Detection Overhead at Project Scale}
\label{sec:results_rq4}


\begin{table*}[htbp]
  \centering
  \caption{Computational overhead of evaluated tools on real-world projects. -- represents tools that do not rely on LLMs.}
  \label{tab:rq4_cost}
  \scriptsize
\renewcommand{\arraystretch}{0.9}
\resizebox{\textwidth}{!}{
  \begin{tabular}{@{}ll*{12}{r}@{}}
  \toprule
  \multirow{2}{*}{\textbf{Category}}
  & \multirow{2}{*}{\textbf{Method}} 
    & \multicolumn{3}{c}{\textbf{Input Tokens (K)}} 
    & \multicolumn{3}{c}{\textbf{Output Tokens (K)}} 
    & \multicolumn{3}{c}{\textbf{\add{Financial (\$)}}} 
    & \multicolumn{3}{c}{\textbf{Time (min)}} \\
  \cmidrule(lr){3-5} \cmidrule(lr){6-8} \cmidrule(lr){9-11} \cmidrule(lr){12-14}
   & & Min & Max & Avg 
    & Min & Max & Avg
    & \add{Min} & \add{Max} & \add{Avg}
    & Min & Max & Avg \\
  \midrule
  \multirow{10}{*}{\makecell{\add{Specialized}\\\add{LLM-based}}}
  &\repoaudit{}\add{$_{\claude{}\sonnett{}}$}
    & 0.00 & 225{,}493.97 & 32{,}634.78 
    & 0.00 & 38{,}078.26 & 4{,}991.27
    & \add{0.00}    & \add{1,247.66}           & \add{172.77}
    & 3.32 & 312.87 & 65.40 \\
  &\add{\repoaudit{}$_{\openai{}\gptfive{}}$}
    & \add{0.00} & \add{143{,}791.47} & \add{21{,}118.02}
    & \add{0.00} & \add{13{,}799.98} & \add{2{,}157.24}
    & \add{0.00}    & \add{566.48}           & \add{85.15}
    & \add{2.77}	& \add{894.63} &	\add{196.71} \\
  &\knighter{}\add{$_{\openai{}\othree{}}$}
    & 61.41 & 185.25 & 124.73 
    & 16.48 & 47.92 & 36.55
    & \add{0.14}    & \add{0.41}           & \add{0.34}
    & 37.55 & 221.51 & 144.09 \\
  &\add{\knighter{}$_{\openai{}\gptfive{}}$}
    &  \add{103.31} &  \add{201.95} &  \add{165.71}
    &  \add{27.39} &  \add{373.73} &  \add{151.72}
    & \add{0.67}    & \add{6.09}           & \add{2.36}
    &  \add{35.20} &  \add{232.57} &  \add{104.57} \\
  &\iris{}\add{$_{\openai{}\gptfour{}}$}
    & 8.47 & 1{,}752.77 & 508.77 
    & 1.45 & 267.13 & 103.60
   & \add{0.04}    & \add{8.39}           & \add{2.59}
    & 2.55 & 2{,}089.47 & 213.44 \\
  &\add{\iris{}$_{\openai{}\gptfive{}}$}
    & \add{19.40}& \add{1{,}011.07} & \add{310.73} 
    & \add{3.17} & \add{274.94} &\add{83.15} 
    & \add{0.10}    & \add{5.82}           & \add{2.02}
    & \add{4.72} & \add{122.38} & \add{35.69} \\
  &\llmdfa{}\add{$_{\openai{}\gptfour{}}$}
    & 2{,}535.12 & 38{,}310.33 & 17{,}445.71
    & 69.50 & 2{,}646.89 & 1{,}199.05 
    & \add{7.91}    & \add{135.48}           & \add{61.60}
    & 44.00 & 4{,}638.00 & 2{,}000.00 \\
 & \add{\llmdfa{}$_{\openai{}\gptfive{}}$}
    & \add{2{,}565.84}    & \add{28{,}561.13}           & \add{16{,}269.34}
    & \add{52.87}    & \add{2{,}600.17}           & \add{1{,}174.38}
    & \add{9.54}    & \add{110.41}           & \add{58.29}
    & \add{57.58}    & \add{3{,}460.48}           & \add{1{,}504.01} \\
  &\inferroi{}\add{$_{\openai{}\gptfour{}}$}
    & 142.50 & 2{,}871.49 & 1{,}193.45
    & 64.51 & 1{,}408.23 & 575.84
    & \add{1.32}    & \add{28.30}           & \add{11.62}
    & 20.25 & 176.27 & 78.10 \\
  &\add{\inferroi{}$_{\openai{}\gptfive{}}$}
    & \add{153.61}    & \add{2{,}933.26}           & \add{1{,}230.17}
    & \add{82.95}    & \add{1{,}565.81}           & \add{660.73}
    & \add{1.63}    & \add{30.82}           & \add{12.99}
    & \add{12.57}    & \add{97.14}           & \add{43.74} \\
 \midrule
  \multirow{2}{*}{\add{General-Agents}}
  &\add{\codex{}$_{\openai{}\gptfive{}}$}
    & \add{359.25}    & \add{5{,}999.23}           & \add{1{,}958.57}
    & \add{13.26}    & \add{79.99}           & \add{32.39}
    & \add{1.21}    & \add{16.20}           & \add{5.38}
    & \add{5.22}    & \add{32.70}           & \add{14.80} \\
 & \add{\cc{}$_{\claude{}\sonnetfourr{}}$}
    & \add{72.41}    & \add{14{,}620.30}           & \add{3{,}043.48}
    & \add{3.56}    & \add{127.44}           & \add{36.18}
    & \add{0.40}    & \add{45.48}           & \add{9.67}
    & \add{1.94}    & \add{49.54}           & \add{13.19} \\
 \midrule
  \multirow{4}{*}{\add{Traditional}}&\codeql{}
    & -- & -- & --
    & -- & -- & --
    & -- & -- & --
    & 0.21 & 378.95 & 21.29 \\
 & \semgrep{} 
    & -- & -- & --
    & -- & -- & --
    & -- & -- & --
    & 0.07 & 2.43 & 0.64 \\
  &\add{\csa{}}
    & -- & -- & --
    & -- & -- & --
    & -- & -- & --
    & \add{0.25} & \add{123.05} & \add{60.92} \\
 & \add{\spotbugs{}}
    & -- & -- & --
    & -- & -- & --
    & -- & -- & --
    & \add{0.72} & \add{10.90} & \add{3.11} \\
  \bottomrule
  \end{tabular}
  }
\end{table*}

\begin{figure*}[htbp]
    \centering
    \includegraphics[width=0.9\linewidth]{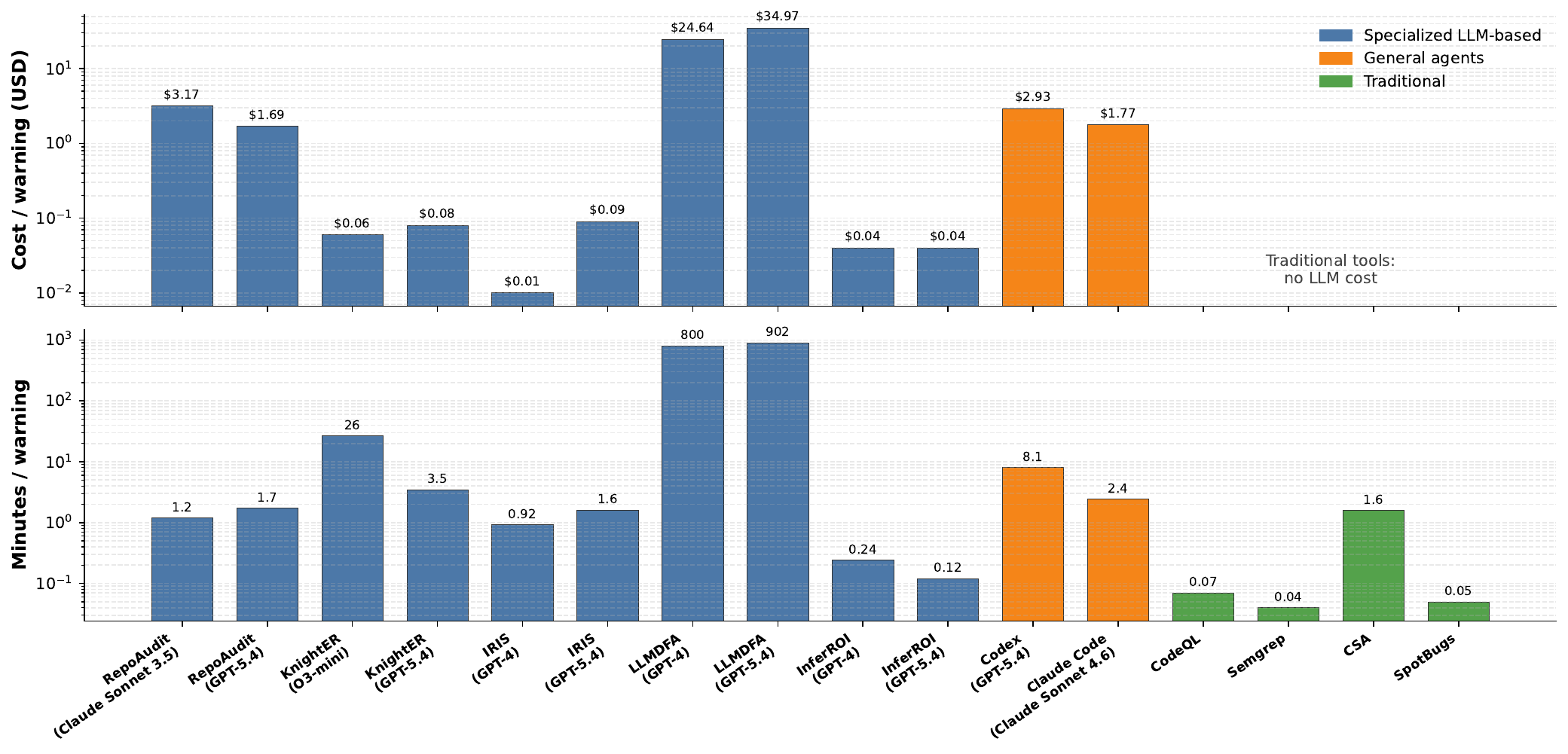}
    \caption{\add{Financial and time cost per warning for each tool on real-world projects/modules.}}
    \label{fig:warning_level_cost}
\end{figure*}

One primary concern for LLM-based methods is their cost and efficiency. In this research question, we measure the token consumption and time cost of the selected tools. Table~\ref{tab:rq4_cost} reports the average overhead of these methods, \add{Fig.~\ref{fig:warning_level_cost} further reports their warning-level cost,} and more detailed statistics are provided in 
Appendix~C.
Overall, \add{compared to traditional tools, }LLM-based vulnerability detection methods impose \textbf{substantial computational costs, raising practical concerns for project-scale adoption.} \add{Specifically,} \del{Among all LLM-based methods}\add{among specialized LLM-based methods, multi-agent methods such as} \repoaudit{} and \llmdfa{} exhibit the highest computational overhead on C/C++ and Java projects\del{, respectively}. \repoaudit{} can consume more than \textbf{225 million} input tokens and more than \textbf{38 million} output tokens for a single project, resulting in an \add{average cost of \$172.77 and a} maximum analysis time of over \textbf{\del{448}\add{800} minutes} (more than \del{7}\add{13} hours) per project. Similarly, \llmdfa{} also incurs extremely high token consumption, up to \textbf{38 million} input tokens, and requires \add{\$61.60 and }\textbf{2000 minutes} (more than 33 hours) on average to analyze a single Java project. These results indicate that current LLM-based approaches face a severe scalability challenge when applied to large real-world repositories. \add{At the warning level, \llmdfa{} remains the most expensive method, requiring more than \textbf{\$24} and over \textbf{800 minutes} per warning under both model settings. These results indicate that current LLM-based approaches face a severe scalability challenge when applied to large real-world repositories/modules.}

Hybrid approaches such as \iris{} and \inferroi{}, which integrate LLMs with static analysis tools or methods, show more moderate overhead, though still requiring hundreds of thousands to millions of tokens\add{, incur non-trivial financial costs,} and \del{multi-hour runtimes for}\add{can take multiple hours on} large projects. \add{General-agent methods also incur visible warning-level overhead: \codex{} costs \textbf{\$2.93} and \textbf{8.1 minutes} per warning, while \cc{} costs \textbf{\$1.77} and \textbf{2.4 minutes} per warning. } These results indicate that current LLM-based vulnerability detection methods remain computationally expensive, and their scalability is highly sensitive to prompting strategies and detection workflows, posing practical challenges for project scale adoption. \minoradd{Backbone selection offers another opportunity to reduce costs. For example, lower-priced models such as \dsfourflash{}~\cite{deepseek-v4-flash} could reduce API charges, but their overall cost-effectiveness requires evaluating token consumption, runtime, and detection effectiveness together.}

In contrast, traditional tools (\codeql{}\del{and}\add{,} \semgrep{}\add{, \csa{}, and \spotbugs{}}) incur negligible overhead, completing analysis within minutes to hours. This difference emphasizes a significant scalability gap: while LLM-based methods provide stronger semantic reasoning capabilities, they do so at the cost of significantly higher computational resources. These results suggest that improving cost efficiency, through better context pruning, incremental analysis, or lighter-weight LLM pipelines, remains essential before LLM-based vulnerability detectors can be deployed reliably at the project scale.

\begin{findingbox}
\textbf{Finding \del{4}\add{5}:} Project-scale LLM-based vulnerability detection is computationally expensive and time-consuming. A single project can require up to hundreds of millions of tokens and a multi-day running time, making scalability a critical bottleneck.
\end{findingbox}

\section{Discussion}
\label{sec:discussion}

\subsection{Implications}
\label{sec:discussion_implication}

Our comprehensive analysis across five \add{specialized} LLM-based \add{methods, two general agents,}\del{vulnerability detectors} and \del{two}\add{four} traditional tools reveals that, although LLMs offer promising semantic reasoning capabilities and the potential to generalize beyond manually modeled security patterns, they remain far from mature enough for reliable vulnerability analysis in real-world scenarios. Despite demonstrating an ability to infer higher-level intentions that traditional static tools often miss, current specialized LLM-based methods suffer from several practical limitations that restrict their effectiveness:

\noindent\textbf{\add{(1) Workflow choices create distinct trade-offs.}}

\add{Our results show that different LLM-based workflows exhibit different strengths and limitations. Specialized methods decompose detection into structured subtasks, such as source/sink inference, path extraction, checker generation, or path validation. However, even with the same backbone model, they may underperform general-agent methods, suggesting that their task formulation can constrain how fully the model's reasoning capability is used. For example, \repoaudit{} mainly uses the LLM to extract path information and then applies fixed logic for vulnerability judgment, so path-extraction errors directly affect the final result. General-agent methods inspect project context more flexibly and achieve stronger recall, but they still struggle with complex files and large code contexts. Future detectors should therefore combine the flexibility of general agents with specialized program-analysis support, such as explicit path constraints, intermediate representations, and validation modules that check and revise LLM-generated paths or judgments.}

\noindent \textbf{(\del{1}\add{2}) Dataflow reasoning is shallow and incomplete.} 

Both general agent-based and specialized LLM-based methods typically follow a source-first detection workflow: they first search for potential source program points and then initiate subsequent analysis from these identified locations. Consequently, risk is often inferred from local evidence without the global execution context, causing models to overlook sanitization or correct sink behavior that occurs further along the program path. Shallow dataflow is therefore a fundamental bottleneck: when the full path from ``input $\rightarrow$ propagation $\rightarrow$ output'' is not analyzed holistically, real vulnerabilities may be missed, while safe code is frequently misclassified as risky. \add{One direct mitigation is to add a post-processing stage that further validates generated warnings, as in approaches such as LLM4PFA~\cite{du2025minimizing} and BugLens~\cite{li2025towards}. Such methods can eliminate some false positives, but they mostly focus on branch-condition feasibility and path reachability, while providing limited modeling of program values during execution. A more promising direction is to integrate LLMs with formal analysis during detection. For example, LLMs could be used to infer constraints over program value spaces, while symbolic execution or other formal techniques check vulnerability reachability under global execution context.}

\noindent \textbf{(\del{2}\add{3}) \minordel{Imprecise source-sink identification.}\minoradd{Incomplete or incorrect API modeling.}} Based on our experiments, a considerable portion of FNs and FPs is due to the incomplete and incorrect modeling of project-specific APIs. In real-world projects, \minordel{sources may be wrapped or abstracted behind project-specific APIs, propagate through multiple layers before reaching sinks, or be triggered via reflection, callbacks, and framework-driven lifecycle events}\minoradd{vulnerability-relevant operations may be hidden behind wrappers, callbacks, reflection, or framework-managed lifecycles}. Although several LLM-based approaches (\iris{}, \llmdfa{}) attempt to infer project-specific or vulnerability-specific sources/sinks via few-shot prompting, the inferred results are often unreliable in practice (e.g., \iris{} mislabels \texttt{Map.put()} as a sink in Java projects, resulting in an amount of FPs). Thus, robust and context-aware \minordel{source-sink}\minoradd{API-role} inference remains an unresolved challenge. \add{Future work should move from few-shot \minordel{source/sink}\minoradd{API-role} prediction toward evidence-grounded specification inference. One direction is to combine LLMs with program-analysis signals, data-dependence relations, API documentation, and usage patterns, \minordel{to distinguish true security-relevant sources/sinks from benign helper APIs}\minoradd{to infer vulnerability-relevant API roles and usage requirements. For example, SpecAuditor~\cite{lin2026specauditor} extracts audit specifications from historical patches and extends them to semantically related entities using documentation.} Another direction is to introduce iterative validation: inferred \minordel{source/sink}\minoradd{API} specifications should be checked against observed dataflows, known sanitizers, and project-specific API contracts before being used for detection. \minoradd{For example, agentic property-based testing~\cite{maaz2025agentic} infers properties from code and documentation, generates executable tests, and analyzes test failures to assess potential bugs.} Building reusable project-level \minordel{source/sink}\minoradd{API} summaries could further reduce repeated inference errors across related modules and versions.}

\noindent \textbf{(\del{3}\add{4}) Semantic understanding in complex contexts remains limited.} As shown in Section~\ref{sec:results_rq3}, a substantial portion of false positives produced by LLM-based vulnerability detectors stems from the misinterpretation of code semantics. In many cases, models overlook key program points along a dataflow (e.g., a sanitizer, a deferred release, a null-check, or a conditional exit), or fail to infer the actual execution behavior of code due to language-specific constructs such as try-with-resources, destructors, template expansion, or macro-controlled branching. These semantic gaps often cause LLMs to reason along incorrect control paths or assume unsafe behavior where none exists, ultimately misclassifying benign code as vulnerable. We suggest future work strengthen semantic reasoning through hybrid workflows, augmenting LLMs with intermediate program representations, graph-based context retrieval, and static analysis signals to provide more complete execution context~\cite{dhulshette2025hierarchical, lu2024grace}. \add{In particular, detectors should explicitly model key semantic events, such as ownership transfer, sanitizer effects, error-handling paths, and resource lifecycle operations, rather than relying only on natural-language reasoning over code snippets.}

\noindent \textbf{(\del{4}\add{5}) Prompt-alignment and task compliance are inconsistent.} In our evaluation, we observe that methods such as \minordel{\repoaudit{}}\minoradd{\codex{}} and \iris{}, which rely on LLMs to perform reasoning over extracted dataflows, may still fail to adhere to the intended vulnerability definition or CWE scope in the prompt. Even when the reasoning is correct at a syntactic or control-flow level, the model occasionally drifts from the task objective, producing conclusions that do not align with the vulnerability type under detection—reflecting semantic drift and task-misalignment. To improve this, future systems may require structured multi-agent collaboration, self-verification modules, or constraint-guided reasoning to enforce consistent detection goals~\cite{bouzenia2024repairagent, hong2024closer}.

\noindent \textbf{(\del{5}\add{6}) Scalability remains \minordel{the}\minoradd{a} bottleneck.} Project-scale analysis demands enormous computational overhead. As shown in Section~\ref{sec:results_rq4}, \repoaudit{} and \llmdfa{} consume up to hundreds of millions of tokens and require hours to days to complete a single project, making them unsuitable for continuous integration. This observation aligns with both the authors' discussions in their papers and the feedback we received from them, who confirmed that the current implementations may incur substantial computational overhead when analyzing large projects. Hybrid approaches (\iris{}, \inferroi{}) reduce, but do not eliminate, the high cost. \del{To achieve scalable detection, future work may adopt hierarchical analysis, reuse intermediate reasoning through caching, and minimize redundant LLM calls}\add{Future work should prioritize cost-aware detection designs. Promising directions include hierarchical analysis, caching and reusing intermediate reasoning results, incremental analysis across code revisions, and deduplicating similar path queries to avoid redundant LLM calls~\cite{dhulshette2025hierarchical}. Another direction is adaptive invocation, where expensive LLM reasoning is reserved for uncertain or high-risk candidates, while simpler cases are handled by lightweight static checks.} \minoradd{Alongside these workflow optimizations, adopting LLMs with efficient architectures and lower inference costs, such as \dsfourflash{}, could further reduce monetary costs. Their effects on token consumption, runtime, and detection effectiveness require further evaluation.}

\subsection{Threats to Validity}
\label{sec:discussion_threats}

\noindent \textbf{Internal Validity.} \minoradd{Our controlled evaluation focuses on labeled vulnerabilities, so warnings about other genuine vulnerabilities may be counted as FPs. Due to the substantial evaluation cost, we ran each LLM-based configuration once. Although some tools' performance (e.g., \iris{} and \inferroi{}) is close to that reported in their original studies, LLM nondeterminism may still affect our results.} Our analysis involves expert judgment. Although two independent authors conducted labeling and reconciliation to reduce subjectivity (Section~\ref{sec:results_rq2} and \ref{sec:results_rq3}), classification accuracy may still be influenced by personal experience, interpretation of code semantics, or reasoning about exploitability. Misreading or misinterpretation of complex code flows may introduce bias into the taxonomy of FP causes.\minordel{The strength of evidence supporting our root-cause findings also varies. For example, we directly evaluate the impact of source/sink mismatch through a dedicated ablation study, whereas other findings rely primarily on manual and session analysis. Regarding data-leakage issues, for experiments on the in-house dataset, since these vulnerabilities have public CVE identifiers, the evaluated LLMs may have been exposed to related descriptions, patches, or discussions during pretraining. To mitigate explicit leakage, we retained and manually inspected the LLM sessions and found no evidence that the tools used CVE identifiers, vulnerability descriptions, patches, or other ground-truth information during detection. For experiments on the real-world dataset, the tools were run directly on the latest project versions. Although this setting reduces explicit leakage risk, and our session inspection found no evidence that the tools used repository-specific issue reports, CVE records, patches, or other ground-truth information, it still cannot guarantee the absence of pretraining contamination.} Moreover, our large-scale warning classification relies partly on \codex{}-generated labels. Although stratified manual validation shows high accuracy, these labels may still contain errors or model biases. To overcome these threats, the two authors adopt an inductive open-coding procedure with double coding~\cite{denzin2011sage}, independently validated sampled cases, and resolved any disagreements through discussion. We also released all experimental artifacts, including evaluation scripts, prompts, taxonomy labels, and detailed statistics for external inspection and replication.

\noindent \minoradd{Another concern is about data-leakage. Pretraining contamination remains possible for both datasets used in this paper. For the controlled dataset, LLMs may have encountered public vulnerability descriptions, patches, or discussions during training. For the real-world dataset, 21 of the 24 project/module entries use commits postdating the reported August 2025 knowledge cutoffs of \gptfive{}~\cite{gpt-5.4} and \sonnetfour{}~\cite{claude46}. However, these versions may retain older code, and knowledge cutoffs do not guarantee exclusion from training. Although pretraining contamination cannot be ruled out, we manually inspected the LLM sessions and found no evidence of explicit reliance on CVE records, vulnerability descriptions, patches, or other ground-truth information during detection.}

\noindent \textbf{External Validity.} Our evaluation considers 5 \add{specialized }LLM-based \add{methods, 2 general agents, }and \del{2}\add{4} traditional vulnerability detectors, selected because they are open-source\add{/publicly available}, runnable at the project scale, and representative of emerging design workflows. However, they may not fully represent all possible architectures, languages, or prompting paradigms. As the field evolves rapidly, future models or customized enterprise-scale deployments may demonstrate different capabilities. Moreover, our evaluation covers a set of real-world repositories \add{and modules} across two major programming languages (Java, C/C++). \add{Although our evaluation includes 265 known vulnerabilities and 24 real-world projects/modules, it may not capture the full diversity of software systems and programming patterns. Some C/C++ modules are Linux subsystems from the same repository and commit. While they differ in APIs, functionality, and vulnerability patterns, their shared development context may reduce the diversity represented in our evaluation, potentially limiting external validity. Moreover, the eight evaluated CWE types represent only part of the broader vulnerability landscape. \minordel{Because other vulnerability classes may require different control-flow, data-flow, API-semantic, or domain-specific reasoning capabilities, o}\minoradd{Other vulnerability classes, including concurrency vulnerabilities and type confusion, are outside our scope and may require different control-flow, data-flow, or semantic analyses. O}ur findings may not generalize uniformly to them.} \del{, but }\add{Finally, our findings} may not generalize uniformly to systems written in Rust, Go, TypeScript, or large multi-language projects. \minoradd{Future work should broaden the evaluation across these settings, adapting detection methods where necessary.}
\section{Conclusion}
\label{sec:conclusion}

In this paper, we conduct the first comprehensive study of specialized LLM-based vulnerability detectors at the project scale, \del{analyzing 5 representative LLM-based tools and comparing them against}\add{evaluating five specialized LLM-based methods, two general-purpose agents, and} \del{2}\add{4} traditional static analysis tools. \del{Our evaluation begins with 219 known vulnerabilities spanning 8 CWE types, where we observe that current approaches often fail to identify these vulnerabilities, primarily due to inaccurate source-sink identification. We then scale our analysis to 24 open-source projects and manually examine 385 sampled reports. From this investigation, we construct a taxonomy of false positive causes and quantify how different tools are affected. Our findings reveal that although LLM-based methods exhibit stronger performance than traditional static analyzers, they still face fundamental challenges in dataflow construction, source-sink identification, and deep semantic understanding. Furthermore, our runtime and token usage measurements show that scalability remains a core bottleneck, especially for end-to-end LLM workflows. Finally, to support future progress in this domain, we distill our observations into 5 implications for the research community. We hope this work can inspire future designs that make automated security detection both more accurate and more practical in real-world scenarios.}\add{Our evaluation covers 265 known vulnerabilities across eight CWE types and 24 open-source projects/modules. On the \minordel{in-house}\minoradd{controlled} benchmark, general-purpose agents achieve the highest recall, whereas specialized LLM-based detectors exhibit limited recall but provide partial complementary coverage. Our API-guided experiment in Section~\ref{sec:results_rq1} further confirms that \minordel{inaccurate source/sink identification}\minoradd{incomplete API modeling} is a key factor limiting detection performance. On real-world projects/modules, many specialized LLM-based and traditional methods exhibit high false discovery rates, while general-purpose agents reduce warning volume but remain susceptible to false positives. Based on Codex-assisted analysis of 6,442 sampled warnings with stratified manual validation, manual examination of 355 reports, and further classification of 5,896 false positives, we construct a taxonomy that identifies \minordel{inaccurate source/sink selection}\minoradd{incorrect program-point classification}, truncated interprocedural context, and semantic misunderstanding as prominent failure patterns. Our measurements of token consumption, monetary cost, and runtime further demonstrate that scalability remains a major obstacle, particularly for end-to-end LLM workflows. Finally, we distill these findings into six implications for developing more accurate, reliable, and cost-effective project-scale vulnerability detection methods.}

%



\bibliographystyle{IEEEtran}
\bibliography{reference}

@inproceedings{
guo2025repoaudit,
title={RepoAudit: An Autonomous {LLM}-Agent for Repository-Level Code Auditing},
author={Jinyao Guo and Chengpeng Wang and Xiangzhe Xu and Zian Su and Xiangyu Zhang},
booktitle={Forty-second International Conference on Machine Learning},
year={2025},
url={https://openreview.net/forum?id=TXcifVbFpG}
}

@inproceedings{yang2025knighter,
  title={Knighter: Transforming static analysis with llm-synthesized checkers},
  author={Yang, Chenyuan and Zhao, Zijie and Xie, Zichen and Li, Haoyu and Zhang, Lingming},
  booktitle={Proceedings of the ACM SIGOPS 31st Symposium on Operating Systems Principles},
  pages={655--669},
  year={2025}
}

@inproceedings{li2025iris,
title={LLM-Assisted Static Analysis for Detecting Security Vulnerabilities},
author={Ziyang Li and Saikat Dutta and Mayur Naik},
booktitle={International Conference on Learning Representations},
year={2025},
url={https://arxiv.org/abs/2405.17238}
}

@article{wang2024llmdfa,
  title={LLMDFA: analyzing dataflow in code with large language models},
  author={Wang, Chengpeng and Zhang, Wuqi and Su, Zian and Xu, Xiangzhe and Xie, Xiaoheng and Zhang, Xiangyu},
  journal={Advances in Neural Information Processing Systems},
  volume={37},
  pages={131545--131574},
  year={2024}
}

@inproceedings{wang2025boosting,
  title={Boosting static resource leak detection via llm-based resource-oriented intention inference},
  author={Wang, Chong and Liu, Jianan and Peng, Xin and Liu, Yang and Lou, Yiling},
  booktitle={2025 IEEE/ACM 47th International Conference on Software Engineering (ICSE)},
  pages={668--668},
  year={2025},
  organization={IEEE Computer Society}
}

@misc{csa,
  author       = {{Clang and LLVM}},
  title        = {Clang Static Analyzer},
  howpublished = {\url{https://clang-analyzer.llvm.org/}},
  note         = {Accessed: 2025-10-12}
}

@inproceedings{de2008z3,
  title={Z3: An efficient SMT solver},
  author={De Moura, Leonardo and Bj{\o}rner, Nikolaj},
  booktitle={International conference on Tools and Algorithms for the Construction and Analysis of Systems},
  pages={337--340},
  year={2008},
  organization={Springer}
}

@misc{codeql2025,
  author       = {{GitHub}},
  title        = {CodeQL},
  howpublished = {\url{https://codeql.github.com/}},
  note         = {Accessed: 2025-10-13}
}

@misc{semgrep2025,
  author       = {{The Semgrep platform}},
  title        = {Semgrep},
  howpublished = {\url{https://semgrep.dev/}},
  note         = {Accessed: 2025-10-13}
}

@misc{infer2025,
  author       = {{Facebook}},
  title        = {Infer Static Analyzer},
  howpublished = {\url{https://github.com/facebook/infer}},
  note         = {Accessed: 2025-10-13}
}

@inproceedings{wang2024reposvul,
  title={Reposvul: A repository-level high-quality vulnerability dataset},
  author={Wang, Xinchen and Hu, Ruida and Gao, Cuiyun and Wen, Xin-Cheng and Chen, Yujia and Liao, Qing},
  booktitle={Proceedings of the 2024 IEEE/ACM 46th International Conference on Software Engineering: Companion Proceedings},
  pages={472--483},
  year={2024}
}

@inproceedings{liu2024jleaks,
  title={Jleaks: A featured resource leak repository collected from hundreds of open-source java projects},
  author={Liu, Tianyang and Ji, Weixing and Dong, Xiaohui and Yao, Wuhuang and Wang, Yizhuo and Liu, Hui and Peng, Haiyang and Wang, Yuxuan},
  booktitle={Proceedings of the IEEE/ACM 46th International Conference on Software Engineering},
  pages={1--13},
  year={2024}
}

@inproceedings{thapa2022transformer,
  title={Transformer-based language models for software vulnerability detection},
  author={Thapa, Chandra and Jang, Seung Ick and Ahmed, Muhammad Ejaz and Camtepe, Seyit and Pieprzyk, Josef and Nepal, Surya},
  booktitle={Proceedings of the 38th annual computer security applications conference},
  pages={481--496},
  year={2022}
}

@inproceedings{ullah2024llms,
  title={Llms cannot reliably identify and reason about security vulnerabilities (yet?): A comprehensive evaluation, framework, and benchmarks},
  author={Ullah, Saad and Han, Mingji and Pujar, Saurabh and Pearce, Hammond and Coskun, Ayse and Stringhini, Gianluca},
  booktitle={2024 IEEE symposium on security and privacy (SP)},
  pages={862--880},
  year={2024},
  organization={IEEE}
}

@article{wen2024vuleval,
  title={Vuleval: Towards repository-level evaluation of software vulnerability detection},
  author={Wen, Xin-Cheng and Wang, Xinchen and Chen, Yujia and Hu, Ruida and Lo, David and Gao, Cuiyun},
  journal={arXiv preprint arXiv:2404.15596},
  year={2024}
}

@article{gao2023far,
  title={How far have we gone in vulnerability detection using large language models},
  author={Gao, Zeyu and Wang, Hao and Zhou, Yuchen and Zhu, Wenyu and Zhang, Chao},
  journal={arXiv preprint arXiv:2311.12420},
  year={2023}
}

@inproceedings{lin2025large,
  title={From large to mammoth: A comparative evaluation of large language models in vulnerability detection},
  author={Lin, Jie and Mohaisen, David},
  booktitle={Proceedings of the 2025 Network and Distributed System Security Symposium (NDSS)},
  year={2025}
}

@inproceedings{yildiz2025benchmarking,
  title={Benchmarking llms and llm-based agents in practical vulnerability detection for code repositories},
  author={Yildiz, Alperen and Teo, Sin G and Lou, Yiling and Feng, Yebo and Wang, Chong and Divakaran, Dinil Mon},
  booktitle={Proceedings of the 63rd Annual Meeting of the Association for Computational Linguistics (Volume 1: Long Papers)},
  pages={30848--30865},
  year={2025}
}

@inproceedings{purba2023software,
  title={Software vulnerability detection using large language models},
  author={Purba, Moumita Das and Ghosh, Arpita and Radford, Benjamin J and Chu, Bill},
  booktitle={2023 IEEE 34th International Symposium on Software Reliability Engineering Workshops (ISSREW)},
  pages={112--119},
  year={2023},
  organization={IEEE}
}

@inproceedings{khare2025understanding,
  title={Understanding the effectiveness of large language models in detecting security vulnerabilities},
  author={Khare, Avishree and Dutta, Saikat and Li, Ziyang and Solko-Breslin, Alaia and Alur, Rajeev and Naik, Mayur},
  booktitle={2025 IEEE Conference on Software Testing, Verification and Validation (ICST)},
  pages={103--114},
  year={2025},
  organization={IEEE}
}

@article{steenhoek2024comprehensive,
  title={A comprehensive study of the capabilities of large language models for vulnerability detection},
  author={Steenhoek, Benjamin and Rahman, Md Mahbubur and Roy, Monoshi Kumar and Alam, Mirza Sanjida and Barr, Earl T and Le, Wei},
  journal={CoRR},
  year={2024}
}

@inproceedings{liu2024exploring,
  title={Exploring $\{$ChatGPT's$\}$ capabilities on vulnerability management},
  author={Liu, Peiyu and Liu, Junming and Fu, Lirong and Lu, Kangjie and Xia, Yifan and Zhang, Xuhong and Chen, Wenzhi and Weng, Haiqin and Ji, Shouling and Wang, Wenhai},
  booktitle={33rd USENIX Security Symposium (USENIX Security 24)},
  pages={811--828},
  year={2024}
}

@article{openai2023gpt4,
  title={GPT-4 Technical Report},
  author={ {OpenAI} and Achiam, Josh and Adler, Steven and others },
  journal={arXiv preprint arXiv:2303.08774},
  year={2023},
  url={https://arxiv.org/abs/2303.08774}
}

@misc{anthropic2024claude,
  title={The Claude 3 Model Family: Opus, Sonnet, Haiku},
  author={ {Anthropic} },
  year={2024},
  url={https://www-cdn.anthropic.com/de8ba9b01c9ab7cbabf5c33b80b7bbc618857627/Model_Card_Claude_3.pdf},
  note={Technical Report}
}

@article{rohan2023gemini,
  title={Gemini: A Family of Highly Capable Multimodal Models},
  author={Anil, Rohan and Borgeaud, Sebastian and Alayrac, Jean-Baptiste and others},
  journal={arXiv preprint arXiv:2312.11805},
  year={2023},
  url={https://arxiv.org/abs/2312.11805}
}

@misc{o3mini,
  author       = {{OpenAI}},
  title        = {O3-mini},
  howpublished = {\url{https://platform.openai.com/docs/models/o3-mini}},
  note         = {Accessed: 2025-10-13}
}

@misc{gpt-4,
  author       = {{OpenAI}},
  title        = {GPT-4},
  howpublished = {\url{https://platform.openai.com/docs/models/gpt-4}},
  note         = {Accessed: 2025-10-13}
}

@misc{deepseek-v4-flash,
  author       = {{DeepSeek}},
  title        = {DeepSeek V4 Flash},
  howpublished = {\url{https://api-docs.deepseek.com/quick_start/pricing/}},
  note         = {Accessed: 2025-10-13}
}

@misc{claude35,
  author       = {{Anthropic}},
  title        = {Claude Sonnet 3.5},
  howpublished = {\url{https://platform.claude.com/docs/en/about-claude/models/overview}},
  note         = {Accessed: 2025-10-13}
}

@misc{gpt-5.4,
  author       = {{OpenAI}},
  title        = {GPT-5.4},
  howpublished = {\url{https://platform.openai.com/docs/models/gpt-5.4}},
  note         = {Accessed: 2026-05-08}
}

@misc{claude46,
  author       = {{Anthropic}},
  title        = {Claude Sonnet 4.6},
  howpublished = {\url{https://platform.claude.com/docs/en/models/sonnet-4-6/overview}},
  note         = {Accessed: 2026-05-08}
}

@misc{cc,
  author       = {{Anthropic}},
  title        = {Claude Code},
  howpublished = {\url{https://claude.com/product/claude-code}},
  note         = {Accessed: 2026-05-08}
}

@misc{codex,
  author       = {{OpenAI}},
  title        = {Codex},
  howpublished = {\url{https://developers.openai.com/codex}},
  note         = {Accessed: 2026-05-08}
}

@misc{findsecuritybugs,
  author       = {Philippe Arteau and David Form{\'a}nek and Tom{\'a}{\v{s}} Pole{\v{s}}ovsk{\'y}},
  title        = {{OWASP Find Security Bugs}},
  howpublished = {\url{https://find-sec-bugs.github.io/}},
  note         = {SpotBugs plugin for security audits of Java web applications. Accessed: 2026-05-08}
}

@inproceedings{ahmed2024automatic,
  title={Automatic Semantic Augmentation of Language Model Prompts (for Code Summarization)},
  author={Ahmed, Toufique and Pai, Kunal Suresh and Devanbu, Premkumar and Barr, Earl T},
  booktitle={Proceedings of the 46th International Conference on Software Engineering (ICSE)},
  year={2024},
  url={https://dl.acm.org/doi/10.1145/3597503.3639183}
}

@inproceedings{brown2020language,
  title={Language Models are Few-Shot Learners},
  author={Brown, Tom and Mann, Benjamin and Ryder, Nick and Subbiah, Melanie and others},
  booktitle={Advances in Neural Information Processing Systems (NeurIPS)},
  year={2020},
  url={https://dl.acm.org/doi/abs/10.5555/3495724.3495883}
}

@inproceedings{wei2022chain,
  title={Chain-of-Thought Prompting Elicits Reasoning in Large Language Models},
  author={Wei, Jason and Wang, Xuezhi and Schuurmans, Dale and others},
  booktitle={Advances in Neural Information Processing Systems (NeurIPS)},
  year={2022},
  url={https://proceedings.neurips.cc/paper_files/paper/2022/file/9d5609613524ecf4f15af0f7b31abca4-Paper-Conference.pdf}
}

@inproceedings{wu2024autogen,
  title={AutoGen: Enabling Next-Gen LLM Applications via Multi-Agent Conversation},
  author={Wu, Qingyun and Bansal, Gagan and Zhang, Jieyu and Wu, Yiran and others},
  booktitle={Proceedings of the 12th International Conference on Learning Representations (ICLR)},
  year={2024},
  url={https://openreview.net/forum?id=tEAF9LBdgu}
}

@misc{spotbugs,
  author       = {{SpotBugs Team}},
  title        = {SpotBugs: Find Bugs in Java Programs},
  howpublished = {\url{https://spotbugs.github.io/}},
  note         = {Accessed: 2025-12-10}
}

@misc{sonarqube,
  author       = {{SonarSource}},
  title        = {SonarQube},
  howpublished = {\url{https://www.sonarqube.org/}},
  note         = {Accessed: 2025-12-10}
}

@misc{codesonar,
  author       = {{GrammaTech}},
  title        = {CodeSonar},
  howpublished = {\url{https://www.grammatech.com/codesonar-cc}},
  note         = {Accessed: 2025-12-10}
}

@misc{codeguru,
  author       = {{Amazon Web Services}},
  title        = {Amazon CodeGuru},
  howpublished = {\url{https://aws.amazon.com/codeguru/}},
  note         = {Accessed: 2025-12-10}
}

@inproceedings{yamaguchi2014modeling,
  title={Modeling and Discovering Vulnerabilities with Code Property Graphs},
  author={Yamaguchi, Fabian and Golde, Nico and Arp, Daniel and Rieck, Konrad},
  booktitle={Proceedings of the 2014 IEEE Symposium on Security and Privacy (S\&P)},
  year={2014},
  url={https://ieeexplore.ieee.org/document/6956589}
}

@article{chess2004static,
  title={Static analysis for security},
  author={Chess, Brian and McGraw, Gary},
  journal={IEEE security \& privacy},
  volume={2},
  number={6},
  pages={76--79},
  year={2004},
  publisher={IEEE}
}

@inproceedings{smith2015questions,
  title={Questions developers ask while diagnosing potential security vulnerabilities with static analysis},
  author={Smith, Justin and Johnson, Brittany and Murphy-Hill, Emerson and Chu, Bill and Lipford, Heather Richter},
  booktitle={Proceedings of the 2015 10th Joint Meeting on Foundations of Software Engineering},
  pages={248--259},
  year={2015}
}

@article{li2025automated,
  title={Automated Static Vulnerability Detection via a Holistic Neuro-symbolic Approach},
  author={Li, Penghui and Yao, Songchen and Korich, Josef Sarfati and Luo, Changhua and Yu, Jianjia and Cao, Yinzhi and Yang, Junfeng},
  journal={arXiv preprint arXiv:2504.16057},
  year={2025}
}

@article{cao2023learning,
  title={Learning to detect memory-related vulnerabilities},
  author={Cao, Sicong and Sun, Xiaobing and Bo, Lili and Wu, Rongxin and Li, Bin and Wu, Xiaoxue and Tao, Chuanqi and Zhang, Tao and Liu, Wei},
  journal={ACM Transactions on Software Engineering and Methodology},
  volume={33},
  number={2},
  pages={1--35},
  year={2023},
  publisher={ACM New York, NY}
}

@inproceedings{charoenwet2024empirical,
  title={An empirical study of static analysis tools for secure code review},
  author={Charoenwet, Wachiraphan and Thongtanunam, Patanamon and Pham, Van-Thuan and Treude, Christoph},
  booktitle={Proceedings of the 33rd ACM SIGSOFT international symposium on software testing and analysis},
  pages={691--703},
  year={2024}
}

@inproceedings{zhang2024prompt,
  title={Prompt-enhanced software vulnerability detection using chatgpt},
  author={Zhang, Chenyuan and Liu, Hao and Zeng, Jiutian and Yang, Kejing and Li, Yuhong and Li, Hui},
  booktitle={Proceedings of the 2024 IEEE/ACM 46th International Conference on Software Engineering: Companion Proceedings},
  pages={276--277},
  year={2024}
}

@inproceedings{halder2025funcvul,
  title={FuncVul: An Effective Function Level Vulnerability Detection Model using LLM and Code Chunk},
  author={Halder, Sajal and Ahmed, Muhammad Ejaz and Camtepe, Seyit},
  booktitle={European Symposium on Research in Computer Security},
  pages={166--185},
  year={2025},
  organization={Springer}
}

@article{jiang2024stagedvulbert,
  title={Stagedvulbert: Multi-granular vulnerability detection with a novel pre-trained code model},
  author={Jiang, Yuan and Zhang, Yujian and Su, Xiaohong and Treude, Christoph and Wang, Tiantian},
  journal={IEEE Transactions on Software Engineering},
  year={2024},
  publisher={IEEE}
}

@inproceedings{sun2024gptscan,
  title={Gptscan: Detecting logic vulnerabilities in smart contracts by combining gpt with program analysis},
  author={Sun, Yuqiang and Wu, Daoyuan and Xue, Yue and Liu, Han and Wang, Haijun and Xu, Zhengzi and Xie, Xiaofei and Liu, Yang},
  booktitle={Proceedings of the IEEE/ACM 46th International Conference on Software Engineering},
  pages={1--13},
  year={2024}
}

@inproceedings{fu2022linevul,
  title={Linevul: A transformer-based line-level vulnerability prediction},
  author={Fu, Michael and Tantithamthavorn, Chakkrit},
  booktitle={Proceedings of the 19th International Conference on Mining Software Repositories},
  pages={608--620},
  year={2022}
}

@article{risse2025top,
  title={Top score on the wrong exam: On benchmarking in machine learning for vulnerability detection},
  author={Risse, Niklas and Liu, Jing and B{\"o}hme, Marcel},
  journal={Proceedings of the ACM on Software Engineering},
  volume={2},
  number={ISSTA},
  pages={388--410},
  year={2025},
  publisher={ACM New York, NY, USA}
}

@inproceedings{croft2023data,
  title={Data quality for software vulnerability datasets},
  author={Croft, Roland and Babar, M Ali and Kholoosi, M Mehdi},
  booktitle={2023 IEEE/ACM 45th International Conference on Software Engineering (ICSE)},
  pages={121--133},
  year={2023},
  organization={IEEE}
}

@article{li2022sysevr,
  title={SySeVR: A Framework for Using Deep Learning to Detect Software Vulnerabilities},
  author={Li, Zhen and Zou, Deqing and Xu, Shouhuai and Ou, Xinyu and Jin, Hai and Wang, Sujuan and Deng, Zhijun},
  journal={IEEE Transactions on Dependable and Secure Computing (TDSC)},
  year={2022},
  volume={19},
  number={4},
  pages={2244--2258},
  url={https://ieeexplore.ieee.org/document/9321538}
}

@article{cheng2021deepwukong,
  title={DeepWukong: Statically Detecting Software Vulnerabilities Using Deep Graph Convolutional Neural Networks},
  author={Cheng, Xiao and Wang, Haoyu and Hua, Jiayi and Xu, Guoai and Sui, Yulei},
  journal={ACM Transactions on Software Engineering and Methodology (TOSEM)},
  year={2021},
  volume={30},
  number={3},
  pages={1--33},
  url={https://dl.acm.org/doi/10.1145/3436877}
}

@article{dong2025survey,
  title={A survey on code generation with llm-based agents},
  author={Dong, Yihong and Jiang, Xue and Qian, Jiaru and Wang, Tian and Zhang, Kechi and Jin, Zhi and Li, Ge},
  journal={arXiv preprint arXiv:2508.00083},
  year={2025}
}

@inproceedings{fu2022vulrepair,
  title={VulRepair: a T5-based automated software vulnerability repair},
  author={Fu, Michael and Tantithamthavorn, Chakkrit and Le, Trung and Nguyen, Van and Phung, Dinh},
  booktitle={Proceedings of the 30th ACM joint european software engineering conference and symposium on the foundations of software engineering},
  pages={935--947},
  year={2022}
}

@article{zhou2019devign,
  title={Devign: Effective vulnerability identification by learning comprehensive program semantics via graph neural networks},
  author={Zhou, Yaqin and Liu, Shangqing and Siow, Jingkai and Du, Xiaoning and Liu, Yang},
  journal={Advances in neural information processing systems},
  volume={32},
  year={2019}
}

@inproceedings{li2021vulnerability,
  title={Vulnerability detection with fine-grained interpretations},
  author={Li, Yi and Wang, Shaohua and Nguyen, Tien N},
  booktitle={Proceedings of the 29th ACM Joint Meeting on European Software Engineering Conference and Symposium on the Foundations of Software Engineering},
  pages={292--303},
  year={2021}
}

@inproceedings{dhulshette2025hierarchical,
  title={Hierarchical repository-level code summarization for business applications using local llms},
  author={Dhulshette, Nilesh and Shah, Sapan and Kulkarni, Vinay},
  booktitle={2025 IEEE/ACM International Workshop on Large Language Models for Code (LLM4Code)},
  pages={145--152},
  year={2025},
  organization={IEEE}
}

@article{lu2024grace,
  title={GRACE: Empowering LLM-based software vulnerability detection with graph structure and in-context learning},
  author={Lu, Guilong and Ju, Xiaolin and Chen, Xiang and Pei, Wenlong and Cai, Zhilong},
  journal={Journal of Systems and Software},
  volume={212},
  pages={112031},
  year={2024},
  publisher={Elsevier}
}

@article{bouzenia2024repairagent,
  title={Repairagent: An autonomous, llm-based agent for program repair},
  author={Bouzenia, Islem and Devanbu, Premkumar and Pradel, Michael},
  journal={arXiv preprint arXiv:2403.17134},
  year={2024}
}

@inproceedings{hong2024closer,
  title={A closer look at the self-verification abilities of large language models in logical reasoning},
  author={Hong, Ruixin and Zhang, Hongming and Pang, Xinyu and Yu, Dong and Zhang, Changshui},
  booktitle={Proceedings of the 2024 Conference of the North American Chapter of the Association for Computational Linguistics: Human Language Technologies (Volume 1: Long Papers)},
  pages={900--925},
  year={2024}
}

@article{manes2019art,
  title={The art, science, and engineering of fuzzing: A survey},
  author={Man{\`e}s, Valentin JM and Han, HyungSeok and Han, Choongwoo and Cha, Sang Kil and Egele, Manuel and Schwartz, Edward J and Woo, Maverick},
  journal={IEEE Transactions on Software Engineering},
  volume={47},
  number={11},
  pages={2312--2331},
  year={2019},
  publisher={IEEE}
}

@inproceedings{livshits2005finding,
  title={Finding Security Vulnerabilities in Java Applications with Static Analysis.},
  author={Livshits, V Benjamin and Lam, Monica S},
  booktitle={USENIX security symposium},
  volume={14},
  pages={18--18},
  year={2005}
}

@misc{homepage,
  author = {{Homepage}},
  year = {2025},
  note = {Available at: \url{https://github.com/Feng-Jay/LLM4Security}},
}

@inproceedings{lekssays2025llmxcpg,
  title={$\{$LLMxCPG$\}$:$\{$Context-Aware$\}$ Vulnerability Detection Through Code Property $\{$Graph-Guided$\}$ Large Language Models},
  author={Lekssays, Ahmed and Mouhcine, Hamza and Tran, Khang and Yu, Ting and Khalil, Issa},
  booktitle={34th USENIX Security Symposium (USENIX Security 25)},
  pages={489--507},
  year={2025}
}

@book{denzin2011sage,
  title={The Sage handbook of qualitative research},
  author={Denzin, Norman K and Lincoln, Yvonna S},
  year={2011},
  publisher={sage}
}

@article{du2025minimizing,
  title={Minimizing False Positives in Static Bug Detection via LLM-Enhanced Path Feasibility Analysis},
  author={Du, Xueying and Yu, Kai and Wang, Chong and Zou, Yi and Deng, Wentai and Ou, Zuoyu and Peng, Xin and Zhang, Lingming and Lou, Yiling},
  journal={arXiv preprint arXiv:2506.10322},
  year={2025}
}

@inproceedings{li2025towards,
  title={Towards More Accurate Static Analysis for Taint-Style Bug Detection in Linux Kernel},
  author={Li, Haonan and Zhang, Hang and Pei, Kexin and Qian, Zhiyun},
  booktitle={2025 40th IEEE/ACM International Conference on Automated Software Engineering (ASE)},
  pages={380--392},
  year={2025},
  organization={IEEE}
}

@article{du2024vul,
  title={Vul-rag: Enhancing llm-based vulnerability detection via knowledge-level rag},
  author={Du, Xueying and Zheng, Geng and Wang, Kaixin and Zou, Yi and Wang, Yujia and Deng, Wentai and Feng, Jiayi and Liu, Mingwei and Chen, Bihuan and Peng, Xin and others},
  journal={ACM Transactions on Software Engineering and Methodology},
  year={2024},
  publisher={ACM New York, NY}
}

@article{yang2025dlap,
  title={Dlap: A deep learning augmented large language model prompting framework for software vulnerability detection},
  author={Yang, Yanjing and Zhou, Xin and Mao, Runfeng and Xu, Jinwei and Yang, Lanxin and Zhang, Yu and Shen, Haifeng and Zhang, He},
  journal={Journal of Systems and Software},
  volume={219},
  pages={112234},
  year={2025},
  publisher={Elsevier}
}

@inproceedings{lin2026specauditor,
  title={SpecAuditor: Generating Audit Specifications for LLM-Driven Bug Detection},
  author={Lin, Miaoqian and Chen, Hao},
  booktitle={2026 IEEE Symposium on Security and Privacy (SP)},
  pages={3396--3413},
  year={2026},
  organization={IEEE}
}

@article{maaz2025agentic,
  title={Agentic property-based testing: Finding bugs across the python ecosystem},
  author={Maaz, Muhammad and DeVoe, Liam and Hatfield-Dodds, Zac and Carlini, Nicholas},
  journal={arXiv preprint arXiv:2510.09907},
  year={2025}
}

@book{cochran1977sampling,
  title={Sampling techniques},
  author={Cochran, William Gemmell and Cochran, William G},
  volume={3},
  number={4},
  year={1977},
  publisher={wiley New York}
}
\clearpage
\setcounter{page}{1}
\end{document}